\documentclass[preprint,authoryear,1p,11pt,8.5x11]{elsarticle}

\usepackage{deluxetable}
\makeatletter
\def\ps@pprintTitle{%
 \let\@oddhead\@empty
 \let\@evenhead\@empty
 \def\@oddfoot{}%
 \let\@evenfoot\@oddfoot}
\makeatother

\newcommand{\simlt}{\lower.5ex\hbox{$\; \buildrel < \over \sim \;$}}

\begin{document}

\begin{frontmatter}
\title{Survey of cometary CO$_2$, CO,  and particulate emissions using the {\it Spitzer} Space Telescope}

\author{William T. Reach}
\address{Stratospheric Observatory for Infrared Astronomy, Universities Space Research Association, MS 232-12, NASA/Ames Research Center, Moffett Field, CA 94035}
\ead{wreach@sofia.usra.edu}

\author {Michael S. Kelley}
\address{Department of Astronomy, University of Maryland, College Park, MD 20742}

\author{Jeremie Vaubaillon}
\address{Institut de M\'ecanique C\'eleste et de Calcul des \'Eph\'em\'erides, 77 avenue Denfert-Rochereau, 75014 Paris, France}

\begin{abstract}

We surveyed 23 comets using the Infrared Array Camera on the {\it Spitzer} Space Telescope in wide filters centered at 3.6 and 4.5 $\mu$m.
Emission in the  3.6 $\mu$m filter arises from sunlight scattered by dust grains; the 3.6 $\mu$m images generally have a coma near the nucleus and a tail in the antisolar direction 
due to dust grains swept back by solar radiation pressure.
The 4.5 $\mu$m filter contains scattered sunlight by, and thermal emission from, the same dust grains, as well as strong emission lines 
from CO$_2$ and CO gas. The 4.5 $\mu$m images are often much brighter than could be explained by dust grains, and they show sometimes distinct
morphologies, in which cases we infer they are dominated by gas.
Based on the ratio of 4.5 to 3.6 $\mu$m brightness, we classify the survey comets as CO$_2$+CO `rich' and `poor'.
This classification is correlated
with previous classifications by A'Hearn based on carbon-chain molecule abundance, in the sense that comets classified as `depleted' in carbon-chain
molecules are also `poor' in CO$_2$+CO.
The gas emission in the IRAC 4.5 $\mu$m images is characterized by a smooth morphology, typically a fan in the sunward hemisphere with a radial profile that varies approximately as the inverse of 
projected distance
from the nucleus, as would apply for constant production and free expansion.
There are very significant radial and azimuthal enhancements in many of the comets, and these are often distinct between the gas and dust, indicating 
that ejection of solid material may be driven either by H$_2$O or CO$_2$.
Notable features in the images include the following. There is a prominent loop of gas emission from 103P/Hartley 2, offset toward the sunward direction; the loop could be due to an outburst of CO$_2$ before the {\it Spitzer} image.
Prominent, double jets are present in the image of 88P/Howell, with one directed nearly toward the Sun and the other closer to the terminator (but still on the daytime hemisphere).
A prominent single jet is evident for C/2002 T7 (LINEAR), 22P/Kopff and 81P/Wild 2.
Spirals are apparent in 29P/Schwassmann-Wachmann 1 and C/2006 W3 (Christensen); we measure a rotation rate of 21 hr for the latter comet.
Arcs (possibly parts of a spiral) are apparent in the images of 10P/Tempel 2, and 2P/Encke.
\end{abstract}


\end{frontmatter}

\section{Introduction}

Comets are the leftover building blocks from formation of the planets beyond the frost line in the early Solar Nebula. Cometary materials comprise the cores of the ice giant planets Uranus and Neptune, and the organic- and ice-rich late veneer of planets and satellites throughout the Solar System \citep{ChybaSagan}. Comets from the Jupiter-Saturn region were ejected into the distant Oort cloud (or into interstellar space), and those from Neptune to the edge of the nebula reside in the present-day Kuiper belt \citep{wiegerttremaine}. Observations of comets provide a unique opportunity to remotely sample primitive outer Solar System material. The only true sample return from a comet, the NASA {\it Stardust} mission, retrieved relatively robust refractory solid material from comet 81P/Wild 2, but it could not retrieve the more delicate and volatile materials that drive cometary activity and constitute much a cometÕs mass. The Rosetta spacecraft en route to comet 67P/Churyumov-Gerasimenko will measure properties of that comet directly, upon landing on its surface in 2014. To study a wide range of comets, and sample properties of the early solar nebula at a range of distances from the Sun, we must still rely on telescopic observations. Observing opportunities for comets are frequent, because gravitational perturbations (by passing stars, for Oort cloud comets, or by the outer planets, for Kuiper belt comets) send a flotilla of comets into the inner Solar System.

Comets are composed primarily of volatile ices and refractory solids, in comparable amounts by mass. Despite the seminal work by Whipple in explaining cometary activity based on the sublimation of water ice, many of the observed properties of cometary activity and composition remain unknown. The presence of water ice is well established through observations of OH. But water ice does not sublimate readily outside 2.5 AU from the Sun, while cometary activity is common at much greater distances, even at 20 AU \citep{meech04}. Other molecules may dominate the activity of some comets. Observations of 9P/Tempel 1 by the Deep Impact spacecraft revealed a CO$_2$ active region on the surface of that comet.  \citet{feaga07} found that the particulate (dust) production of the CO$_2$-enhanced active area was equivalent to the H$_2$O-enhanced active area, even though overall the production rate of water was greater than the production rate of CO$_2$.  These observations were made when the comet was 1.5 AU from the Sun, where conventional knowledge predicts H$_2$O outgassing would dominate dust production.  Thus CO$_2$ could be an important species for levitating solid material from this comet even close to the Sun where H$_2$O sublimation is predicted to dominate. 

The prevailing model for production and distribution of cometary ices defines two types of observable cometary species \citep{festou81}. Parent species are those of which the comet is actually composed. On cosmochemical grounds, abundant parent species are likely to include H$_2$O, CO$_2$, H$_2$CO, CO, CH$_4$, as well as more complicated molecules. Upon sublimation, the vapor is exposed to the solar radiation field and survives until dissociated; daughter species are the dissociation products. Most observable cometary molecules are daughter species. Examples include OH, measurement of which is a proxy for the most abundant species, H$_2$O. Observations of the most abundant parent molecules provide direct measures of the composition of comets, but such measurements are rare. The observations described in this paper will address two abundant species, CO$_2$ and CO, estimated to be the 2nd and 3rd most important species in terms of abundance and ability to drive cometary activity (especially at great distances from the Sun).
Measurements of parent species are important because they directly trace the abundant species and they avoid the model dependence required to interpret daughter species. A growing number of results show important deviations from the standard model. The 1P/Halley encounter missions in 1986 and the C/Hale-Bopp apparition in 1995 provided some surprising results. The Neutral Mass Spectrometer on Giotto demonstrated that CO production from Halley is widespread in the coma, requiring an extended source rather than ejection from the nucleus \citep{eberhardt99}. The extended source could not be identified but was suggested to be photodissociation of H$_2$CO, which was itself produced by an extended source, suggested to be a polymerized form on grains. A far-ultraviolet sounding rocket experiment showed that atomic carbon was also inconsistent with the standard comet model, leading to a `carbon puzzle,' requiring an extended source of `identity unknown' for P/Halley \citep{woods87}; however, this result could also be explained without an extended source \citep{rubin11}. 
The subject has been reviewed by \citet{festou99}. For comet C/Hale-Bopp, the radial profile of OCS relative to H$_2$O indicated an extended emission source for OCS \citep{dellorusso98}. Based on evidence to date, it appears that CO may be both a parent (present not he nucleus) and daughter molecule. 

\begin{figure}
\includegraphics[width=5in]{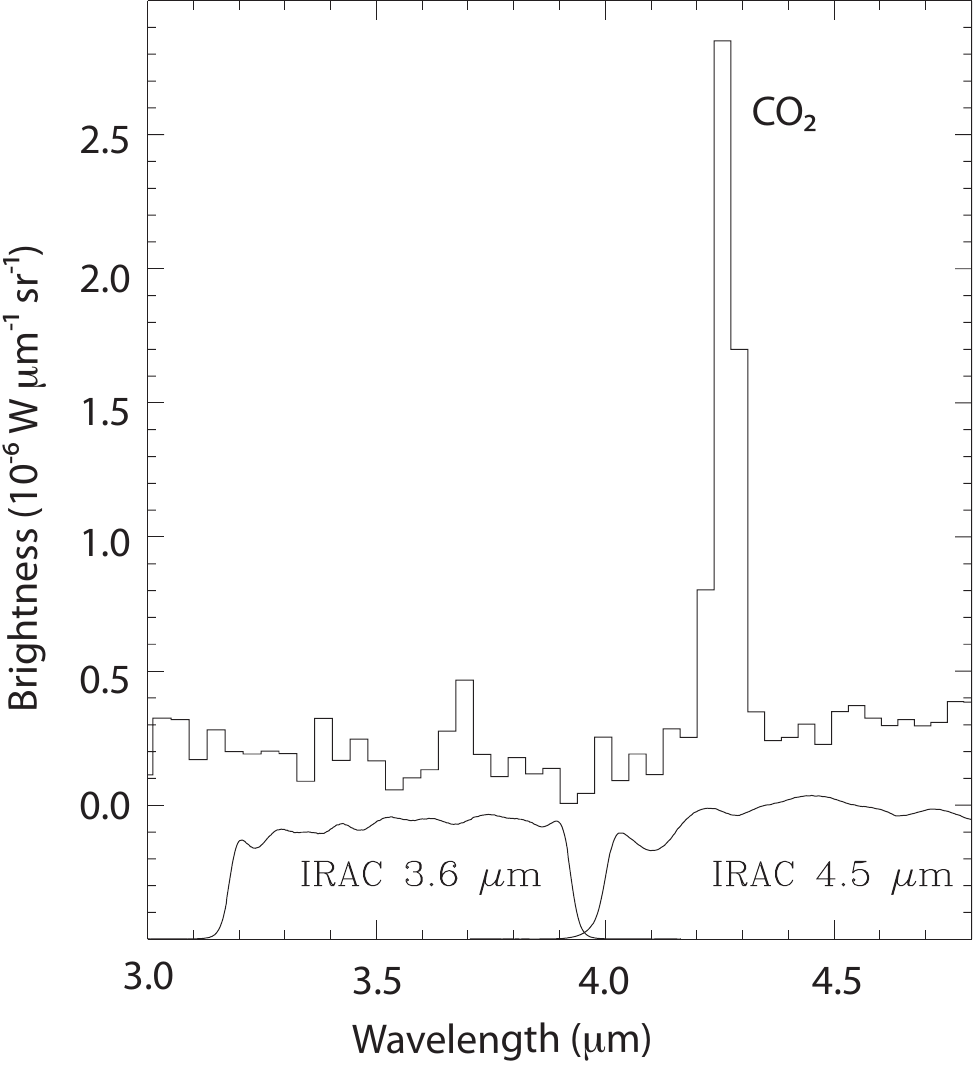}
\caption{\label{hartley2spec}
Spectrum of comet 103P/Hartley 2 obtained with ISOPHOT-S on
the {\it Infrared Space Observatory} \citep{colangeli103P}. The passbands
of the two short-wavelength IRAC channels are shown for comparison. The spectral response for the IRAC 4.5 $\mu$m band ends at 5 $\mu$m (off the right-hand side of the plot).
It is evident that for this comet, the strong CO$_2$ band at 4.3 $\mu$m would dominate the observed brightness in the IRAC 4.5 $\mu$m channel.
}
\end{figure}

Before the {\it Akari} mission \citep{akarimission}, very little was known about gas emission in the mid-infrared spectral range.
Near-infrared images of comet Hale-Bopp in the CO band failed to reveal the expected bright, centrally-peaked CO emission for such an abundant parent species, suggesting CO production is very low or from a very extended source \citep{santossanz97}. The CO and CO$_2$ emission was very bright in the 2.5-5 $\mu$m spectra taken with ISO, indicating a very high CO$_2$ production rate\citep{crovisierHaleBopp}. 
The spectrum of 103P/Hartley 2, obtained from the {\it ISO} archive, is shown in Figure~\ref{hartley2spec} together with the 
bandpasses of the {\it Spitzer}/IRAC 3.6 and 4.5 $\mu$m channels.
A relatively definitive observational study of the 4.7 $\mu$m emission in the fundamental CO lines (exceptionally difficult from the ground) showed that there is both a native source, varying as $r^{-2}$, and an extended source in the coma \citep{disanti01}. The extended source of CO, C, OCS, and H$_2$CO must be due to something that can survive relatively long in the coma and then thermally evaporates. Comparing the H$_2$O and CO production rates for several comets, it appears that CO may be better-correlated with overall activity level in a comet \citep{feldman97}. CO was detected in the coma of 29P/Schwassmann-Wachmann 1 \citep{gunnarsson02}, with a much wider distribution than expected for a parent species in the Haser model, as noted above for Halley and Hale-Bopp. These results suggest that important volatile species are produced in the coma, not directly from the nucleus. 

Mid-infrared imaging of the gaseous emission from comets is challenging due to their faint surface brightness and the challenge of observing extended sources.
In particular, observations of cometary emission from gases present in the Earth's atmosphere is exceptionally difficult from the ground. The {\it Spitzer} Space Telescope has made this possible due to its presence outside the Earth's atmosphere and its ability to image extended emission in a broad filter near 4.5 $\mu$m.
There are two bright lines in this band: the CO $1\rightarrow 0$ fundamental vibrational band at 4.7 $\mu$m and CO$_2$ $\nu_3$ vibrational band at 4.3 $\mu$m; no other important gas lines are present in the mid-infrared from Hale-Bopp or 103P/Hartley 2 observed by ISO \citep{crovisier99hartley2}. 
Evidence for strong contributions of cometary gas to the IRAC 4.5 $\mu$m images has been found from {\it Spitzer}/IRAC imaging. 
For 21P/Giacobini-Zinner, the 4.5 $\mu$m image was shown to be dominated by gas \citep{pittichova08}.
For the split comet 73P/Schwassmann-Wachmann 3, we also found clear evidence for CO+CO$_2$ \citep{reach09sw3}. The IRAC 4.5 $\mu$m image, with a dust contribution removed based on the adjacent 3.6 and 5.8 $\mu$m images, shows the CO+CO$_2$ emission is in a sunward fan with a scale length of $3\times 10^5$ km. The abundance of CO$_2$ was high, requiring CO2/H$_2$O$\sim0.05$ to 0.1 in the major fragments B and C. For comparison, ground-based measurements found 
CO/H$_2$O$=0.005$ \citep{bockeleemorvan00}. Combining the ground- and {\it Spitzer}-based data yields CO$_2$/CO$\sim 20$, making CO$_2$ the second-most-abundant (by far) molecule in the ice exposed on 73P-C and indicating CO$_2$ is abundant even in ice relatively freshly exposed by the major splitting event that transformed 73P into a string of fragments.

Now that comet 103P/Hartley 2 has been visited by the sensitive imaging and spectroscopic instruments as part of the {\it EPOXI} mission \citep{ahearn11epoxi}, there is even more reason to pay attention to the role of CO$_2$ in comets. 
In contrast to the expectation that H$_2$O dominates cometary mass loss, for this comet at least it was evident that many active areas were dominated by CO$_2$, and a new mass-loss mechanism was found: blocks of icy material large enough to be individually observed in a swarm around the nucleus. 
In this paper we present a survey of {\it Spitzer}/IRAC observations to image CO$_2$ and CO production for a sample of Jupiter-family and long-period comets. 
Cometary CO has been studied with infrared spectroscopy by several groups (e.g. Disanti \& Mumma 2009), atmospheric and instrumental constraints have limited these studies to the very brightest comets. The spectroscopic method obtains precise information on CO for these objects, but it cannot yet let us measure CO in a statistically significant sample of ÒtypicalÓ (i.e. intrinsically faint) comets. The abundance of CO$_2$ has been almost completely unstudied in the JFC population due to the difficulty in observing that molecule from the ground. 
This changed with the {\it Akari} spectroscopic survey of 18 comets, which clearly showed CO$_2$ emission at 4.3 $\mu$m is a prominent feature of infrared spectra \citep{ootsubo11}. That survey also simultaneously covered H$_2$O and CO, making it particularly valuable for assessing relative abundances and showing that in the {\it Spitzer} 4.5 $\mu$m waveband, CO$_2$ is the dominant emitter from JFCs while CO is important in Centaurs and some Oort cloud comets. 
The {\it Spitzer} survey described in this paper covers some of the same comets, and it allows for the first time a sensitive imaging of the distribution of extended CO$_2$ emission from comets.

\section{Observations}

\subsection{Target Selection}
Table~\ref{obslog} summarizes the observations.
 The comets in this survey were selected based on their visibility from {\it Spitzer} during 2009-2010. Short-period comets with perihelion passage in 2009-2011 were  included if they could be observed when bright (predicted visible total magnitude $T<17$, solar elongation 85-115$^\circ$) and preferably in more than one {\it Spitzer} visibility window with heliocentric distance less than 5 AU. Long-period (Oort cloud) comets were selected from lists compiled by the well-known amateurs Seiichi Yoshida and the British Astronomical Association, and the Minor Planet Center, all of whom post predictions for bright comets in upcoming years. We did not include comets whose predicted visual magnitude is fainter than 18, since they are likely to be only a bare nucleus in the {\it Spitzer} observations. Two newly discovered comets were added during the survey, based on their potential to be bright and their visibility to {\it Spitzer}, these were P/2009 K5 (McNaught) and P/2010 H2 (Vales). The survey observations took place from 2009 Jul 28 to 2011 Feb 28. Some archival observations were added to the list from other programs due to their suitability; these were 65P/Gunn and 2P/Encke (observed during initial observatory check-out), and C/2002 T7 (LINEAR), 32P/Comas Sola, and 49P/Arend-Rigaux.

\clearpage

\begin{deluxetable}{lclrrrrrc}
\tablecolumns{9}
\tablecaption{Observation Summary\label{obslog}}
\tablewidth{0pt}
\tabletypesize{\footnotesize}

\tablehead{
\colhead{Comet} &
 \colhead{Date} & 
 \colhead{$R$} & 
 \colhead{$\Delta_{S}$} &
  \colhead{Solar} & 
  \colhead{$V$\tablenotemark{a} } & 
  \colhead{$F3.6$} &
  \colhead{$F4.5$} & \colhead{Aperture$^b$}\\
& & \colhead{(AU)} & \colhead{(AU)} & \colhead{Elong.} & & \colhead{(mJy)} & \colhead{(mJy)} & \colhead{($''$)} } 

\startdata
C/2002 T7 (LINEAR)		       & 2003/12/28 & 2.18 & 1.55 &116.5$^\circ$ & 10.6  &   215 & 482 & 30\\  
C/2006 W3 (Christensen) 	       & 2009/08/06 & 3.14 & 2.67 &108.5 & 12.3  & 138.5 & 294.7 & 30\\
C/2006 W3 (Christensen) 	       & 2010/06/09 & 4.45 & 4.26 &  94.6 & 14.7 & 18.2  & 30.6 & 30\\
C/2007 N3 (Lulin)		       		& 2009/11/20 & 4.20 & 4.02 &  92.9 & 16.2  & 0.19  & 0.21 & 6 \\
C/2007 Q3 (Siding Spring)	       & 2010/03/14 & 2.93 & 2.70 &  89.6 & 12.4 & 13.0  & 23.1 & 30\\ 
C/2007 Q3 (Siding Spring)	       & 2010/08/26 & 4.12 & 4.00 &  89.5 & 14.6 & 1.0   & 1.1 & 30\\  
C/2009 K5 (McNaught)		       & 2009/09/21 & 3.20 & 2.78 &104.9 & 14.7  &   2.7: & 2.4: & 6  \\
2P/Encke  			      			& 2003/11/17 & 1.01 & 0.23 &  82.0 &  7.1 & 19 & 75 & 30\\
2P/Encke			       			& 2009/12/24 & 2.93 & 2.31 & 118.1 & 19.3  & 0.04 & 1.2 & 30\\  
2P/Encke			       			& 2010/09/26 & 1.13 & 0.59 &   85.0 & 12.0 &  46.8 & 243.0 & 30\\
10P/Tempel 2			       		& 2009/07/29 & 3.16 & 3.11 &   83.1 & 19.7  &  0.05 &   0.4 & 6\\
10P/Tempel 2			       		& 2010/09/16 & 1.61 & 1.37 &   83.5 & 14.4  & 17.9  & 110.0 & 30\\
10P/Tempel 2			       		& 2011/02/28 & 2.66 & 2.17 & 117.2 & 18.0  & 0.22  & 4.6 & 30 \\
19P/Borrelly			       		& 2009/07/29 & 3.58 & 3.33 &   95.3 & 19.1  &  0.36 & 0.35 & 30\\ 
22P/Kopff			       			& 2009/08/16 & 1.78 & 1.33 &   97.5 & 14.1  & 14.9  & 59.3& 30 \\
22P/Kopff			       			& 2010/01/22 & 2.67 & 2.19 & 107.5 & 17.0  & 0.66  & 2.4  & 30\\
29P/Schwassmann-Wachmann 1&2010/01/25 & 6.19 & 5.82&107.8 & 14.9  & 3.5 & 18.1 & 84 \\	  
32P/Comas Sola                         & 2005/01/19 & 1.96 & 1.31 & 114.8 & 14.5 & 42    & 133 & 84\\
49P/Arend-Rigaux                       & 2004/11/29 & 1.69 & 0.98 & 114.3 & 15.3 & 24    & 123  & 30\\
57P/duToit-Neujmin-Delporte	& 2009/10/06 & 2.88 & 2.80 &   84.6 & 20.9  &  $<0.009$& 0.012: & 6\\
65P/Gunn                                      & 2003/09/21 & 2.58 & 3.25 & 101.3 & 14.5  &   6.8 &  24.2 & 30\\
65P/Gunn			       			& 2009/08/06 & 2.74 & 2.50 &  92.8 & 15.2  &  7.1  & 26.8  & 30\\
65P/Gunn			       			& 2010/07/30 & 2.61 & 2.08 & 110.0 & 14.7  & 3.6   &  18.4 & 30\\
77P/Longmore			       		& 2009/08/13 & 2.32 & 2.00 &   94.4 & 16.0  &   2.7 & 5.3 & 30\\  
77P/Longmore			       		& 2010/08/16 & 3.34 & 2.69 & 122.2 & 19.2 &0.02: &0.04: & 6		\\
81P/Wild 2			       			& 2010/04/12 & 1.68 & 1.33 &  89.2 & 12.1  &80.4  &	322 & 40\\
81P/Wild 2			       			& 2010/09/26 & 2.63 & 2.24 & 114.5 & 16.1  & 1.3   & 4.8 & 6 \\
88P/Howell			       			& 2009/08/24 & 1.47 & 0.84 & 102.8 & 13.4  &  93.0 & 630 & 30 \\
94P/Russell 4			       		& 2010/01/22 & 2.30 & 2.19 &   86.9 & 18.2  & 0.11  & 0.28 & 6 \\
94P/Russell 4			      		& 2010/07/13 & 2.34 & 1.67 & 112.3 & 17.7  & 0.10  & 0.27 & 30\\
103P/Hartley 2  		       		& 2009/09/09 & 3.98 & 3.38 & 118.4 & 20.2  &   0.39:& 0.35: & 6 \\
103P/Hartley 2  		       		& 2011/01/26 & 1.72 & 1.22 &   89.2 & 14.8 & 24	 &156 & 84	\\
116P/Wild 4			       		& 2009/07/28 & 2.18 & 2.01 &  85.2 & 12.9  & 10.3  &  21.3 & 30\\
116P/Wild 4			       		& 2010/06/09 & 3.02 & 2.96 &  84.1 & 16.8  & 0.10  & 0.14 & 6 \\
118P/Shoemaker-Levy 4		& 2009/11/28 & 2.01 & 1.80 & 90.8 & 13.3  & 3.9   & 20.1 & 30 \\
118P/Shoemaker-Levy 4		& 2010/05/08 & 2.21 & 1.50 & 122.3 & 14.0 &  2.9  & 12.2 & 30\\ 
143P/Kowal-Mrkos		       		& 2009/08/13 & 2.57 & 2.17 & 100.7 & 16.8  &   0.15& 0.50 & 6\\
P/2010 H2 (Vales)		       		& 2010/08/16 & 3.18 & 2.63 & 113.2 & 14.0 & $<$0.3	 & 0.47 & 30	\\
\enddata
\tablenotetext{a}{Total brightness (in visible magnitudes) predicted by Horizons, ssd.jpl.nasa.gov. These values were used in planning observations but not for the interpretation of the results.}
\tablenotetext{b}{radius of the circular aperture centered on the nucleus that was used for measuring the flux. Inspection of the curves of growth indicate the flux does not increase more than 20\% beyond these apertures, to the sensitivity of the {\it Spitzer} observations.}
\end{deluxetable}

\clearpage

\subsection{Observing strategy}
The observations were made using the Infrared Array Camera (IRAC, \citep{fazioirac}). The main survey was made after the cryogenic mission of {\it Spitzer} had completed, so that only the 3.6 and 4.5 $\mu$m channels of IRAC were operational. Each array has a $5'$ instantaneous field of view with 256$\times$256 pixels. 
The supplemental images in Table~\ref{obslog} with observing date prior to 2009 Jul were made during the cryogenic mission and additionally had 5.8 and 8 $\mu$m channels. One of the observations, the first of 65P/Gunn, was made before the telescope was focused, but the effect of focus was negligible for the purpose of this project.

To measure the CO and CO$_2$ emission, we made images of sufficient size to allow separation of the dust and gas (i.e. scaling the 3.6 $\mu$m image and subtracting from the 4.5 $\mu$m image) and to extend far enough into the coma that the flux measurement does not suffer from complicated velocity and density distributions in deviation from those of a steady outflow. If the production rate is $Q$, the expansion speed is $v$ and the timescale for dissociation by sunlight is $\tau$, then the density of a parent species is predicted to have a spatial distribution  
\begin{equation}
n = \frac{Q}{4\pi r^2 v} e^{-r/(v\tau)}.
\end{equation}
Images much larger than $v\tau$ will contain all the material and accurately trace the mass regardless of details of the density distribution, allowing us to derive $Q$ without assumptions on $v$ or $\tau$. Practical limitations (observing time) constrain the largest images, but the size scales with $v\tau$. 
For CO and CO$_2$, $\tau$(1 AU)= 15.4 and 5.8 days, respectively at 1 AU \citep{cochran93}, and the expansion velocity for the gas is $\sim 0.4$ km~s$^{-1}$ 
for comets at $\sim 2$ AU from the Sun \citep{reachencke00,bockelee90}, yielding scale lengths for the two molecules of $v\tau\sim4\times 10^6$ and $2\times 10^6$ km, respectively, at 3 AU from the Sun. We computed the 
CO$_2$ scale length for each cometÕs observing circumstances, yielding image size scale lengths of 4$'$ to 14$'$, very well suited to the IRAC field of view. Images smaller than $v\tau$ can be used (as can photometry) if the dust and gas can be accurately separated. As we will show, the CO+CO$_2$ is much brighter than dust for distant comets, so aperture photometry in the comet in one IRAC field of view in each of the 3.6 and 4.5 $\mu$m channels is sufficient. For comets passing closer to Spitzer, maps were 
created from multiple pointings of the telescope to cover a scale out to $v\tau$. 

To determine exposure times, the brightness of each comet was predicted using models for emission from the nucleus and from dust. Nuclear fluxes were predicted using the IAU standard model for reflected light plus the IRAS standard thermal model taking the diameter from the literature (or using 4 km if unknown). The dust coma flux was predicted using the $Af\rho$ from the literature \citep{ahearn95} or using 100 cm at 1 AU if unknown, and scaled by $r_h^{-2}$ to account for heliocentric variation. 
To detect the CO+CO$_2$ emission, we required that the signal-to-noise on the central coma be at least 10, to enable separation of the gas from the dust for relatively inactive comets, using their 3.6/4.5 $\mu$m color. Instantaneous exposure times of 30 or 100 sec were used, and the observations were typically 1 hour per comet.

\section{Flux measurements}


For each image, the relevant comet was identified and verified either by its obvious morphology (for the bright comets, which dominate the images) or its motion during the interval of observation (for faint comets, which can be confused with stars or galaxies). In most cases this was straightforward. For 103P in 2009 Sep and for C/2009 K5 in 2009 Sep, there was significant confusion from the crowded field of unrelated objects, so the flux measurements are approximate and are indicated as such with a `:' in Table~\ref{obslog}. For 57P, it was not possible to locate the comet at 3.6 $\mu$m so only an upper limit is given, and the identification at 4.5 $\mu$m is tentative. 

Fluxes were measured using the Aperture Photometry Tool developed by R. Laher\footnote{http://www.aperturephotometry.org/}. For each comet, a circular aperture was centered on the brightest condensation around the nucleus, and a curve of growth computed, showing the enclosed flux as a function of aperture radius expanding from 3$''$  to 120$''$. The aperture radius was selected so as to include the entire rising portion of the curve of growth. Very extended emission continues beyond the aperture in some comets, but at a very low level and not following the same power-law deviled with the nucleus as the bulk of the emission. We estimate that the enclosed flux is at least 80\% of the total flux in all cases. An aperture radius of $1'$ was used for the large, extended comets, for which the radial profiles always indicate a strong central condensation and an extended component. The smaller comets were measured in an aperture radius of $6''$ in order to avoid including nearby stars and galaxies. Those for which the compact aperture was used are indicated with a footnote in Table~\ref{obslog}. The aperture sizes are generally smaller than the maximum extent of gaseous emission predicted by the dissociation lifetime and expansion speed mentioned above. For this reason, the measurements may underestimate the total fluxes. Inspecting the images generally shows that there {\it is} extended emission at least out to the edge of the aperture. The integrated flux continues to slowly rise as a function of aperture, at a rate that is less steep than linear. We truncate the aperture because further extent is too dependent upon the uncertain removal of interstellar and cosmic background radiation.

\begin{figure}
\includegraphics[width=5in]{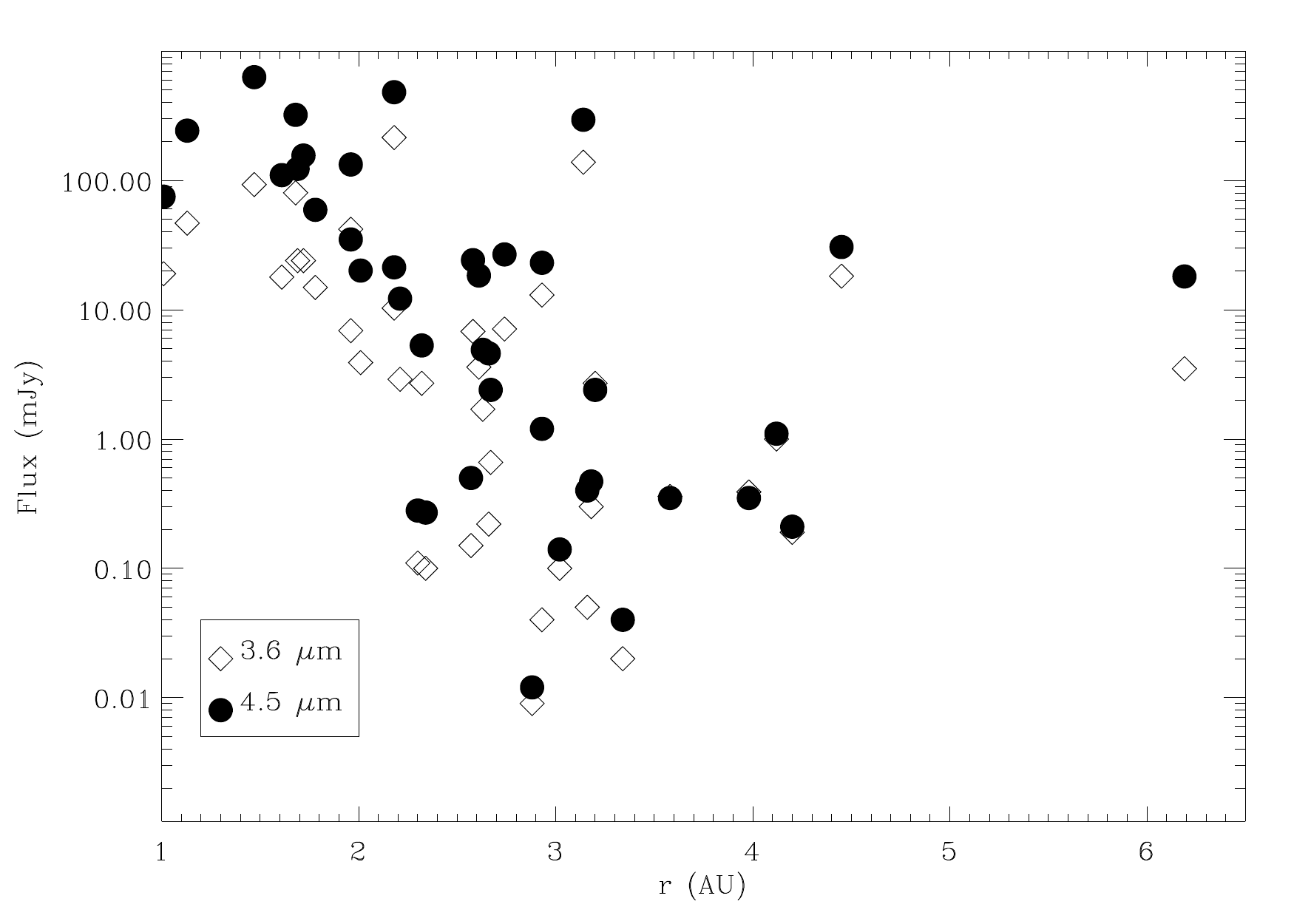}
\caption{Flux densities at 3.6 $\mu$m (open diamonds) and 4.5 $\mu$m (filled circles) for each comet observation, plotted versus the heliocentric distance at which the comet was observed. 
\label{fluxplot}}
\end{figure}

\begin{figure}
\includegraphics[width=5in]{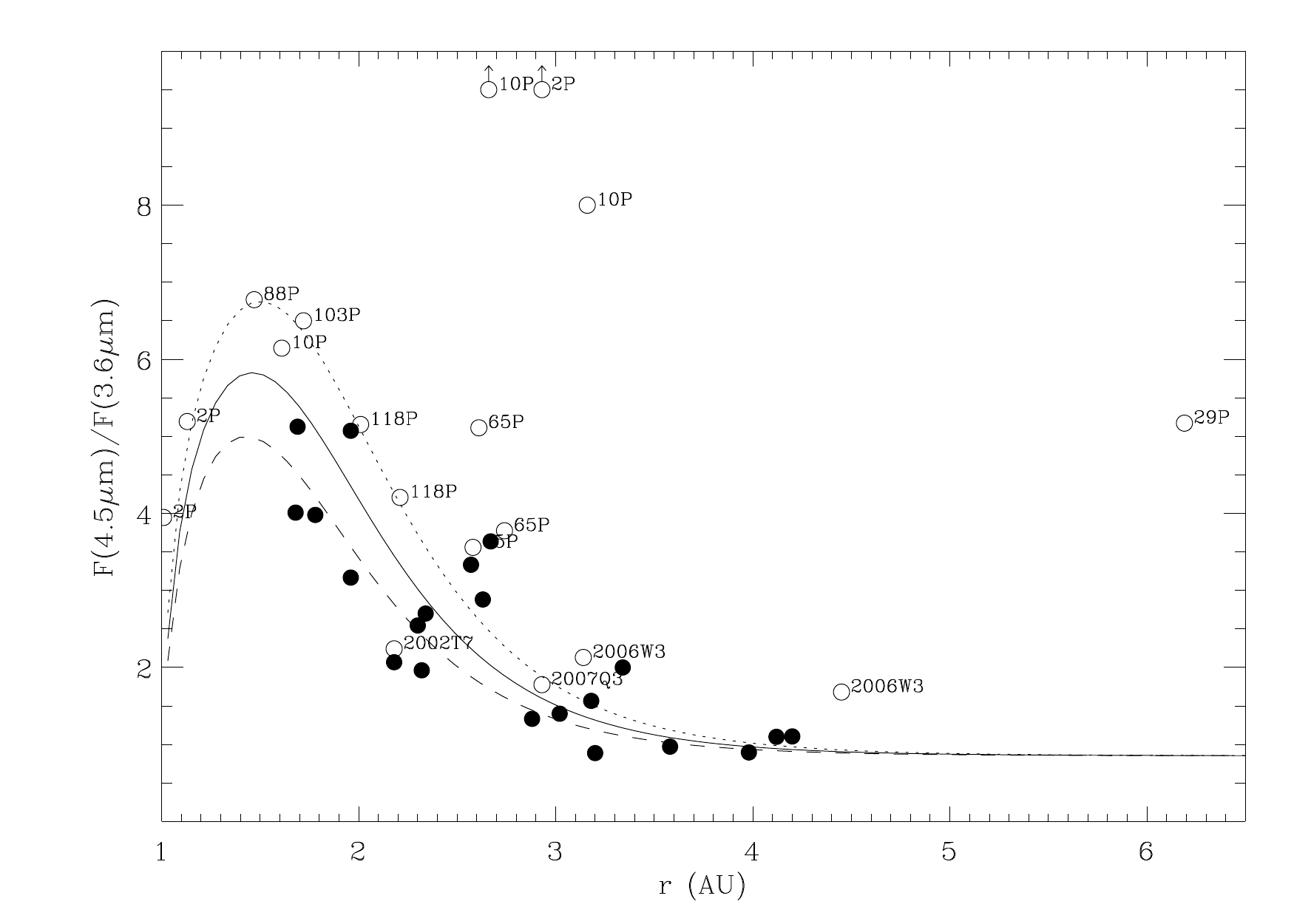}
\caption{Ratio of flux densities at 4.5 $\mu$m to 3.6 $\mu$m from the warm-{\it Spitzer} survey described in this paper. 
Each observation is shown by a circle: the open circles showing the comets that show different morphology at 
4.5 $\mu$m than at 3.6 $\mu$m in their IRAC images, and the filled circles showing those with compact or similar morphology at the
two wavelengths. Two comets whose flux ratio would put them off the top of the plot are shown with arrows; these are not lower limits.
The curves show a simple model prediction of the ratio for scattered light plus thermal emission. 
The solid curve is for isothermal dust particles with temperature enhanced over that of a rapidly-rotating blackbody by a factor of 7\%, and an albedo of 0.32. The dotted curve is for particles with temperature enhanced by 8\% and albedo 0.30. The dashed curve is for particles with temperature enhanced by 6\% and albedo of 0.34.
Comets falling far above the curves have so much 4.5 $\mu$m excess, most likely from strong CO$_2$ emission, that it dominates the images and is brighter than the dust emission, explaining why their 4.5 $\mu$m images are morphologically distinct from those in the dust-dominated 3.6 $\mu$m waveband.
\label{fluxratplot}}
\end{figure}

Figure~\ref{fluxplot} shows the flux densities versus heliocentric distance. Clearly, comets are fainter when further from the Sun, but the plot has a wide scatter due to the different gas and dust production rates for different comets. 
Figure~\ref{fluxratplot} shows the ratio of 4.5 to 3.6 $\mu$m flux for each comet, versus heliocentric distance. This ratio is now insensitive to the total production rate due to different amount of exposed surface ice and different nuclear size of the various comets. The change with heliocentric distance is due to the dust temperature and the dust-to-gas ratio of the material produced by the comet.
For the dust, a variation with heliocentric distance is expected from the strong change in thermal emission on the Wien portion of the thermal emission spectrum. The transition between scattered light and thermal emission occurs around a wavelength of 
$\lambda_r\simeq 3.5$ $\mu$m for material at 1 AU. 
Scattered light decreases with the comet's distance from the Sun as $r^{-2}$, while thermal emission decreases more steeply, being exponentially dependent upon $hc/\lambda k T$ (where $h$ is the Planck and $k$ is the Boltzmann constant and $c$ is the speed of light).
Since the temperature varies as $T\propto r^{-0.5}$, the wavelength where scattering and thermal emission are equal moves to longer wavelengths as $\lambda_r\propto r^{0.5}$ from 1 to 5 AU. Thus the nature of the dust emission changes as a function of heliocentric distance.
For comets more distant than 3.5 AU from the sun, the dust contribution is dominated by scattered light, and for comets within 1.2 AU of the Sun it is mostly thermal emission. 
Between 1.2 and 3.5 AU from the Sun, the 3.6 $\mu$m image is mostly scattered light while the 4.5 $\mu$m image is mostly thermal emission.

These considerations lead to a simple model of comet dust that explains the general trends of the observed cometary fluxes. 
The model is based on the equations from \citet{kelleywooden}. 
The data and three illustrative model curves are shown in in Figure~\ref{fluxratplot}. While the overall trend is explained, it is clear that there is still significant scatter, so that another effect, not related to the dust temperature and albedo, must be contributing to the colors. 
Comparing the shape of the predicted 4.5/3.6 $\mu$m brightness ratio to the observations shows there are significant positive deviants:
many comets are far brighter at 4.5 $\mu$m than the trend versus heliocentric distance and the model predictions could explain. 
Many of the positive deviants are those comets for which we could clearly see that the morphology of the 4.5 $\mu$m emission was different from that at 3.6 $\mu$m. Thus the imaging alone already suggested that an emission mechanism other than dust was contributing.
We show three model predictions in Figure~\ref{fluxratplot}, with each model having a different combination of dust albedo and dust temperature enhancement. The range of values that fit the observations for comets for which the images do {\it not} show a significant difference between
the 3.6 and 4.5 $\mu$m, i.e. for those comets whose images at both wavelengths may be dominated by dust, 
the range of parameters is $0.30 < A < 0.34$ and $1.06 < f_{temp} < 1.08$ where $A$ is the dust albedo and $f_{temp}$ is the temperature enhancement of the dust relative to that expected for rapidly-rotating isothermal grains. 

\begin{figure}
\includegraphics[width=5in]{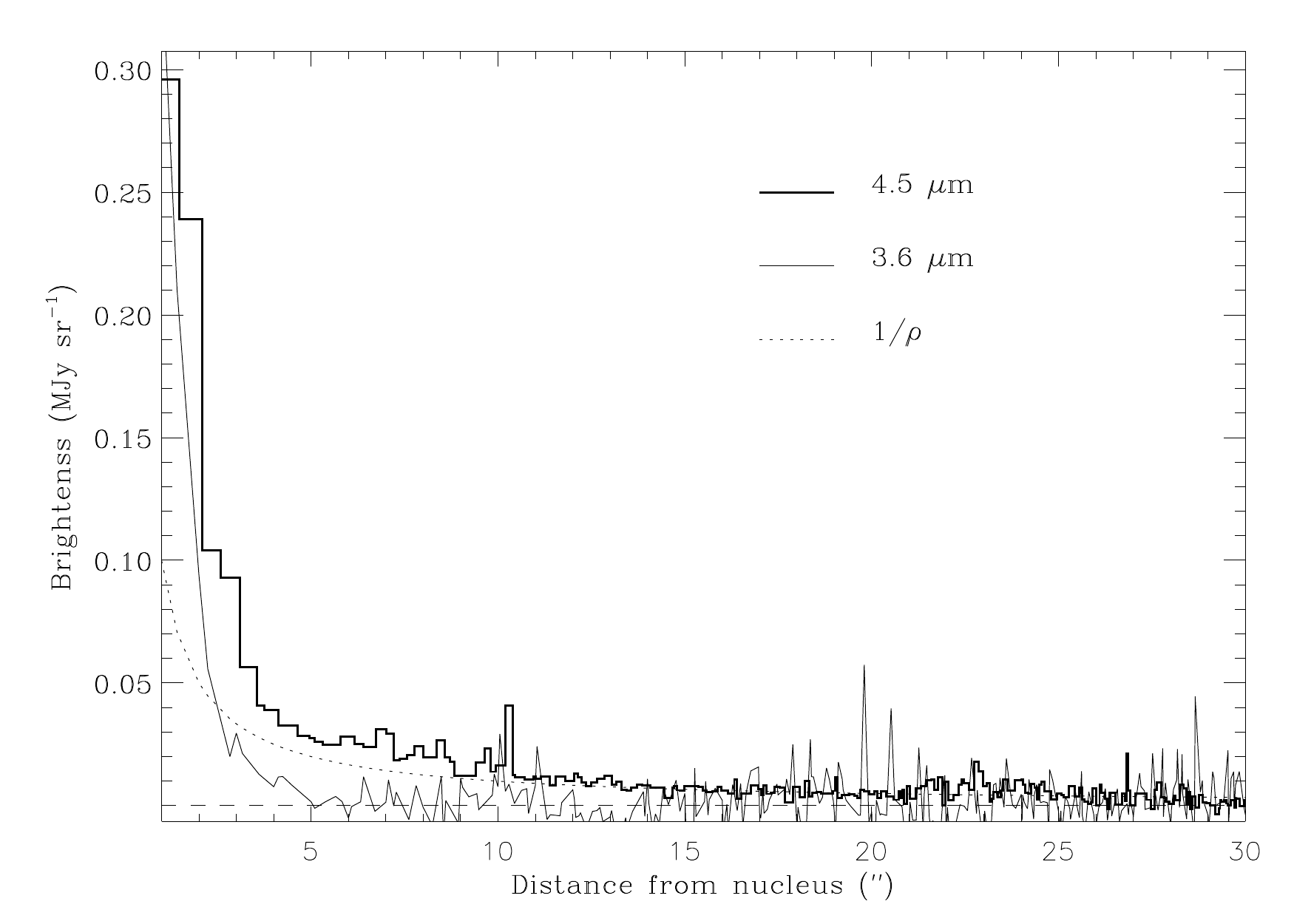}
\caption{Radial surface brightness profile of 10P/Tempel 2 on 2009 July 29. The profile at 4.5 $\mu$m (thick histogram) is significantly broader than that at 3.6 $\mu$m (thin jagged line, scaled by a factor of 5 to approximately match the brightness at 4.5 $\mu$m within the central 1$''$ radial distance from the nucleus).
At both wavelengths, the central condensation (nucleus plus inner coma) is likely due to solid material.
The 4.5 $\mu$m radial profile has an extended coma that approximately matches a $1/\rho$ profile (dotted curve).
This extended coma is likely CO$_2$ gas.
\label{radprof10p}}
\end{figure}

Inspecting the images for those comets that have excess flux at 4.5 $\mu$m confirms that the distribution of material is different
in the two IRAC bands. The most prominent outlier from the dust model is 10P/Tempel 2 observed at 3.2 AU from the Sun. The comet is relatively faint at this large heliocentric distance, but the distinction between the images at the two IRAC wavelengths is clear. At 3.6 $\mu$m the comet is very compact and close to the point spread function of the telescope, with a FWHM of 2.0 pixels (2.4$''$). 
Figure~\ref{radprof10p} shows the radial surface brightness profile.
At 4.5 $\mu$m the image is clearly extended, with a FWHM of 2.8 pixels (3.4$''$) and comprising a compact core plus an extended component with an approximate 1/$\rho$ profile where $\rho$ is the distance from the nucleus. The compact core in the images is the nucleus plus possibly an unresolved dust contribution. The extended emission, not from dust or the nucleus, contributes most to the flux at 4.5 $\mu$m. The radius of the nucleus of 10P is relatively large, at 5.3 km \citep{LamyNuc}.
Using 
a standard thermal model with beaming parameter 0.83 and emissivity 0.94
(Kelley et al., in preparation), the nuclear fluxes are 0.016 and 0.079 mJy, at 3.6 and 4.5 $\mu$m, respectively. This suggests the compact core of the {\it Spitzer} image
 has comparable contributions from the nucleus and the coma, while the diffuse extended emission seen only at 4.5 $\mu$m is almost entirely from gas.

The 3.6 $\mu$m images are assumed to be entirely due to dust, to calculate the dust optical depth using the $A_{3.6}f\rho$ parameterization \citep{ahearn95,kelleywooden}.
In this notation, $A_{3.6}$ is the 3.6 $\mu$m albedo, $f$ is the fraction of the beam filled by dust, and $\rho$ is the distance from the nucleus.
Table~\ref{prodrates} shows the resulting values. These allow an assessment of the dust production rates for the comets; while not the focus of this paper, the
values may be useful for other work. It is possible that the 3.6 $\mu$m albedo has a contribution from the C--H stretch of hydrocarbons at 3.4 $\mu$m wavelength. 
Spectra from {\it Akari} do clearly show this feature, with the amplitude of the feature relative to continuum being less than $\sim 20$\% in the spectra shown 
in \citet{ootsubo2012}. 

\clearpage

 \begin{deluxetable}{llrrcrrr}
 \tablecolumns{8}
\tablecaption{Production rate of CO$_2$ estimated from total flux\label{prodrates}}
\tablewidth{0pt}
\tabletypesize{\footnotesize}
\tablehead{
\colhead{Comet} & 
\colhead{$R$} &
 \colhead{$Af\rho$\tablenotemark{a}} & 
 \colhead{$\log Q_{OH}$}\tablenotemark{a} &  
 \colhead{Class\tablenotemark{b}} &  
 \colhead{$A_{3.6}f\rho$} & 
 \colhead{$\log Q($CO$_2)$\tablenotemark{c}}  & 
 \colhead{$f_{4.5}$\tablenotemark{d}}\\
& \colhead{(AU)} & \colhead{(cm)} &  \colhead{(s$^{-1}$)}  & &  \colhead{(cm)} &  \colhead{(s$^{-1}$)}
} 
\startdata
  C/2002 T7 (LINEAR) 	  & 2.18 &        &  28.90 &    & 9500  &  $<24.5$ &...\\
 C/2006 W3 (Christensen)  & 3.14 &       & 28.66 &        &     20000 &  28.18 $\pm$  0.20 & 0.47\\
 C/2006 W3 (Christensen)  & 4.45 &       & 27.87 &     & 7600  &  27.47 $\pm$  0.20 & 0.52\\
 C/2007 N3 (Lulin)  	  & 4.20 &        & 27.13  &    & 69    &  690 $\pm$  0.20 & 0.26\\
 C/2007 Q3 (Siding Spring)& 2.93 &     0  &        &    & 1720  &  26.78 $\pm$  0.20 & 0.27\\
 C/2007 Q3 (Siding Spring)& 4.12 &        &  27.11 &    & 350   &  25.65 $\pm$  0.20 & 0.24\\
  2P/Encke 		  & 2.93 &     2  &  26.55 &  T & 43   &  26.05 $\pm$  0.20 & 0.96\\
 2P/Encke  		  & 1.13 &    17  &  27.58 &  T & 250   &  27.11 $\pm$  0.20 & 0.35\\
10P/Tempel 2  		  & 3.16 &     6  &  26.32 &  T &  86  & 25.58 $\pm$  0.20 & 0.86\\
 10P/Tempel 2  		  & 2.66 &     9  &  26.58 &  T &  17   & 26.53 $\pm$  0.20 & 0.92\\
 10P/Tempel 2  		  & 1.61 &    25  &  27.14 &  T &  430  & 26.78 $\pm$  0.20 & 0.29\\
 19P/Borrelly  		  & 3.58 &   101  &  27.07 &  D &  82   & 24.25 $\pm$  0.57 & 0.04\\
  22P/Kopff  		  & 1.78 &   504  &  28.32 &  T & 420   &  25.56 $\pm$  1.27 & 0.02 \\
 22P/Kopff  		  & 2.67 &   220  &  28.68 &  T & 60    &  26.04 $\pm$  0.20 & 0.56\\
29P/Schwassmann-Wachmann 1& 6.19 &        & 27.64  &    & 5300  & 27.65 $\pm$  0.20 & 0.85\\
32P/Comas Sola  	  & 1.96 &        &    &    & 220   &  \\
 49P/Arend-Rigaux  	  & 1.69 &    98  &  27.11 &  T & 110   &  26.13 $\pm$  0.35 & 0.19\\
 65P/Gunn		  & 2.58 &    35  &  27.35 &  D & 870   &   26.97 $\pm$  0.20 & 0.51\\
 65P/Gunn  		  & 2.74 &    31  &  27.27 &  D & 780   &  27.14 $\pm$  0.20 & 0.60\\
 65P/Gunn    		  & 2.61 &    34  &  27.33 &  D & 300   &  26.98 $\pm$  0.20 & 0.69 \\
 77P/Longmore  		  & 3.34 &     0  &        &    & 31   & $<24.5$ & ...\\
 81P/Wild 2  		  & 1.68 &    65  &  27.73 &  D & 1170  & $<24.5$ & ...\\
 88P/Howell  		  & 1.47 &   201  &  27.48 &  T & 1150  &  27.50 $\pm$  0.20 & 0.33\\
 81P/Wild 2  		  & 2.63 &   191  &  27.44 &  D & 120   &  26.32 $\pm$  0.20 & 0.42\\
 94P/Russell 4	  	  & 2.30 &    32  &  27.37 &  D & 80   &  24.17 $\pm$  0.42 & 0.09 \\
 94P/Russell 4  	  & 2.34 &   100  &  27.04 &  D & 5.5   &  24.47 $\pm$  0.20 & 0.17\\
 103P/Hartley 2  	  & 3.98 &    17  &  26.61 &  T & 1040   &  24.45 $\pm$  0.44 & 0.05 \\
 103P/Hartley 2  	  & 1.72 &    91  &  27.84 &  T & 420   & 27.17 $\pm$  0.20 & 0.37 \\
 116P/Wild 4  	 	  & 2.18 &    75  &  27.30 &  D & 620   &   $<25$ & ...\\
116P/Wild 4   		  & 3.02 &    39  &  26.87 &  D & 150    &  24.27 $\pm$  0.20 & 0.13\\
  118P/Shoemaker-Levy 4   & 2.21 &        &  27.06 &    & 120   &  26.41 $\pm$  0.20 & 0.38\\
 118P/Shoemaker-Levy 4    & 2.01 &        & 27.16  &    &  180  & 26.52 $\pm$  0.20 & 0.39\\
 P/2010 H2 (Vales)   	  & 3.18 &        &        &    &  $<40$     & $<25.2$ & ...\\
 \enddata
 \tablenotetext{a}{$Q(OH)$ is the production rate from \citet{ahearn95} supplemented by other sources for recent comets. The values of $Q(OH)$ were scaled from the 
heliocentric distance of their observation to that at the time of each {\it Spitzer} observation as described in the text. 
$Af\rho$ is $Q(OH)\times (Af\rho/Q(OH))$ where the ratio of the dust production rate $Af\rho$ to the OH production rate is from \citet{ahearn95}. For C/2002 T7, $Q$ here is for H$_2$O in C/2007 T7 from \citet{combi09}. For C/2006 W3, $Q$ is based on OH from \citet{bockelee10}. 
For C/2007 N3, $Q$ is based on OH from \citet{carter12}. For 29P and 118P, $Q$ is based on H$_2$O from \citet{ootsubo2012}.}
\tablenotetext{b}{From \citet{ahearn95}; taxonomic class D=depleted (lower abundance of carbon-chain molecules, and low CN/OH), T=typical (in terms of carbon-bearing molecules per unit OH)}
\tablenotetext{c}{This is the CO$_2$ production rate under the {\it assumption} that the excess flux at 4.5 $\mu$m above the dust continuum is due to CO$_2$. In some cases (in particular, for 29P), this excess is almost certainly due to CO. In some other cases (for example, the long-period comets C/*) the production rate of CO$_2$ is somewhat overestimated because part of the 4.5 $\mu$m excess could be due to CO.}
\tablenotetext{d}{Fraction of the 4.5 $\mu$m flux in Table 1 that is attributed to CO$_2$+CO}
\end{deluxetable}

\clearpage

\section{Separating the Gas and Dust contributions to the 3.6 and 4.5 $\mu$m images}

Using the two IRAC images at 3.6 $\mu$m and 4.5 $\mu$m, a partial separation of the contributions of dust and gas can be made. This is possible because the 3.6 $\mu$m filter is in a portion of the spectrum with no strong gas lines. The {\it Akari} spectral survey of 18 comets shows this clearly \citep{ootsubo2012}. The only feature evident in the {\it Akari} spectra that would contribute to the 3.6 $\mu$m IRAC band is the C-H stretching of organics which is at 3.3-3.5 $\mu$m. This feature is weak in comets, and in any event may be due to organic dust grains rather than free-floating molecules. If we assume that the 3.6 $\mu$m channel is due only to dust, then we can directly subtract a scaled 3.6 $\mu$m image from the 4.5 
$\mu$m image to obtain an image only of the non-dust contribution to the 4.5 $\mu$m. The simplest approach would be to scale the 3.6 $\mu$m image as high as possible such that the subtracted image has no negative artifacts. This method works if the 3.6 $\mu$m image has no gas, and if the dust color is the same throughout the coma. A constant dust color seems a plausible assumption.
For the comets imaged during the cryogenic {\it Spitzer} mission, it was possible to better-characterize the 4.5 $\mu$m excess by using three wavelengths that are dominated by dust (3.6, 5.8, and 8 $\mu$m) to measure the color temperature and interpolate the dust contribution at 4.5 $\mu$m. This was done for 21P/Giacobini-Zinner \citep{pittichova08} and for 73P/Schwassmann-Wachmann 3 \citep{reach09sw3}. 

The gas production rates were estimated as follows.
First, we estimate the dust contribution to the 4.5 $\mu$m flux by scaling the observed 3.6 $\mu$m flux.
We use the model described above and shown in Figure~\ref{fluxratplot} to predict the ratio of 4.5 to 3.6 $\mu$m flux density at each heliocentric distance, then scale the observed 3.6 $\mu$m flux by the predicted ratio and subtract from the 4.5 $\mu$m flux. 
The remaining 4.5 $\mu$m flux density is then converted to in-band flux by multiplying by the IRAC 4.5 $\mu$m bandwidth 
($\Delta\nu=0.23\nu$). To convert the in-band flux to CO$_2$ production rate, we use the same scaling as was used in the  {\it Akari} spectral survey 
 \citep{ootsubo2012}; specifically, we used their Tables to derive the ratio of production rate to flux as a function of heliocentric distance, and we scaled our fluxes 
 by this ratio at the heliocentric distance of the {\it Spitzer} observation.
Note that we are {\it assuming} that all of the 4.5 $\mu$m excess is due to CO$_2$ when making these calculations. For comets where the contribution from CO is important (e.g. C/2006 W3) or dominant (29P/Schwassmann-Wachmann 1), the CO$_2$ production rates determined from this method are overestimates or irrelevant.

Table~\ref{prodrates} shows the resulting CO$_2$ production rates. The uncertainties in these values are relatively high because they result from subtraction of two comparable numbers. We used a 10\% uncertainty for the fluxes, but also a minimum uncertainty in the production rate of 26\% linear (0.1 in log)  to account for  systematics. 
Even in cases where the 4.5 $\mu$m excess is clearly detected in the image, such as for 2P/Encke at 1.13 AU from the Sun, the ratio of 4.5 $\mu$m to 3.6 $\mu$m flux 
alone may not lead to a convincing detection because the precise dust temperature is not known, and the thermal emission is very sensitive to temperature in the Wien tail of the Planck function (see the 2P left-most circle in Fig.~\ref{fluxratplot}).
In fact, it is only with detailed imaging or spectroscopy that any but the most extreme cases of CO$_2$ brightness (relative to the dust continuum) can be detected convincingly.

\begin{figure}
\includegraphics[width=5in]{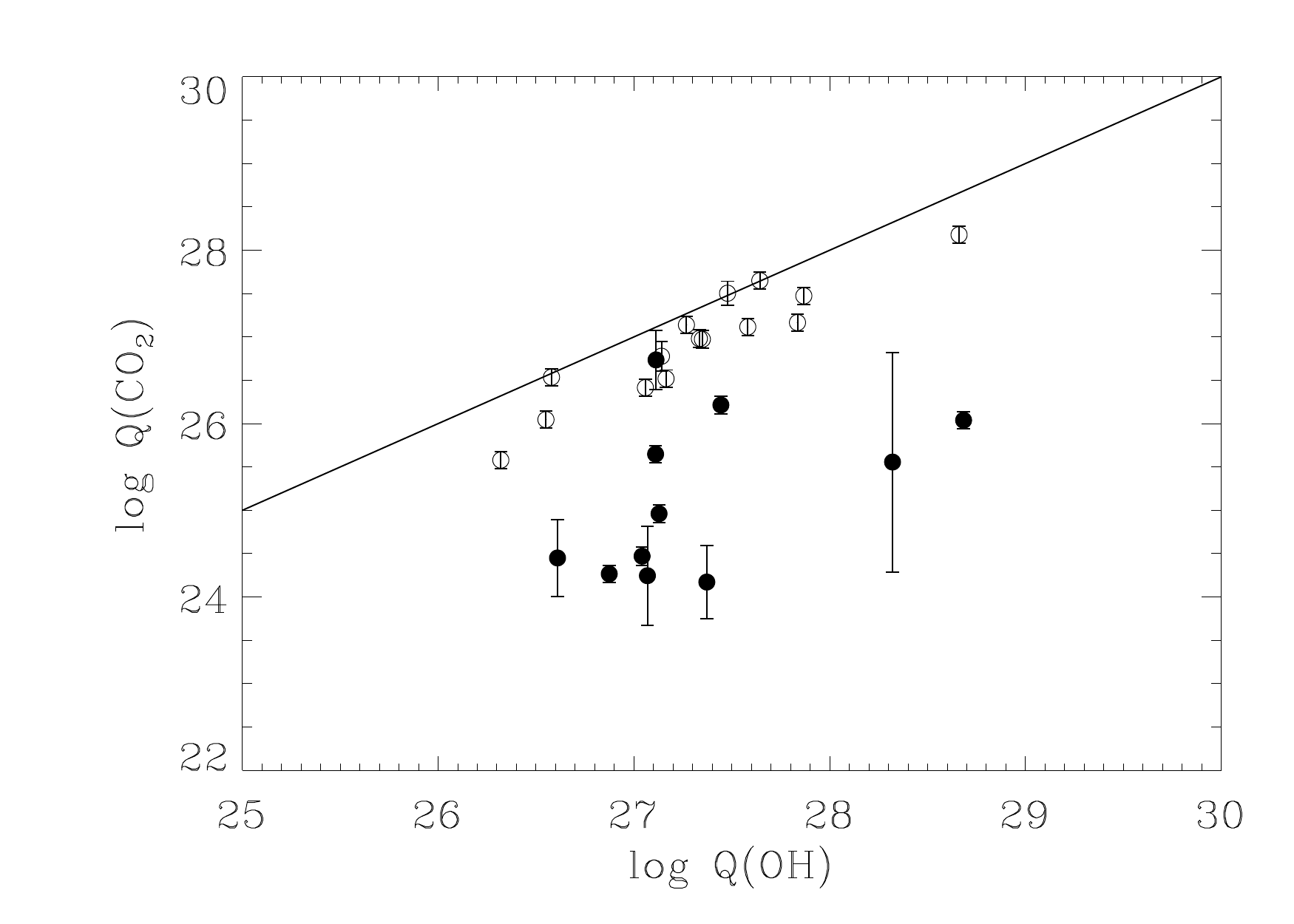}
\caption{Production rates of OH and CO$_2$ compared. The OH production rates are from \citet{ahearn95} scaled to the observing date for the {\it Spitzer} observation as described in the text. The CO$_2$ rates are from the 4.5 $\mu$m flux excess over that expected for dust based on the 3.6 $\mu$m emission, as described in the text. Symbols are as in Fig.~\ref{fluxratplot} (open circles are those with significant extended 4.5 $\mu$m emission). The straight line is a 1:1 proportion. Three of the comets have as much CO$_2$ as OH production.
\label{prodrate}}
\end{figure}

\begin{figure}
\includegraphics[width=5in]{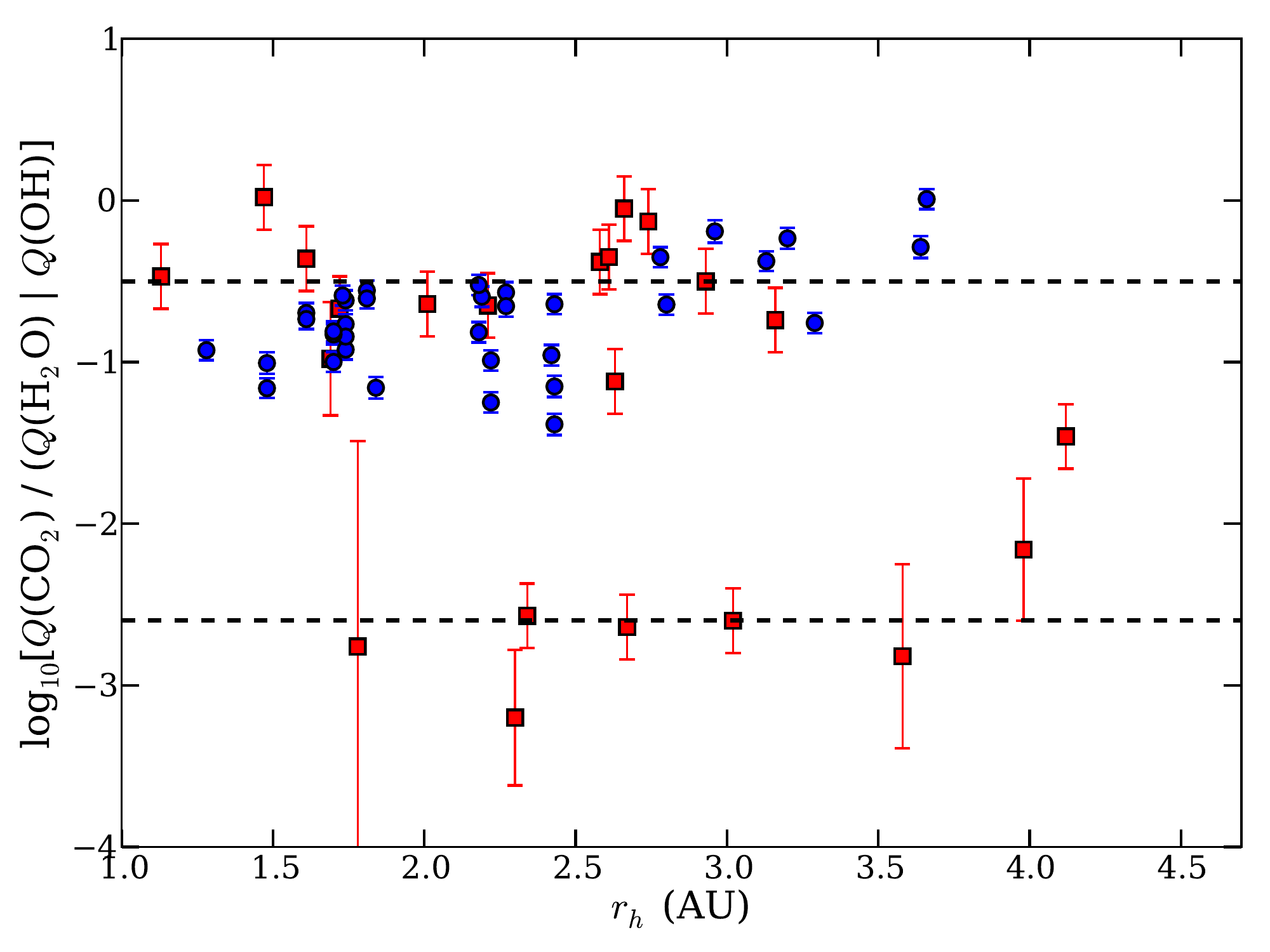}
\caption{Ratio of CO$_2$ to OH production rate versus heliocentric distance from the {\it Spitzer} survey (this paper; red squares) and from the {\it Akari} survey (Ootsubo et al. 2011; blue circles). There is no strong trend of this ratio with heliocentric distance, indicating that the dispersion in values is due to diversity of the comets themselves, not a reflex of the amount of solar heating. A mild trend is present in the {\it Akari} results (which are more precise because they are spectroscopic measurements  of H$_2$O and CO$_2$  simultaneously), such that
CO$_2$ production is relatively higher at larger heliocentric distance.
The horizontal dotted lines indicate typical values of the CO$_2$ rich and poor classes. 
\label{prodrateversusr}}
\end{figure}

Figure~\ref{prodrateversusr} shows the CO$_2$ production rate versus that of OH for direct comparison. The CO$_2$ rates are from the IRAC data in this paper, assuming all the 4.5 $\mu$m excess above dust is due to CO$_2$ (though some of it is from CO in some comets). 
The OH production rates are from \citet{ahearn95}\footnote{data from the Planetary Data System, Small Bodies Node, dataset EAR-C-PHOT-3-RDR-LOWELL-COMET-DB-PR-V1.0}, after scaling to the heliocentric
distances of the {\it Spitzer} observations.
The heliocentric distance scaling was made using a simple fit
\begin{equation}
 Q({\rm OH}) \propto R^{-2} e^{-(R/3.7)^3}
 \end{equation}
(where $R$ is the heliocentric distance in AU), which reproduces the heliocentric trend of the water production rate versus heliocentric distance calculated
using the methods of \citet{cowanahearn} and distributed by the Small Bodies Node of the Planetary Data System.
A significant caveat to the OH production rates used here is that they were measured on previous apparitions of the comets, and given the changes in comets from
one orbit to the next, they may not yield reliable estimates at the present time. We proceed to use them with the understanding that we are only looking for general trends in the survey, 
in hope that variations in some cases will be averaged out by the sample size.

It is notable that the CO$_2$ production rates are as high as those of OH (and by inference, the parent molecule of OH, H$_2$O) for three comets, and comparable for another 5. This indicates CO$_2$ is a very significant component of the cometary ices and may drive a significant fraction of cometary activity.
The production rate of CO$_2$ is about 10 times lower than that of OH for 2 comets, and it is $>100$ times lower for another 7 comets. 
The comparison between CO$_2$ inferred from our imaging survey and that inferred from a spectroscopic survey (Fig.~\ref{prodrateversusr}) shows that the two independent types of measurement are in
statistical agreement. The spectroscopic survey is more precise, and even shows a mild trend of CO$_2$/H$_2$O versus heliocentric distance.

The wide spread in production rates testifies to the intrinsic diversity of comets.
Figure~\ref{prodrateversusr} shows the ratio of CO$_2$ to OH production rate versus heliocentric distance.
If the variation in ratios were due to solar heating, e.g. because of the difference in temperatures at which CO$_2$ or
H$_2$O ice sublimate, then there would be a trend versus $R$. 
The {\it Akari} spectroscopic survey also measured the CO$_2$ to water ratio, using the strengths of lines of each molecule in the same spectrum 
and having cleanly separating the gas-line emission from the dust continuum \citep{ootsubo2012}. 
We include their data in Figure~\ref{prodrateversusr} for comparison to our results.
The \citet{ootsubo11} show an increase in CO$_2$/H$_2$O
for $R>2.8$ AU.  For the CO$_2$-rich comets in our survey, there are only
a few measurements at $R>2.8$ AU.  The lack of a trend similar to the
{\it Akari} results can easily be obscured by low-number statistics,
CO$_2$/OH uncertainties, and the intrinsic scatter of the survey
data.
All of the comets detected by {\it Akari} would be classified as `CO$_2$ rich'. 
There are discrepancies for individual comets when comparing in
detail. We ascribe the discrepancies in the CO$_2$ abundance estimates primarily to our use of the difference between two comparable numbers (the fluxes in the IRAC bands); the water abundances used in our survey are also just estimates (using scaled values from 
the \citet{ahearn95} survey from previous apparitions of the comets), as opposed to direct measurements of the water line in the {\it Akari} spectra.

There is an apparent dichotomy of comets into two classes which we labeled as `CO$_2$ rich' and `poor'.
The CO$_2$ rich comets are
 65P/Gunn (thrice),
 49P/Arend-Rigaux, 
 10P/Tempel 2 (thrice),
 88P/Howell, 
 2P/Encke (twice), 
 29P/Schwassmann-Wachmann 1,
 118P/Shoemaker-Levy 4 (twice), and
 C/2006 W3 (twice).
The `poor' comets are  
19P/Borrelly, 
 22P/Kopff (twice),
 103P/Hartley 2 ,
 81P/Wild 2 (twice),
 94P/Russell 4 (twice), 
  116P/Wild 4 (twice), 
  C/2007 Q3,
  C/2007 N3, and
  C/2002 T7.
 In this classification 29P/Schwassmann-Wachmann 1 is called `CO$_2$ rich' but in fact from the spectroscopy survey
 we know that the 4.5 $\mu$m excess for this comet is from CO , with abundance ratio CO/CO$_2$$>$80 \citep{ootsubo2012} so it is actually `CO rich' (with very little CO$_2$).
  Also, in this classification C/2006 W3 (Christensen) is called `CO$_2$ rich' but the spectroscopic survey shows its proportions of
 H$_2$O:CO$_2$:CO were 10:10:36 and 10:4:10 in observations at 3.66 and 3.13 AU from the Sun, respectively; therefore C/2006 W3 is 
 more properly `CO$_2$+CO rich.'
 
 It is noteworthy that many of the comets were observed more than once, and that, with one exception, {\it each time the same comet was observed
 it was in the same class}, despite the observations being at different heliocentric distance. The exception to this rule was 103P/Hartley 2,
 which was  `CO$_2$ poor' when observed at 3.98 AU, but it was apparently
 `CO$_2$ rich' by the time it was observed at 1.72 AU from the Sun.
 For this comet, our characterization is consistent with the recent paper by \citet{meech11hartley}, who showed 
 that early activity is driven by H$_2$O, while CO$_2$ becomes the dominant driver of activity closer to the Sun.
 Such behavior can be explained by seasonal effects, where some parts of the nucleus are more exposed to sunlight than others depending
 on their location relative to the rotational pole and the comet's position in its orbit.
 
 The class we infer from the infrared-derived CO$_2$ production partially correlates with the taxonomic classes 
 defined by \citet{ahearn95} based on carbon-chain molecule depletions. Of the `CO$_2$ rich' class, there are
 5 comets of `Typical' carbon-chain abundance and 1 `Depleted'; the `Depleted' but `CO$_2$ rich' comet is 65P/Gunn.
 Of the `CO$_2$ poor' class, there are 2 comets of `Typical' carbon-chain abundance and 4 `Depleted'.
 While the correlation is far from one-to-one, it does appear that there are relatively more comets of the `CO$_2$ poor'
 class that are `Depleted' in carbon-chain molecules. This makes some intuitive sense since both classifications
 involve carbon and the carbon-chain `Depleted' and `CO$_2$ poor' classifications go in the same direction of 
 having fewer carbon-bearing molecules.

\section{Morphology of CO+CO$_2$ Gas Emission from Comets}

One value of the {\it Spitzer}/IRAC survey, in comparison to the recently completed {\it Akari}/IRC spectroscopic survey
\citep{ootsubo2012}, is that we can 
assess the spatial distribution of material from comets, including both the dust and, when present, the excess 4.5 $\mu$m emission due to CO$_2$ and CO gas. Many of the IRAC images show extended structure, including radial variations significantly deviating from $1/\rho$ and 
non-axisymmetric azimuthal distributions, that are distinct between 4.5 $\mu$m and 3.6 $\mu$m wavebands, indicating a significantly different 
distribution of gas and dust. In the remainder of this section we describe the features seen in some of the survey comets and present their images in 
such a way that readers can assess the distributions themselves.

\subsection{Jupiter-family comets with distinct gas emission}

\subsubsection{10P/Tempel 2}
\begin{figure}
\includegraphics[width=5in]{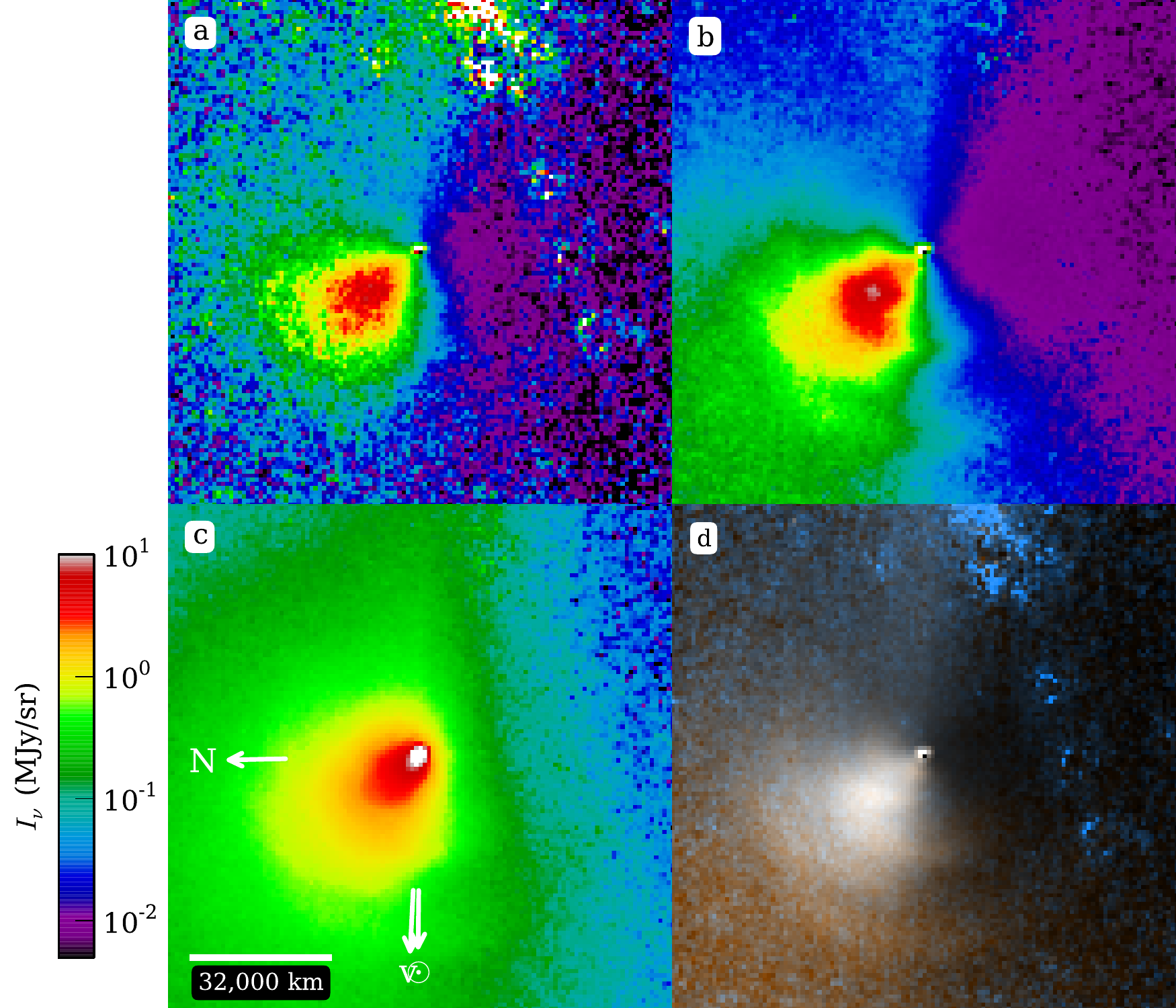}
\caption{
IRAC images of 10P/Tempel 2 on 2010 Sep 16. {\it (a,b) Top row:} Images at 3.6 $\mu$m (left) and 4.5 $\mu$m (right), multiplied by the projected distance from the nucleus, $\rho$, to enhance deviations from the $\rho^{-1}$ dependence expected for spherical expansion at constant velocity, and to enhance structures distant from the nucleus.
{\it (c)} The 4.5 $\mu$m image, with labels indicating the projected direction toward the Sun and the projected direction of the comets velocity vector. A horizontal scale bar shows the image size, and the color bar to the left of the panel shows the surface brightness range.
{\it (d)} `True-color' combination of the 3.6 $\mu$m image in blue, the average of the 3.6 and 4.6 $\mu$m images in green, and the 4.5 $\mu$m image in red.
The images show a prominent sunward-sector fan. The 4.5 $\mu$m image has pronounced asymmetries, most like due to gas emission,
in the form of arcs within the sunward fan.
\label{tempel21009}}
\end{figure}

Comet 10P/Tempel 2 was observed three times during this survey. 
Each time, the 4.5 $\mu$m emission was significantly brighter than would be expected for
dust emission, based on the 3.6 $\mu$m brightness. 
Figure~\ref{tempel21009} shows the images when observed at closest range in 2010 Sep.
The image at 4.5 $\mu$m showed a set of arcs, on the sunward side of the comet.
The arcs and most of the emission occupy a fan-shaped
region that has a full opening angle of $62^\circ$ width, and the center of the emission fan is directed $29^\circ$ East of North.
We measure the sizes of the arcs by overlaying circles centered on the nucleus and quoting the radii of the circles.
The closet such arc is $\sim 6"$ from the nucleus (but is difficult to discern; the outer arcs are at $15"$, $28"$, and approximately $63"$ from the nucleus. 
The pattern could be due to a strong active region on the rotating nucleus, or to a set of outbursts. If due to outbursts, and the expansion velocity is
1 km~s$^{-1}$, then they happened at $\sim 1$, 3.3, 6.1, and 14 hr before the image. The pattern is not linear, and indicates either acceleration or
nonuniformity (i.e. not all outbursts are of the same amplitude.  

10P was covered in the {\it Wide-Field Infrared Survey Explorer} (WISE) \citep{wrightWISE} all-sky survey on 2010 Apr 27.  We obtained the images from the Infrared Science Archive\footnote{http://irsa.ipac.caltech.edu/Missions/wise.html} and measured the fluxes using the same aperture photometry procedure (including curve of growth analysis) that we had applied to the {\it Spitzer} data. The WISE images are substantially less sensitive than the {\it Spitzer} images, but the coma was clearly detected by WISE at all 4 of its wavelengths,
 with fluxes at 3.4, 4.6, 12, and 22 $\mu$m of 5.0, 36, 500, and 2100 mJy, respectively. The 12 and 22 $\mu$m images are clearly dominated by a dust tail, while at 3.4 and 4.6 $\mu$m the images show only the coma. The relative
 fluxes are consistent with dust emission at 12 and 22 $\mu$m; the estimated dust emission at 4.5 $\mu$m based on the color temperature is less than 0.5 mJy. The 3.4 $\mu$m flux is most likely scattered sunlight, from which we estimate the
 scattered sunlight at 4.6 $\mu$m is 3 mJy. That means that the bulk (more than 85\%) of the 4.6 $\mu$m flux is not from dust, in agreement with our inference based on the {\it Spitzer} 2-band images.

\clearpage

\subsubsection{2P/Encke}
\begin{figure}
\includegraphics[width=5in]{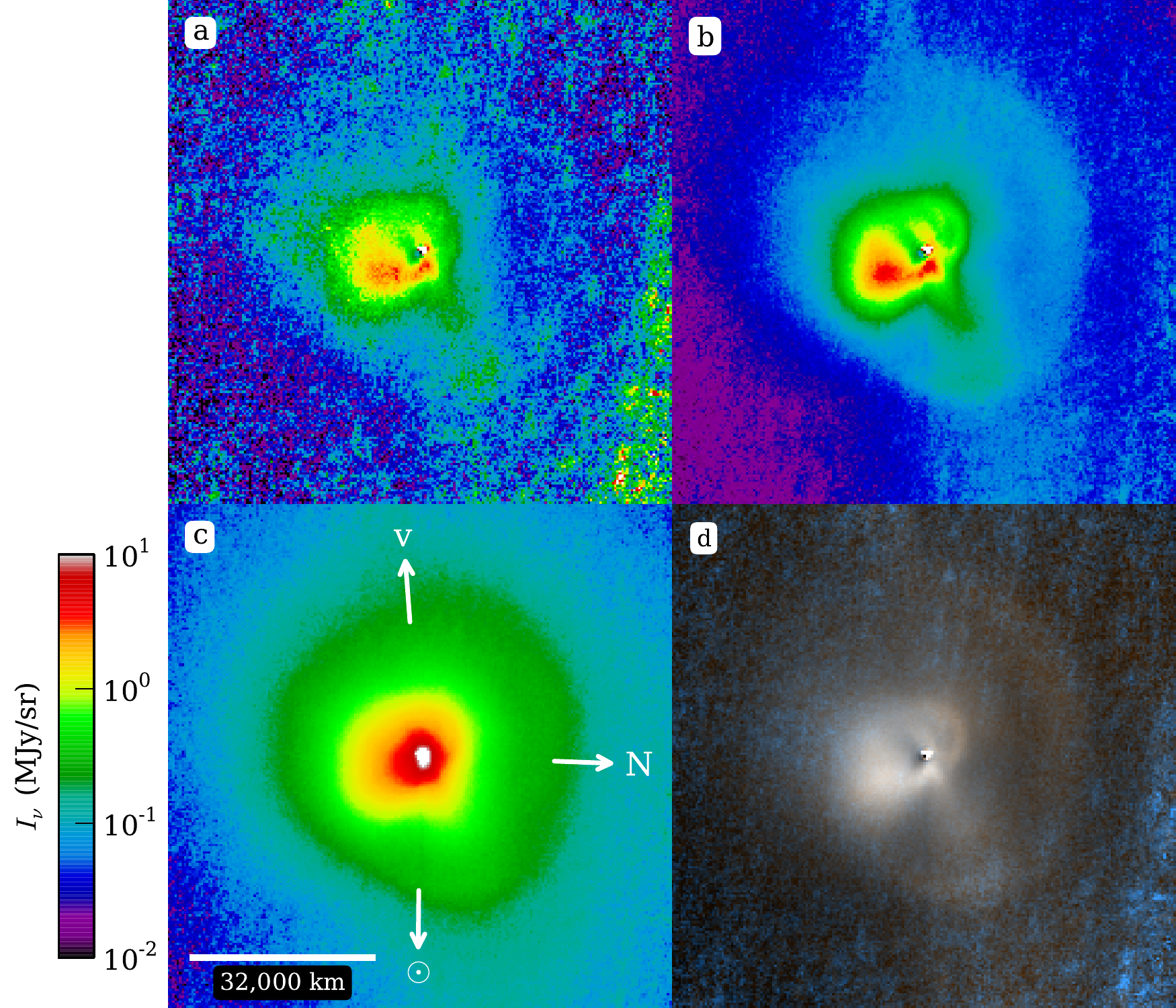}
\caption{
IRAC images of 2P/Encke on 2010 Sep 26. The panels are in the same format as in Figure~\protect\ref{tempel21009}.
The images show a thin arc, possibly a limb-brightened hemispherical shell, to the west of the nucleus, together with a complicated coma shape with arcs potentially due to the dominant jet that becomes active only around perihelion.
\label{encke1009}}
\end{figure}

 Figure~\ref{encke1009} shows the images of 2P/Encke in 2010 Sep, when the comet was 1.1 AU from the Sun, 36 days after perihelion. This is the 
 highest-resolution image from our deep survey, because the comet was only 0.59 AU from {\it Spitzer} at the time of observation.
The arcs near the nucleus are present in both dust and gas and are due to the dominant active region that becomes active 25 days after perihelion \citep{reachencke00}.
An outer arc is present at a distance $78"$ (33,000 km) from the nucleus only in the gas image.
The outer arc can be traced from
approximately 82$"$ near the top of the image (as displayed in
Fig.~\ref{encke1009}) down to 73$"$ near the bottom (3.5 to 3.1$\times10^{4}$
km); this implies that the comet is rotating clockwise with respect
to Spitzer's vantage point.
The 3.6 $\mu$m and 4.5 $\mu$m images are similar, especially when the lower signal-to-noise ratio of the 3.6 $\mu$m image is taken into account.
Encke's debris trail is found along the projected orbital velocity
vector in our 4.5 $\mu$m images.

\begin{figure}
\includegraphics[width=5in]{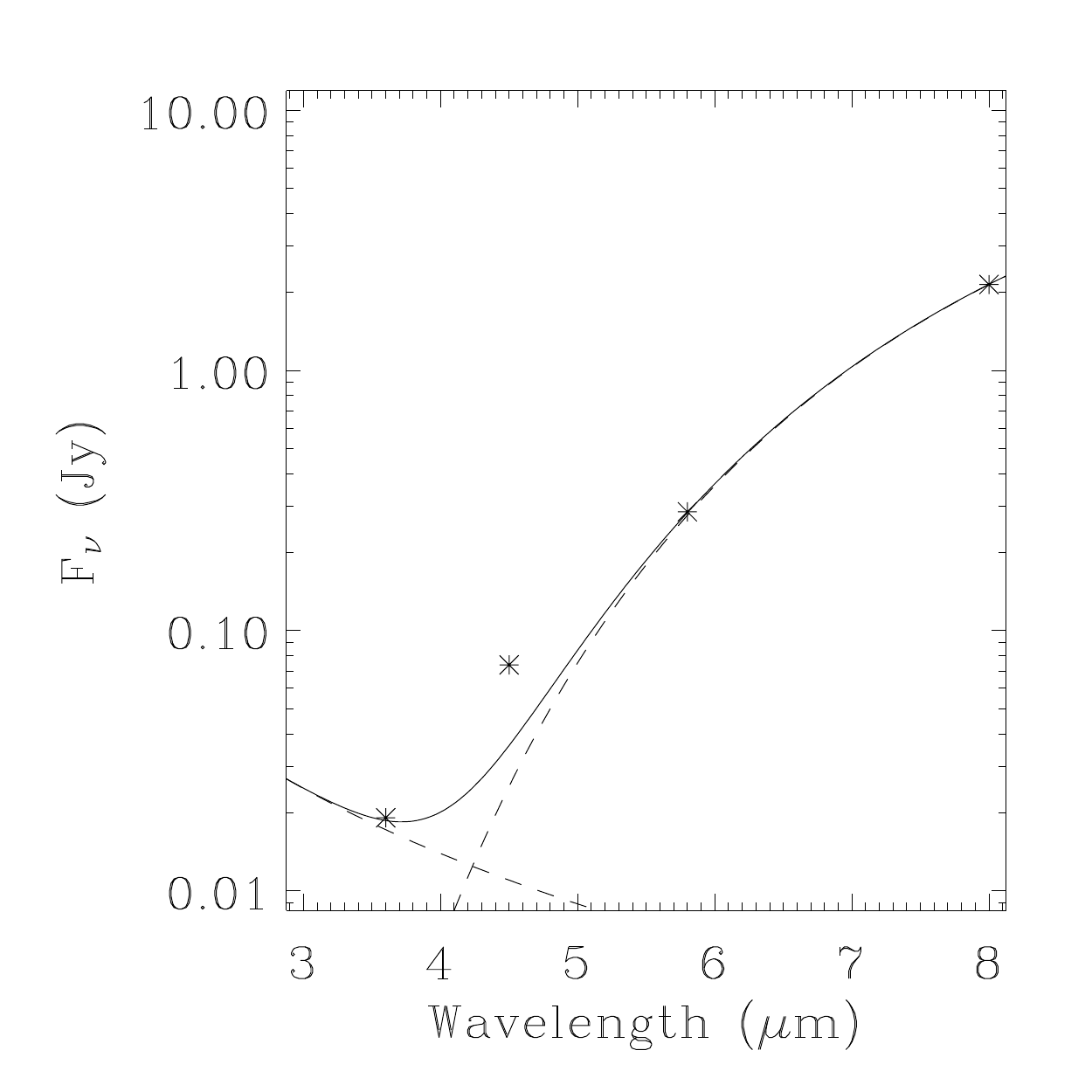}
\caption{\label{enckespec}
Spectral energy distribution of the 2P/Encke coma from the IRAC pre-perihelion 
observation on 2003 Nov 17. The flux in an annulus from 5--15 pixels
(each $1.2''$ in size),
after removing the background in an annulus from 20--60 pixels, was
summed in each channel. 
The fluxes at 5.8 $\mu$m and 8 $\mu$m were divided by
extended-source aperture corrections of 0.69 and 0.83, respectively.
The dashed curves show a blackbody with temperature
222 K (normalized to the 8 $\mu$m flux) and a $\nu^2$ power-law 
(normalized to the 3.6 $\mu$m flux); these curves represent thermal
emission and scattering, respectively, from the dust grains in the
coma. The 4.5 $\mu$m excess is readily evident.
}
\end{figure}

\begin{figure}
\includegraphics[width=5in]{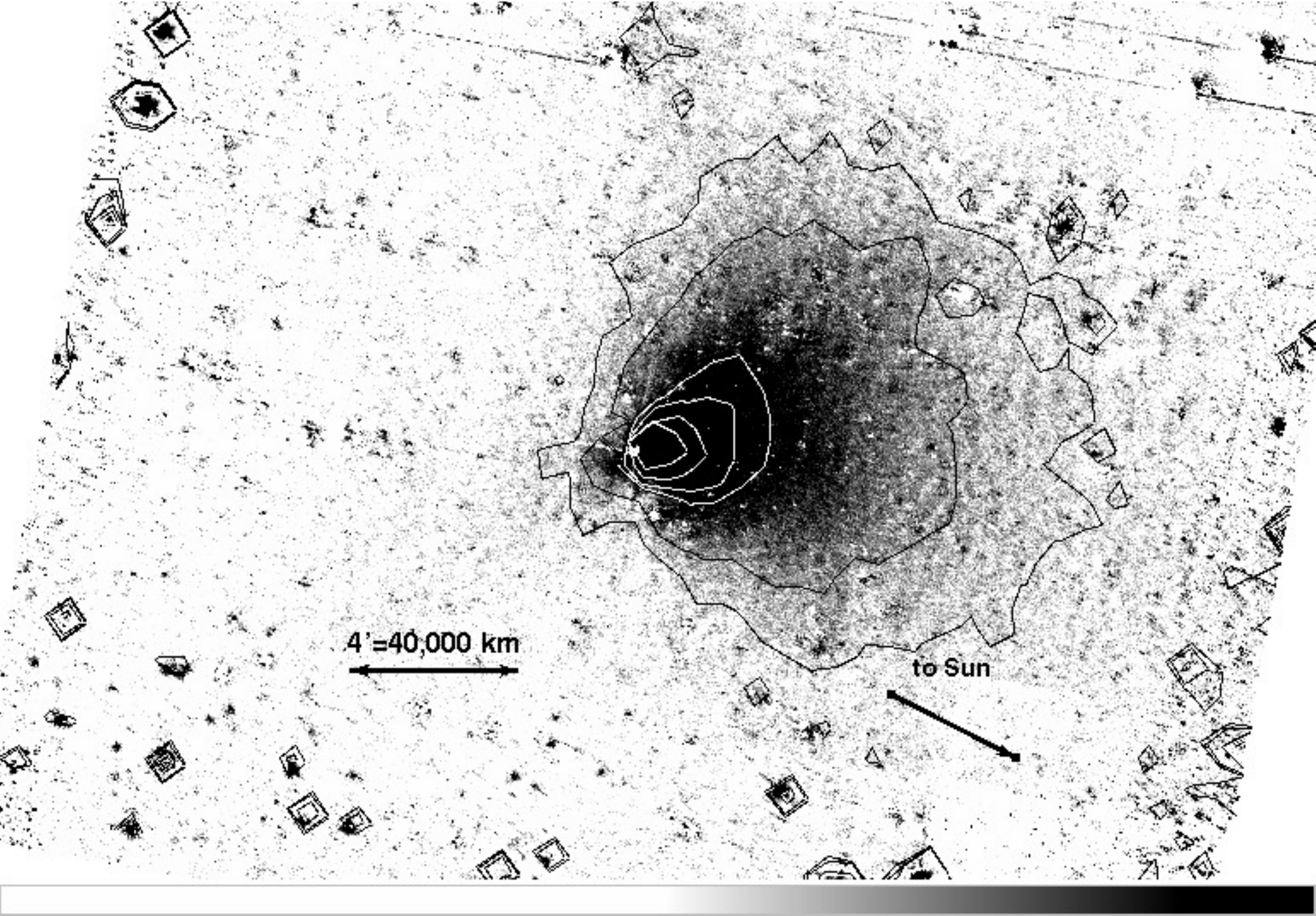}
\caption{\label{enckegasmap}Image of the CO+CO$_2$ emission from 2P/Encke,
derived from the 2003 Nov IRAC 4.5 $\mu$m image after removing 
scaled versions of the 3.6 $\mu$m and 8 $\mu$m images as templates for
dust scattering and thermal emission, respectively.
Contours are drawn at 0.01, 0.05, 0.16, 0.27, 0.41, and 0.6 MJy~sr$^{-1}$,
with the fainter two contours drawn in black for clarity. The dark and bright
spots are point sources that are under-subtracted or over-subtracted.}
\end{figure}

In 2003 Nov, 2P/Encke was observed as part of the cryogenic {\it Spitzer} mission, and the flux density was measured at 3.6, 4.5, 5.8, and 8 $\mu$m. Figure~\ref{enckespec} 
shows the spectral energy distribution together with a simple fit to the dust emission. It is evident that while dust contributes some of the 4.5 $\mu$m brightness in the IRAC
images, that channel is in excess of pure dust emission has a significant gas contribution. The presence of the excess was already suggested by the ratio of 4.5 to 3.6 $\mu$m
fluxes (Fig.~\ref{fluxratplot}, left-most data point), but it is even more clear from the images, which are at relatively high angular resolution with the comet at only 0.23 AU from 
the observatory. Figure~\ref{enckegasmap} shows a gas-only image, generated by subtracting a weighted combination of the 3.6 and 5.8 $\mu$m images, with the weights based on the extrapolated spectral
energy distribution for scattered sunlight and thermal emission, respectively, using the model from Figure~\ref{enckespec}.
The gas emission can be roughly described as a fan-shaped region in the sunward hemisphere (but not directly centered on the sunward vector). 
The brightness of the gas emission decreases approximately as $\rho^{-3/2}$ where $\rho$ is the projected distance from the nucleus. This is somewhat steeper than the $\rho^{-1}$
expected for constant and free expansion, suggesting that the gas production is not steady or the molecules are destroyed or otherwise lost as they travel further from 
the nucleus. 
The lifetime against photodissociation and ionization for CO and CO$_2$ 
molecules at 1 AU from the Sun are $t_{CO}\sim 15$ days and 
$t_{CO_2}\sim 8$ days, respectively \citep{huebner}.
If dissociation occurs with constant probability during the time of flight from 
the nucleus, and the trajectories are relatively straight, then from
the continuity equation the radial profile becomes
\begin{equation}
n = \frac{Q}{v r^2} e^{-r/r_p}.
\end{equation}
This radial profile of CO$_2$ matches the observed projected profile if the 
gas expands from the nucleus with a speed $v_{ej}\simeq 0.2$ km~s$^{-1}$.

The expansion velocity inferred from the {\it Spitzer} image of 2P/Encke can be compared to completely independent measurements.
A compilation of observational measurements of OH line profiles velocities infers H$_2$O expansion velocities in the 0.5--1 km~s$^{-1}$ range for a low-water-production-rate comet 
near 1 AU\citep{tseng07}. As a heavier molecule, CO$_2$ would expand more slowly than H$_2$O.
The gas expansion velocity depends mainly on the physical properties of the gas, its sublimation temperature and the way it cools as it escapes the nucleus. 
The hydrodynamic model from \citet{crifo97}, using a gas temperature provided by \citet{radionov02} applied to comet Encke in the circumstances of the observation,
with nuclear radius 2.4 km \citep{fernandez05,lowry07} and heliocentric distance 1.1 AU, provides a gas expansion terminal velocity of 0.56 km s$^{-1}$ for pure water and 0.36 km s$^{-1}$ for pure CO$_2$. 
Thus the observed CO$_2$ image size is in not quite in agreement with expectations from theory. The observed image size would require a lower expansion velocity than
expected based on independent measurements.

\clearpage

\clearpage

\subsubsection{22P/Kopff}
\begin{figure}
\includegraphics[width=5in]{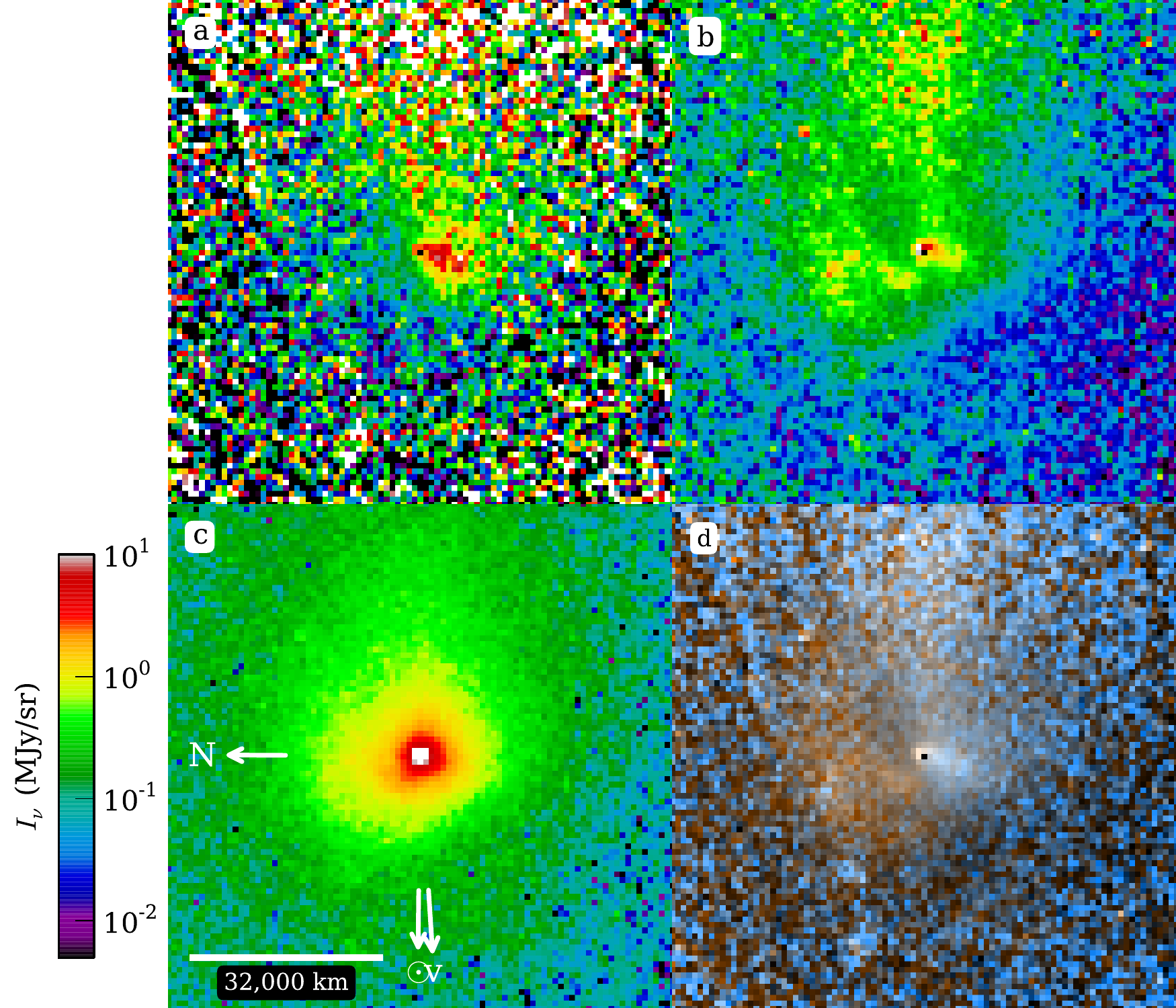}
\caption{
IRAC images of 22P/Kopff on 2009 Aug 16. The panels are in the same format as in Figure~\protect \ref{tempel21009}.
The images show a dust coma together with a feature at 4.5 $\mu$m image in the sunward direction.
\label{kopff0908}}
\end{figure}

\begin{figure}
\includegraphics[width=5in]{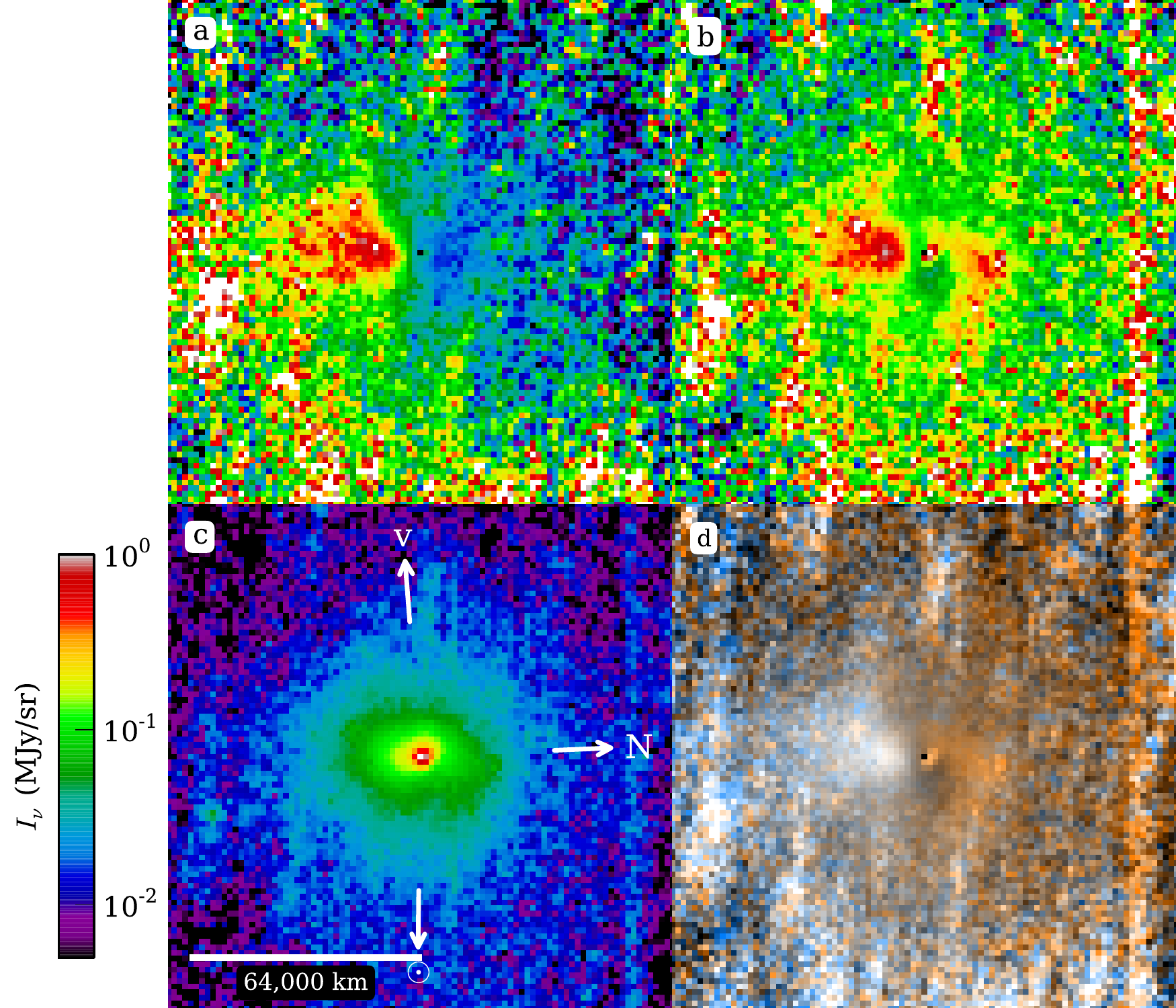}
\caption{
IRAC images of 22P/Kopff on 2010 Jan 22.The panels are in the same format as in Figure~\protect \ref{tempel21009}.
The images show a dust coma together with distinct (complicated) structure only seen at 4.5 $\mu$m.
\label{kopff1001}}
\end{figure}

Figure~\ref{kopff0908} shows the images of 22P/Kopff in 2009 Aug, and
Figure~\ref{kopff1001} shows the images of 22P/Kopff in 2010 Jan.
In both cases, the 3.6 $\mu$m and 4.5 $\mu$m are significantly different. In 2009 Aug, there is apparent gas-only emission that is on the sunward side (but not directly toward the Sun, it is offset to the left in FIg.~\ref{kopff0908}). In the 2010 Jan image, there is apparent gas-only emission roughly perpendicular to the sunward direction. This gas-only emission is as expected
from an active region, possibly the same one at the two epochs (though we have not verified
this with a pole orientation model), that is CO$_2$ rich.

\clearpage

\subsubsection{65P/Gunn}
\begin{figure}
\includegraphics[width=5in]{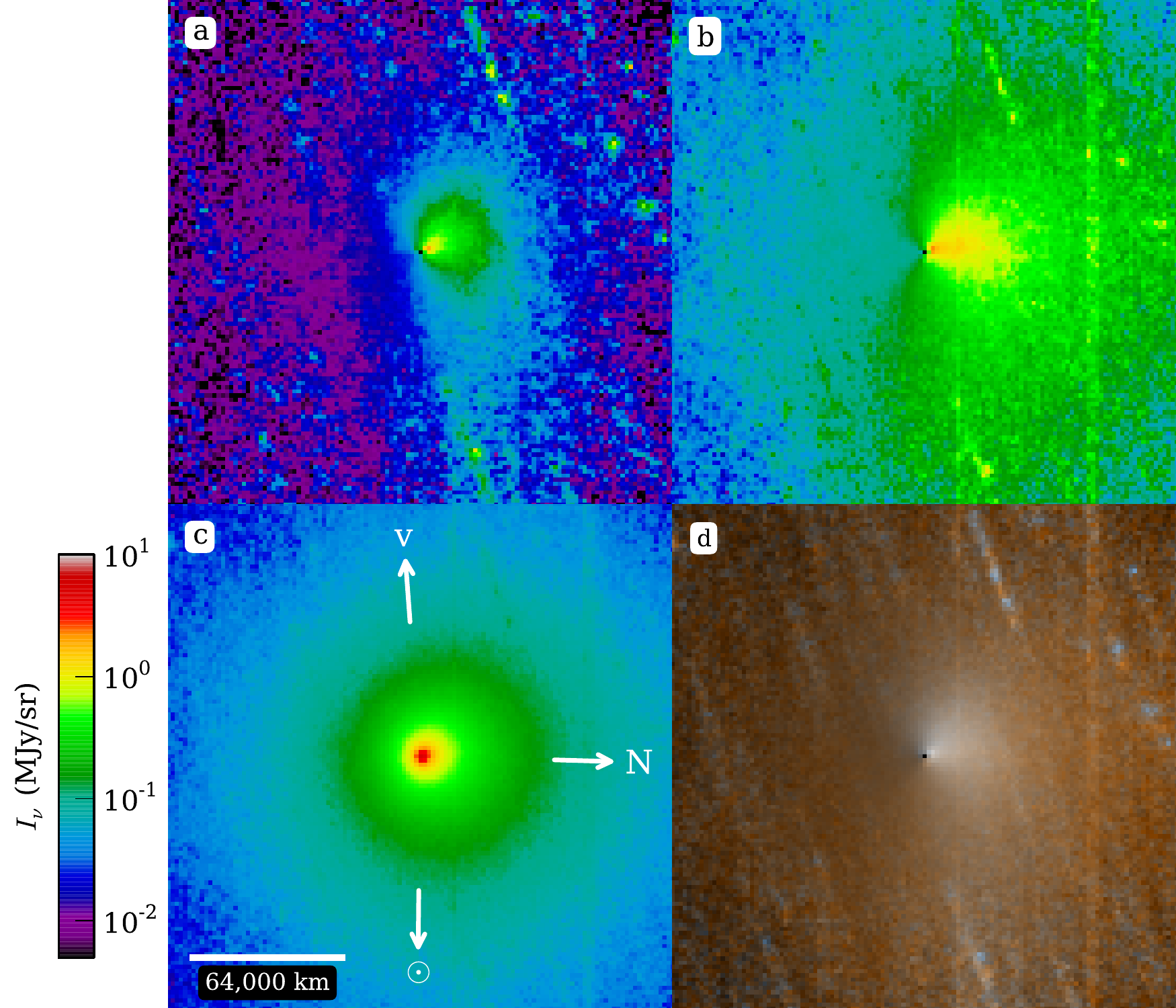}
\caption{
IRAC images of 65P/Gunn on 2009 Aug 6, pre-perihelion. The panels are in the same format as in Figure~\protect \ref{tempel21009}.
The image show a characteristic feature of many of the survey comets, with a roughly spherical diffuse coma at 4.5 $\mu$m.
\label{gunn0908}}
\end{figure}

\begin{figure}
\includegraphics[width=5in]{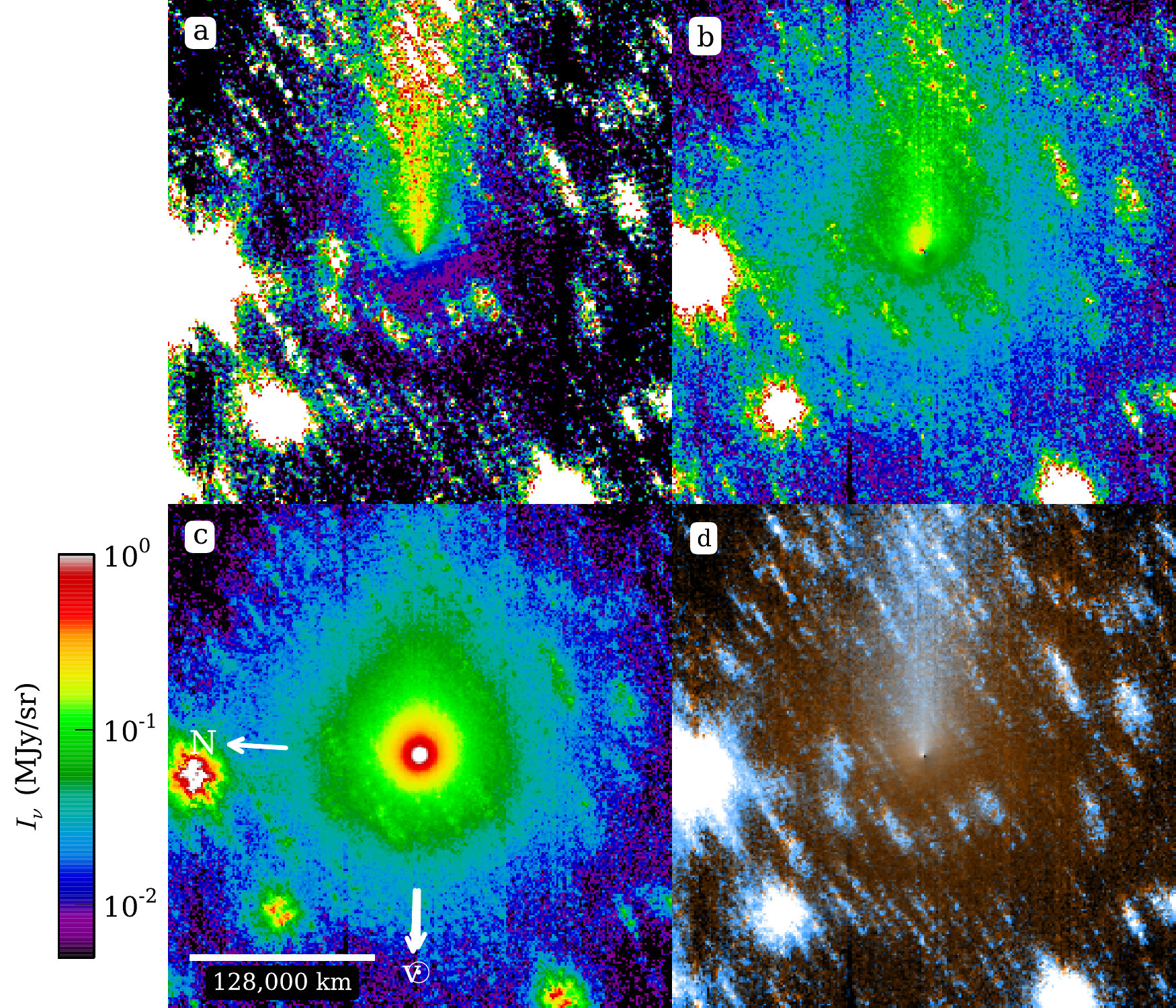}
\caption{
IRAC images of 65P/Gunn on 2010 Jul 30, post-perihelion.The panels are in the same format as in Figure~\protect \ref{tempel21009}.
The images show the characteristic features  of many of the survey comets including the pre-perihelion image of this comet in Figure~\ref{gunn0908}, 
together with a narrower, antisolar tail of dust that is relatively more prominent at 3.6 $\mu$m.
\label{gunn1007}}
\end{figure}

Figure~\ref{gunn0908} shows the images of 65P/Gunn in 2009 Aug, and 
Figure~\ref{gunn1007} shows the images of 65P/Gunn in 2010 Jul.
In both images there is a dust feature that is present at both 3.6 and 4.5 $\mu$m plus a larger, diffuse region that is present at 4.5 $\mu$m only.
The dust feature in 2009 Aug is fan-shaped and is not directed antisunward; indeed, it is almost perpendicular to the sunward vector. This leads us to 
suspect this is not a dust tail but rather is an active jet. In 2010 Jul, the dust feature is consistent with being a dust tail (narrow and directed antisunward). The gas feature is roughly a spherical region offset from the nucleus. In 2009 Aug, the gas feature is offset from the nucleus in the same direction as the dust feature though it is more diffuse. We suspect the emission in 2009 Aug is mostly from the same active region producing both dust and CO$_2$.

65P was covered in the WISE all-sky survey on 2010 Apr 24. We obtained the images from the Infrared Science Archive and measured the fluxes using the same aperture photometry procedure (including curve of growth analysis) that we had applied to the {\it Spitzer} data. While the image is substantially less sensitive than the {\it Spitzer} image, the coma was clearly detected by WISE at all 4 of its wavelengths,
 with fluxes at 3.4, 4.6, 12, and 22 $\mu$m of 9.4, 31, 810, and 5900 mJy, respectively. The 12 and 22 $\mu$m images are clearly dominated by a dust tail, while at 3.4 and 4.6 $\mu$m the images show only the coma. The relative
 fluxes are consistent with dust emission at 12 and 22 $\mu$m; the estimated dust emission at 4.5 $\mu$m based on the color temperature is less than 1 mJy. The 3.4 $\mu$m flux is most likely scattered sunlight, from which we estimate the
 scattered sunlight at 4.6 $\mu$m is 5 mJy. That means that the bulk (more than 80\%) of the 4.6 $\mu$m flux is not from dust, in agreement with the inference based on the {\it Spitzer} 2-band images.

\clearpage

\subsubsection{88P/Howell}
\begin{figure}
\includegraphics[width=5in]{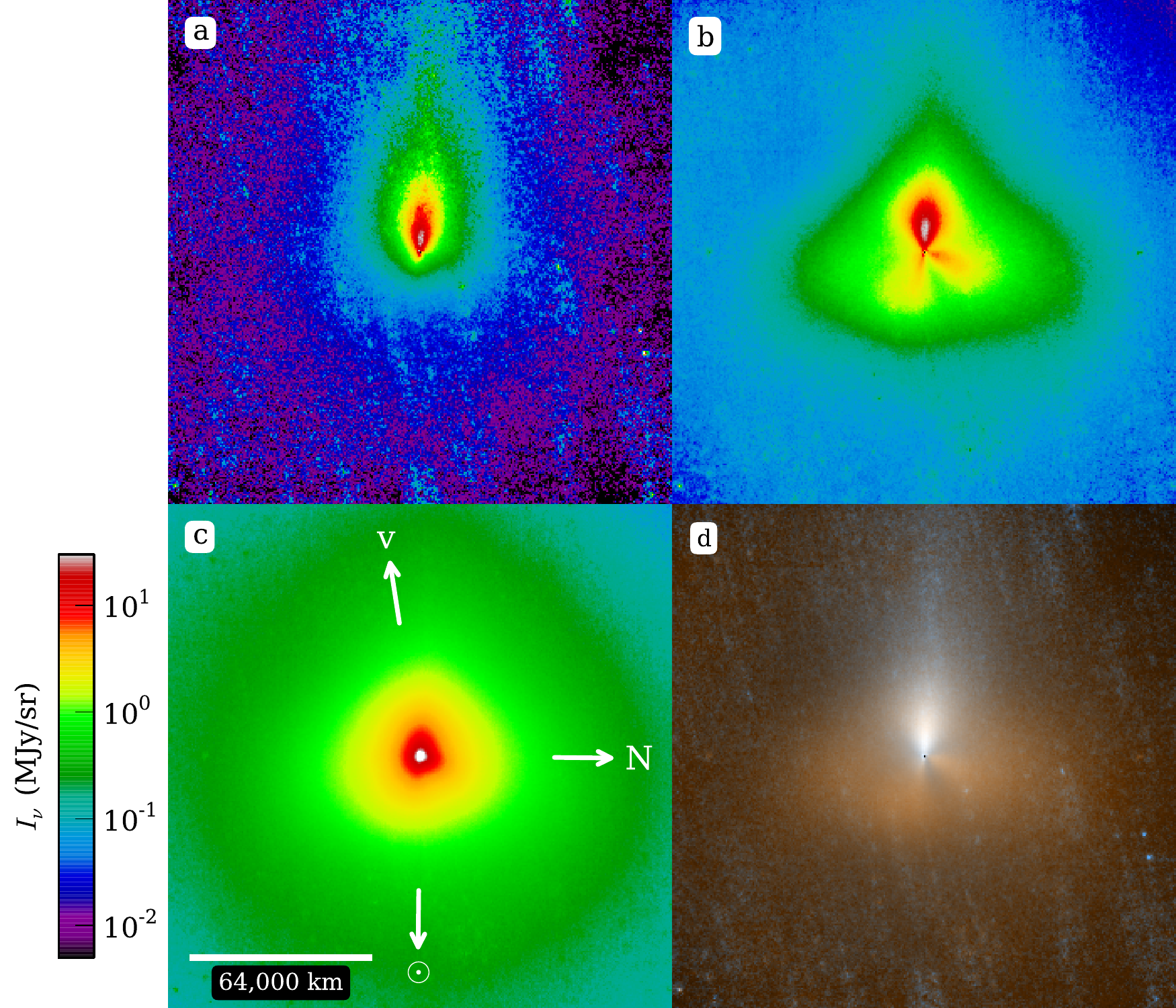}
\caption{
IRAC images of 88P/Howell on 2009 Aug 24.The panels are in the same format as in Figure~\protect \ref{tempel21009}.
The images show a dust tail plus a remarkable pair of jet-like structures in the sunward direction. The jet-like structures are
only present in the 4.5 $\mu$m image and are likely CO$_2$-gas only.
\label{howell0908}}
\end{figure}

Figure~\ref{howell0908} shows the images of 88P/Howell in 2009 Aug. The 4.5 $\mu$m image has perhaps the most remarkable morphology of 
any of the images from the survey. While it is not as complicated as the high-quality images of 10P/Tempel 2 and 2P/Encke, the gas emission from
88P/Howell is clearly distinct and shows two very prominent jets that attach to the nucleus. A third ray emanating from the comet is the dust tail, present at both 3.5 and 4.5 $\mu$m and directed away from the Sun.
The two jets are at position angles $350^\circ$ and $273^\circ$ E of Celestial N.
Based on its 4.5 $\mu$m and 3.6 $\mu$m fluxes, this comet had the highest CO+CO$_2$ production rate among the short-period comets we observed, and it also had the highest CO+CO$_2$ abundance relative to H$_2$O (approximately 1:1).
For comparison, the comet was observed by {\it Akari} on 2009 Jul 3 (52 days before the {\it Spitzer} observation), and the CO$_2$ production rate was lower, with proportions 4:1 of H$_2$O:CO$_2$ \citep{ootsubo2012}. It is likely that the comet's activity actually changed significantly during the 52 days between the two observations.
We suspect that the two prominent jets evident in the {\it Spitzer} images are CO$_2$ rich, and they may be only activated between the times when the two
space telescopes observed them.

\clearpage

\subsubsection{103P/Hartley 2}
\begin{figure}
\includegraphics[width=5in]{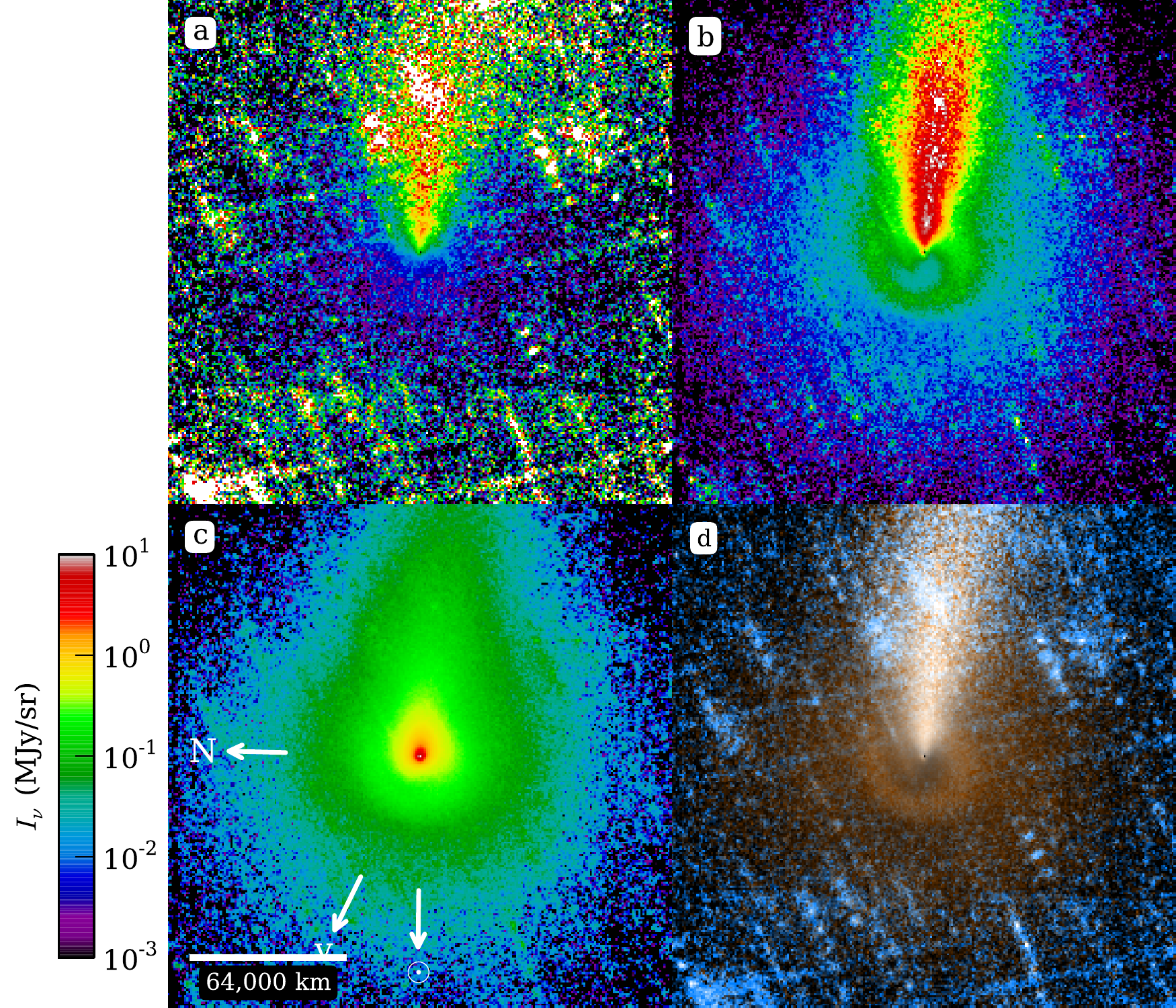}
\caption{
IRAC images of 103P/Hartley 2 on 2011 Jan 26.The panels are in the same format as in Figure~\protect\ref{tempel21009}.
The images show a dust tail (both 3.6 and 4.5 $\mu$m), a diffuse spherical brightness roughly centered on the nucleus at 4.5 $\mu$m only,
plus a sunward, loop-like structure that is in gas (4.5 $\mu$m) only.
\label{hartley21101}}
\end{figure}

Figure~\ref{hartley21101} shows the post-perihelion image of 103P/Hartley 2.  The comet is significantly brighter at 4.5 $\mu$m than at 3.6 $\mu$m, and it shows an unusual loop-like feature on the sunward side that is apparently gas-only (not present at 3.5 $\mu$m). As the EPOXI mission target, 103P is one of the few comets with detailed nucleus images; the near-infrared instruments also discovered that the surface activity of this comet is driven by CO$_2$, in particular from the jets, with a more distributed source of H$_2$O from ice chunks that form a swarm around the nucleus \citep{ahearn11epoxi,kelley13}. The {\it Spitzer} images support a strong CO$_2$ presence, and furthermore shed some light on the wider spatial distribution of the CO$_2$ around the nucleus. The morphology of a loop is unique among comets surveyed. In addition to the loop, there is a lower-amplitude, wide distribution of 4.5 $\mu$m emission with an approximate $\rho^{-1}$ distribution. This wider skirt
of emission is consistent with a steady free expansion of CO$_2$, with the loop superposed as an enhancement.
The loop could be an edge-brightened shell form an impulsive event (outburst). If so, and the ejecta velocity is $v_{ej}$ in km~s$^{-1}$, then the outburst occurred approximately $2 v_{ej}^{-1}$ hr before the image was taken. The loop could also be part of a spiral (seen at an angle) from the dominant active region at the end of one of the two lobes of the nucleus that was found to be extremely active during the EPOXI encounter on 2010 Nov 4. The {\it Spitzer} image was taken 83 days later than the EPOXI flyby.

\clearpage

\subsubsection{118P/Shoemaker-Levy 4}
\begin{figure}
\includegraphics[width=5in]{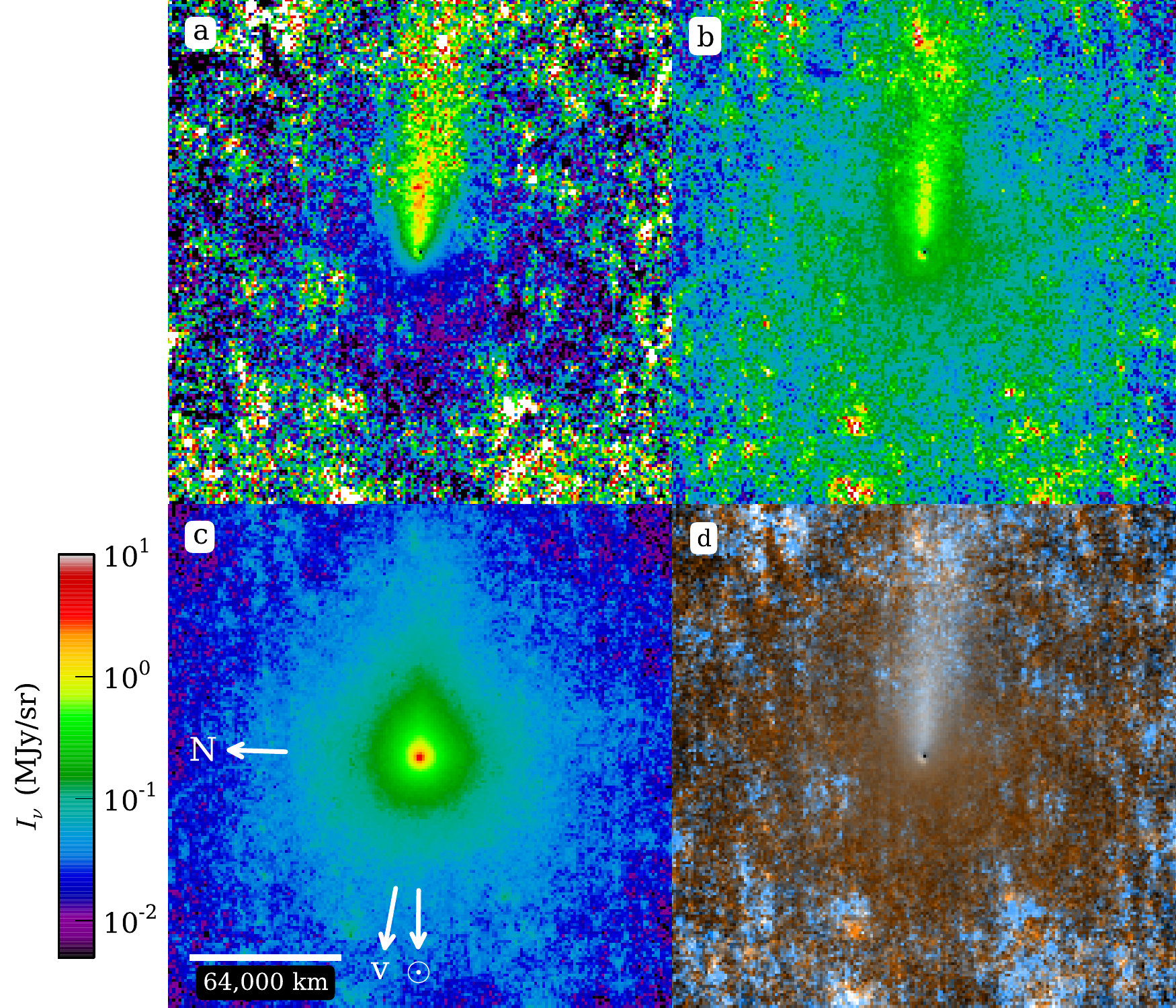}
\caption{
IRAC images of 118P/Shoemaker-Levy 4 on 2009 Nov 28.The panels are in the same format as in Figure~\protect\ ref{tempel21009}.
The images show a dust tail (both 3.6 and 4.5 $\mu$m), plus a sunward, spherical coma that is in gas (4.5 $\mu$m) only.
\label{sl40911}}
\end{figure}

Figure~\ref{sl40911} shows the images of 118P/Shoemaker Levy 4 in 2009 Nov. The images are uncannily similar to those of 65P/Gunn in 2010 Jul,
with a dust tail (seen at both 3.6 and 4.5 $\mu$m and directed antisunward) and roughly spherical diffuse gas coma (seen only at 4.5 $\mu$m). 

\clearpage

\subsection{Centaur}

\subsubsection{29P/Schwassmann-Wachmann 1}

\begin{figure}
\includegraphics[width=5in]{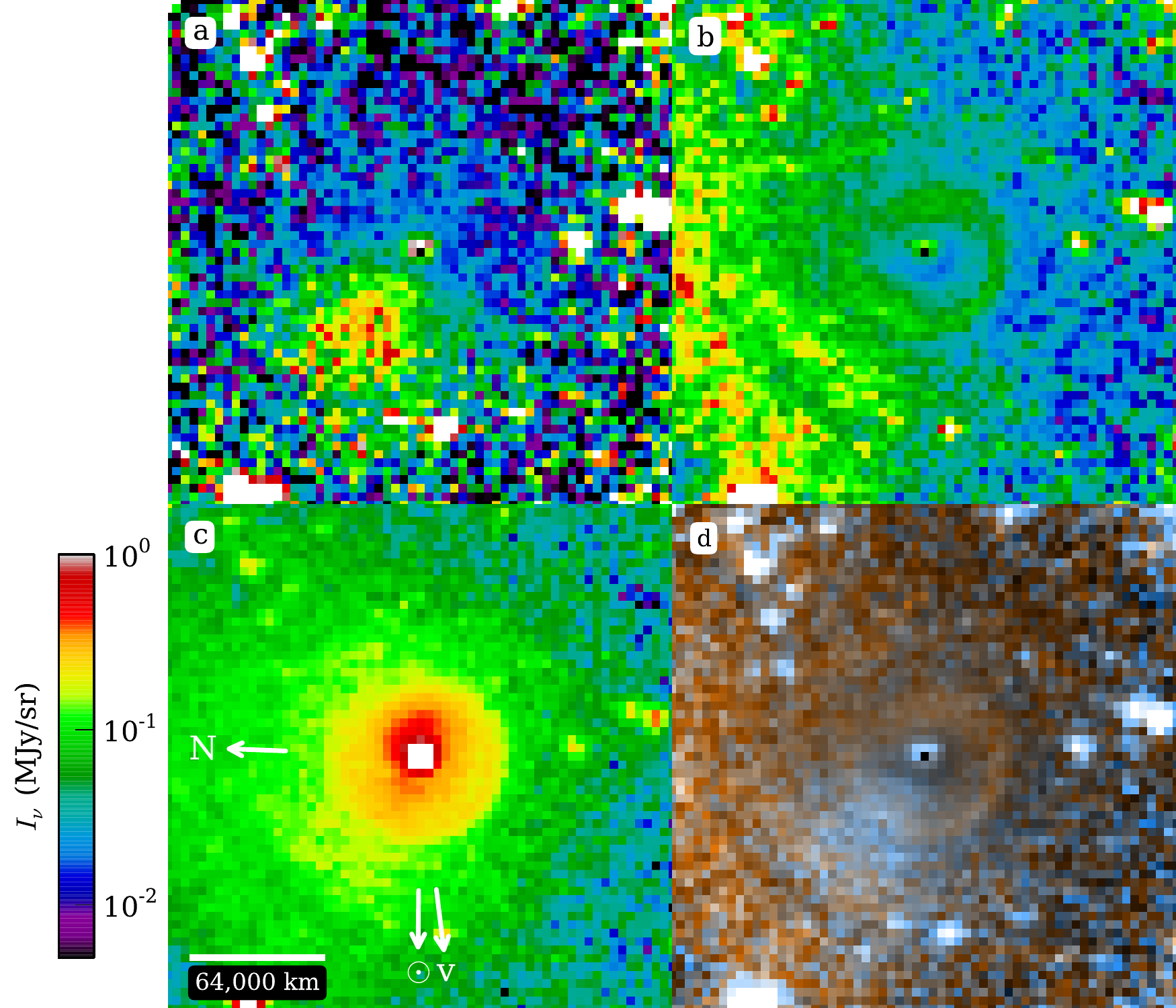}
\caption{
IRAC images of 29P/Schwassmann-Wachmann 1 on 2010 Jan 27. The panels are in the same format as in Figure~\protect\ref{tempel21009}.
The image shows concentric rings, in particular at 4.5 $\mu$m, and a possible limb-brightened hemispherical shell.
\label{sw11001}}
\end{figure}

Figure~\ref{sw11001} shows the images of 29P/Schwassmann-Wachmann 1 in 2010 Jan.
The most obvious feature is a set of arcs in the 4.5 $\mu$m image. These can be seen with effort even in the raw image (Fig.~\ref{sw11001}a), 
and they are seen easily in the image
that was flattened by multiplication by distance to the nucleus (Fig.~\ref{sw11001}b) and the color combination image (Fig.~\ref{sw11001}d).
The arcs appear to form an Archimedean spiral. For a single jet to generate the spiral, if we assume constant expansion at velocity $v$, then
the distance of a particle from the nucleus at time $t$ since its launch is $r = v t$,
and the azimuthal angle (if the jet is perpendicular to the plane of the sky) is $\theta = 2\pi/P t$ where $P$ is the period. The
spiral is then 
\begin{equation}
r = \frac{v P}{2\pi} \theta ,
\end{equation}
which we can compare to the image to measure the product of expansion speed and rotation period. We find the jet is best fit if it
is inclined by $6^\circ$ from the plane of the sky. Taking this into account, a combination of $v=1$ km~s$^{-2}$ and
$P=8.5$ hr works. An equally good fit using the rotation period of 14 hr from \citet{meech93} , and an expansion velocity of 0.5 km~s$^{-1}$.
The observing strategy repeated the images three times, with the second and third epochs separated by 7.83 and 48.63 hr, respectively, from the first epoch. A spiral can fit all three epochs with the rotational phase of the nucleus being -30$^\circ$, and -131$^\circ$ at the second and third epochs, relative to an assumed zero phase at the first epoch. 

\def\commentbymike{ ???
[IÕve now realized that the sense of rotation that I had in my spiral fitting was opposite the image, but even after this correction, IÕm still having difficulty including the inner-most spiral into
the fit. It would help if it were easier to see the arcs on the right hand side of the image. Maybe
the 1/rho normalization is distorting the radial positions of the inner-arms? Instead, I think we
should take the same approach as you did for 06W3, and look at the distance between arcs on the
left hand side}

\clearpage

\subsection{Jupiter-family comets whose 4.5 $\mu$m emission is dominated by dust}

\subsubsection{81P/Wild 2}
Comet 81P was observed twice in this survey.
The total flux was consistent with dust emission
and no need for significant amounts of CO$_2$. 
Even in the closer-range images taken on 2010 Apr 12, there was no distinct gas emission evident morphologically.
The image at 4.5 $\mu$m is much higher contrast compared to the crowded field of stars and galaxies, making it possible to detect the tail all the way to the edge of the image, $7'$ from the nucleus. The
brightness of the tail more than $3'$ from the nucleus is at least 4 times greater at 4.5 $\mu$m than at 3.6 $\mu$m. But this
increased brightness at 4.5 $\mu$m is explained for this observation by dust emission. Referring to Figure~\ref{fluxratplot}, at
1.7 AU, the dust emission is predicted to be more than 5 times greater at 4.5 than at 3.6 $\mu$m, a consequence of the steep Wien 
portion of the blackbody curve. The images at the two IRAC wavelengths are similar, which further supports the idea that the
infrared emission from this comet can be explained by dust alone.

\begin{figure}
\includegraphics[width=5in]{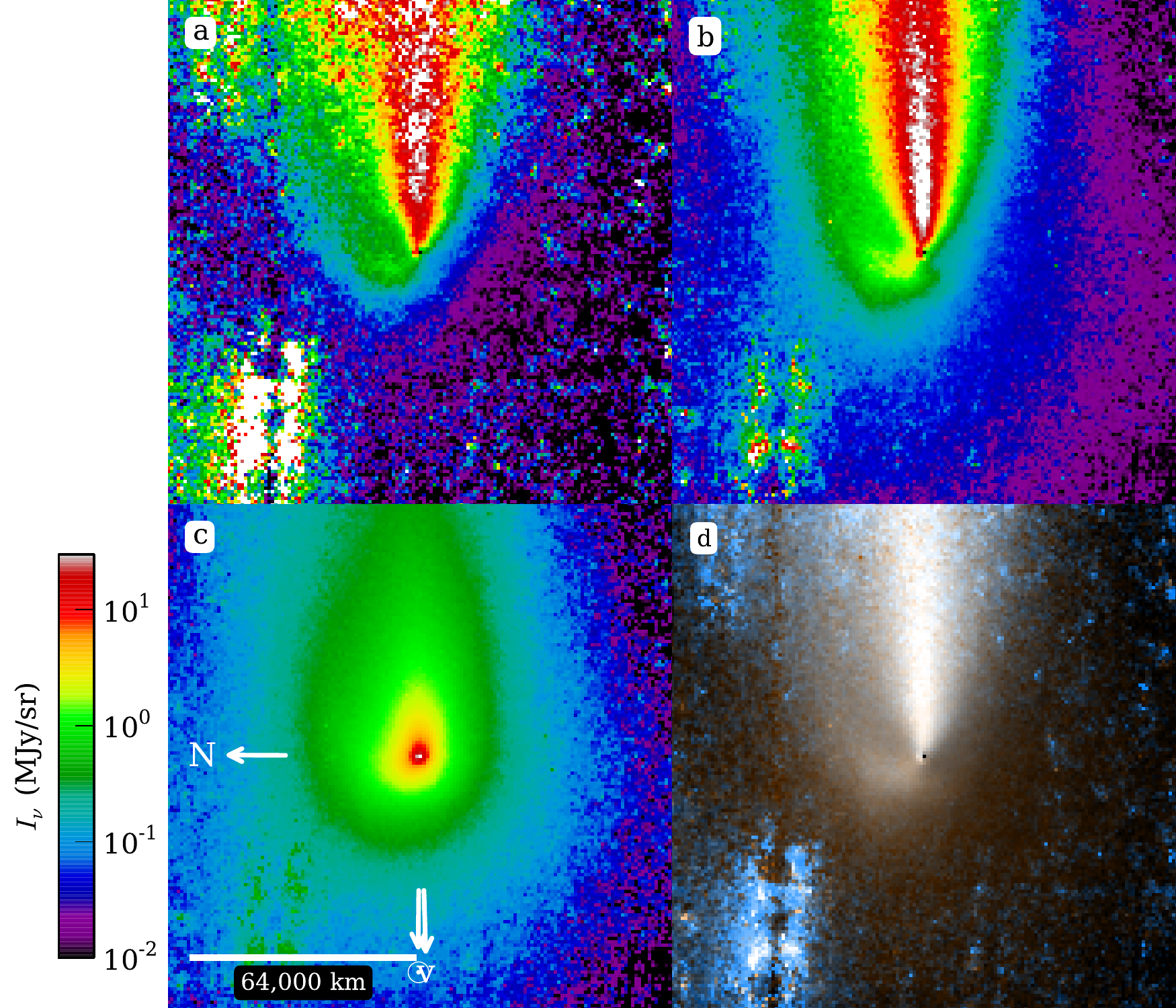}
\caption{
IRAC images of 81P/Wild 2 on 2010 Apr 12.The panels are in the same format as in Figure~\protect \ref{tempel21009}.
The images show a dust tail and a jet-like feature at both 3.6 and 4.5 $\mu$m. The morphologies are similar so the emission
appears to be dominated by dust. The jet-like feature is not a dust tail, because it is nearly in the sunward direction.
\label{wild21004}}
\end{figure}

\begin{figure}
\includegraphics[width=5in]{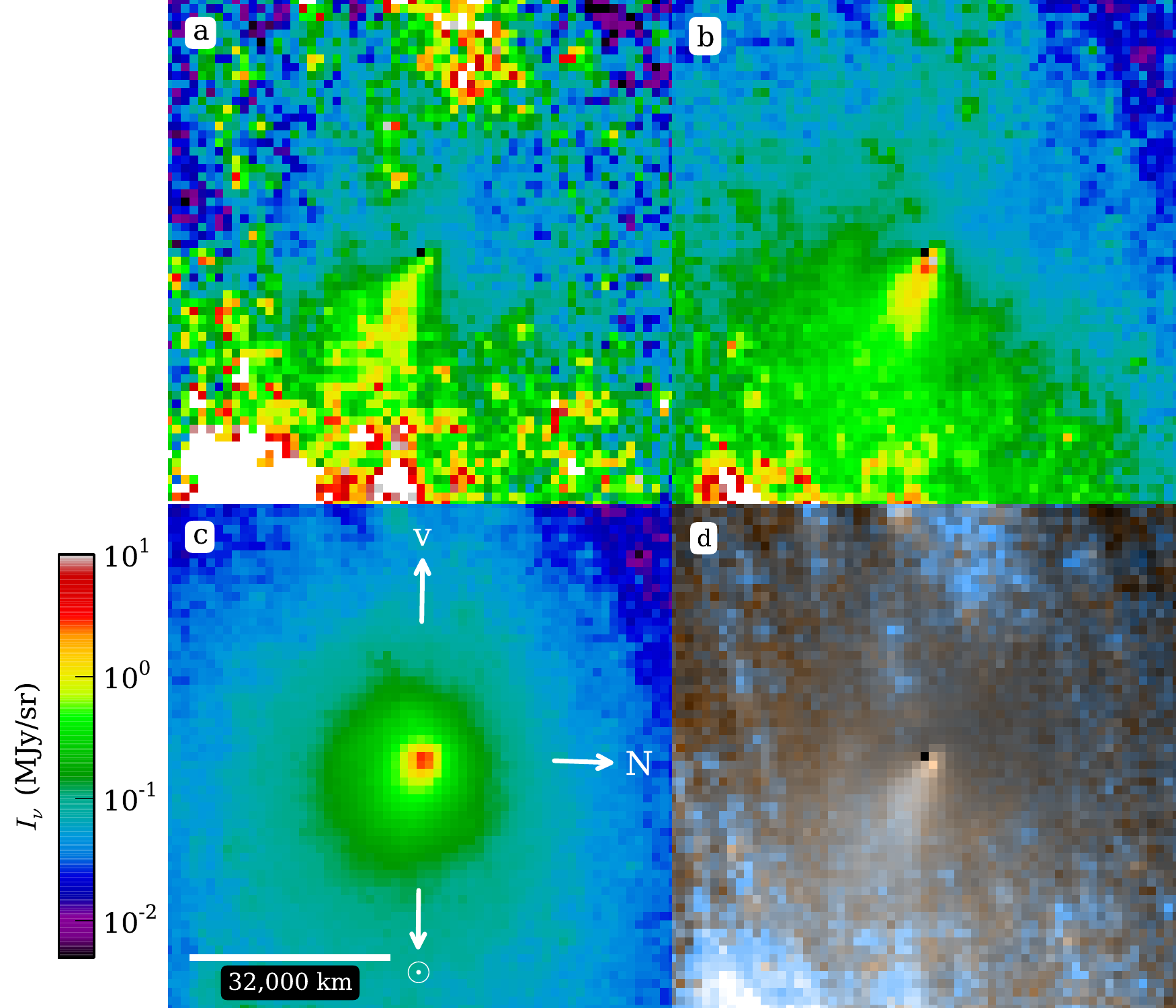}
\caption{
IRAC images of 81P/Wild 2 on 2010 Sep 26.The panels are in the same format as in Figure~\protect \ref{tempel21009}.
The images show a coma and a jet-like feature at both 3.6 and 4.5 $\mu$m. The morphologies are similar so the emission
appears to be dominated by dust. The jet-like feature is not a dust tail, because it is nearly in the sunward direction.
\label{wild21009}}
\end{figure}

The images from 2010 Apr and 2010 Sep are shown in Figures~\ref{wild21004} and \ref{wild21009}. The 2010 Apr image clearly shows an antisolar dust tail due to small dust grains. That tail is not present in the 2010 Sep image. In both images, there is a jet-like feature extending from near the nucleus out to at least 30,000 km. While this feature looks like a dust tail in morphology and from the fact that it is similar at both 3.6 and 4.5 $\mu$m, in fact it cannot be a dust tail. This is obvious from the 2010 Apr image, where
the actual dust tail is clearly seen in the antisunward direction.
Radiation pressure drives dust particles to the antisunward direction, which is labeled in each image. 
The jet-like feature is at a position angle $246^\circ$ E of N, which nearly {\it toward} the Sun (the sunward direction is $280^\circ$ E of N). 
It is remarkable that the jet-like structure has approximately the same position angle relative to the sunward direction in the 2010 Sep image. 
The jet is somewhat brighter at 4.5 $\mu$m (compared the expectation for dust) and may be CO$_2$-rich.
It is notable that while the tail is only present in the 2010 Apr image at 1.7 AU from the Sun, the jet is present in both that image and the one at 2.5 AU from the Sun.
Therefore the driver of activity for the small grains in the tail may have been activated only close to the Sun, possibly due to water ice sublimation that was subsiding by 
the time the comet had reached 2.5 AU. The driver of the jet, on the other hand, may have more CO$_2$.

\clearpage

\subsubsection{77P/Longmore}
\begin{figure}
\includegraphics[width=5in]{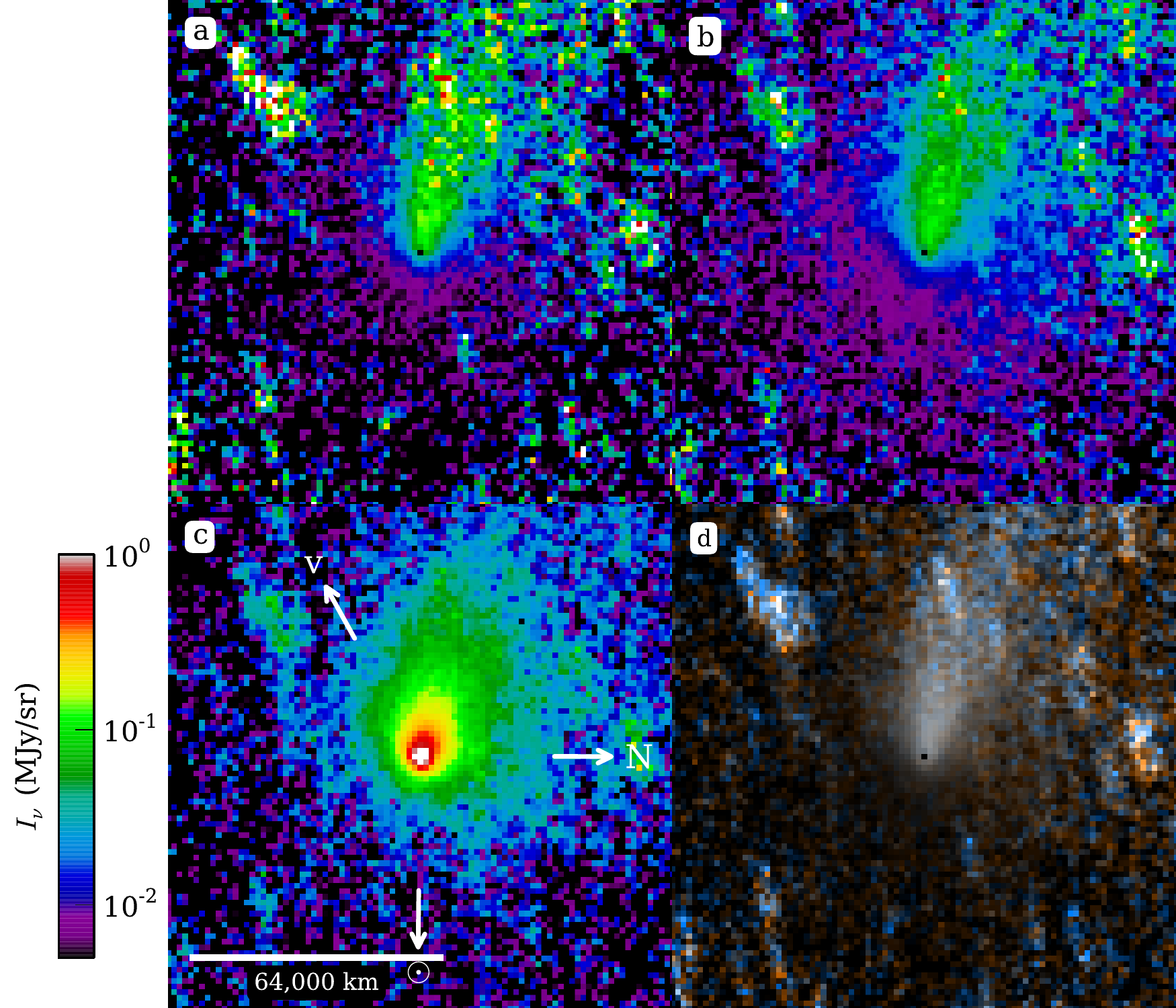}
\caption{
IRAC images of 77P/Longmore on 2009 Aug 13. The panels are in the same format as in Figure~\protect \ref{tempel21009}.
The images show a dust tail only.
\label{longmore0908}}
\end{figure}

Figure~\ref{longmore0908} shows the images of comet 77P/Longmore in 2009 Aug. The 3.6 and 4.5 $\mu$m images are similar, with a roughly antisolar dust tail.

\subsubsection{116P/Wild 4}
\begin{figure}
\includegraphics[width=5in]{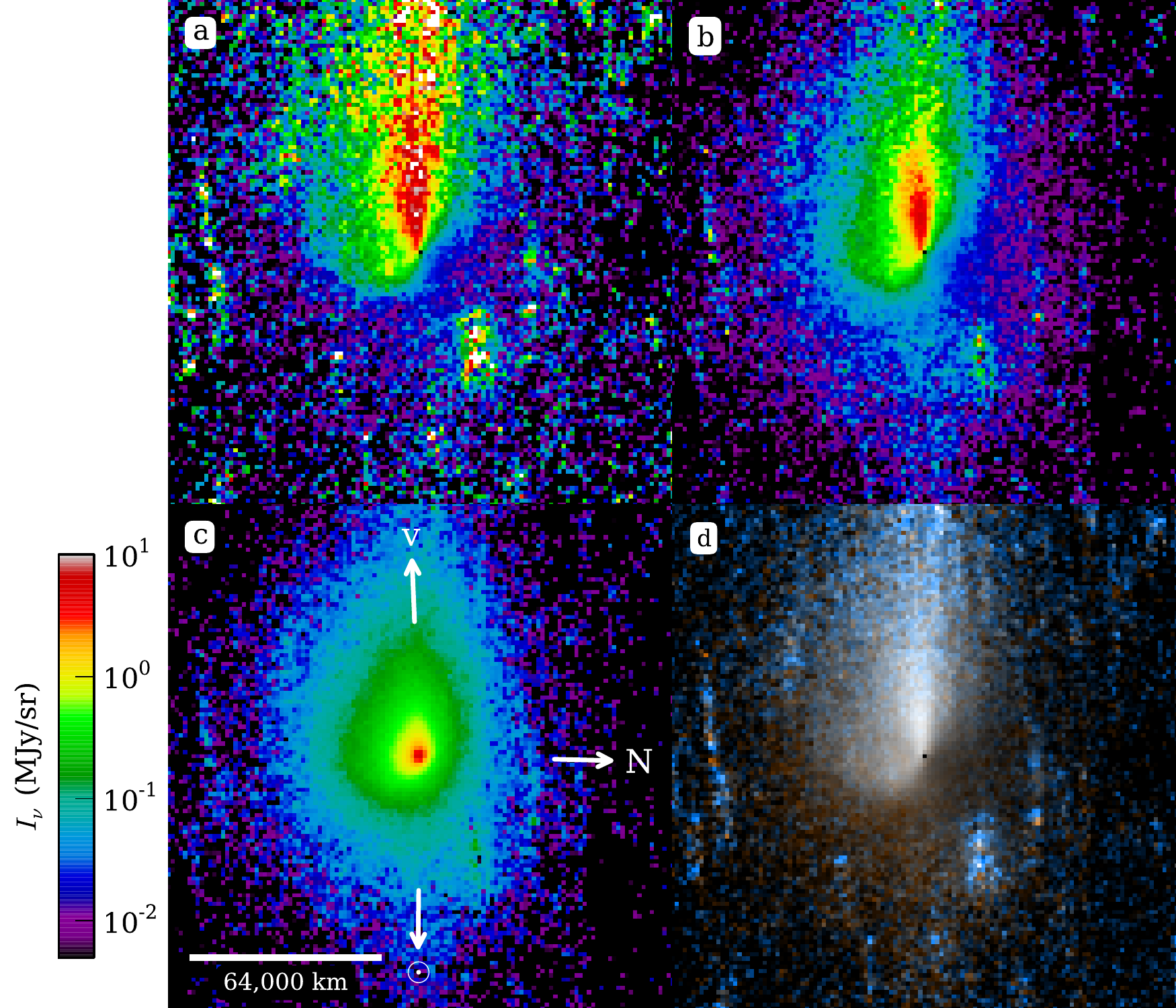}
\caption{
IRAC images of 116P/Wild 4 on 2009 Jul 28.The panels are in the same format as in Figure~\protect \ref{tempel21009}.
The images show a dust tail and coma only.
\label{wild40907}}
\end{figure}

Figure~\ref{wild40907} shows the images of comet 116P/Wild 4 in 2009 Jul. The 3.6 and 4.5 $\mu$m images are similar, with a roughly antisolar 
dust tail plus a jet-like feature more on the sunward side.  The images are very similar to those if 81P/Wild 2 in 2010 Apr, with the jet-like feature
at nearly the same position angle relative to the sunward direction.

\subsubsection{32P/Comas Sola}
\begin{figure}
\includegraphics[width=5in]{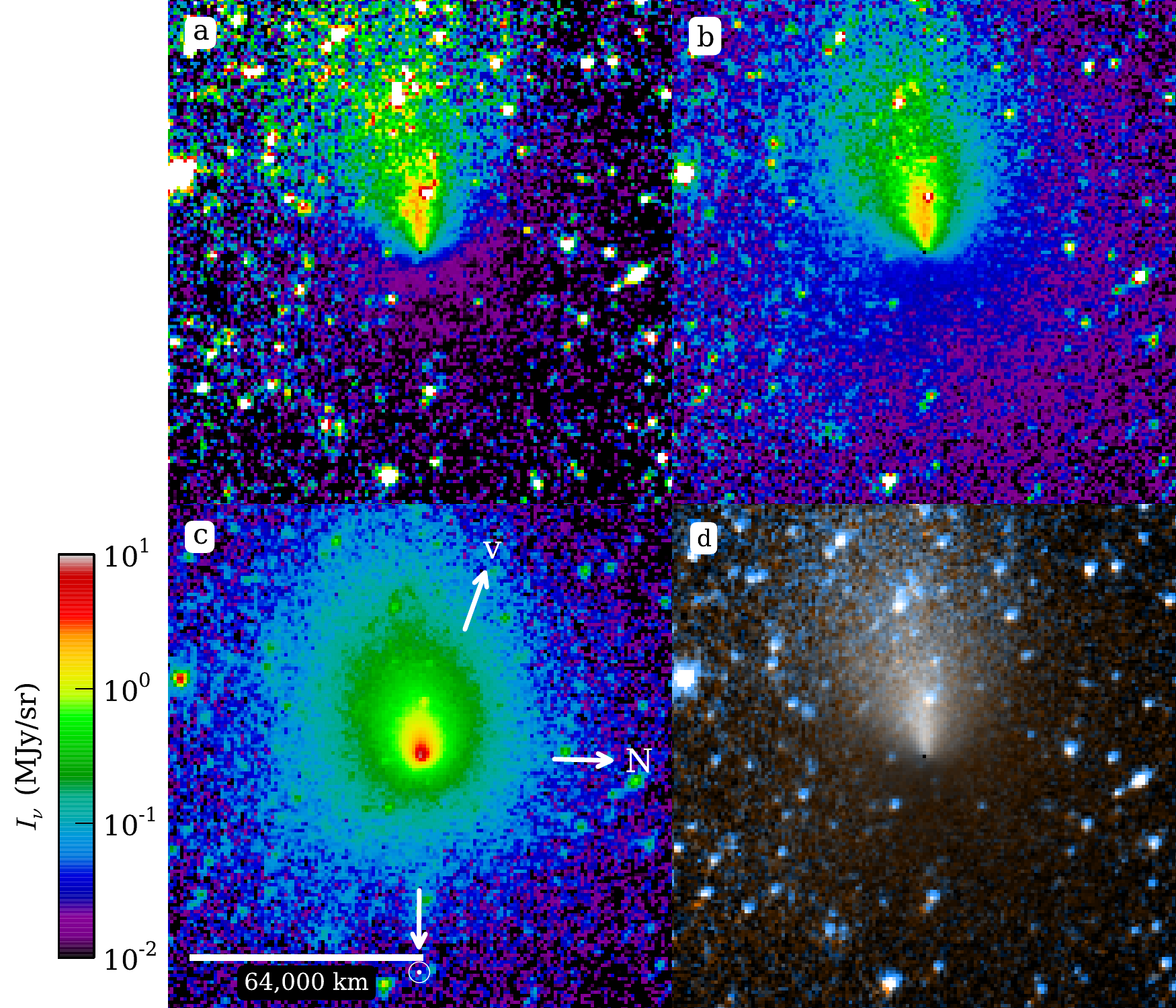}
\caption{
IRAC images of 32P/Comas Sola 4 on 2005 Jan 19.The panels are in the same format as in Figure~\protect \ref{tempel21009}.
The images show a dust tail and coma only.
\label{comassola0501}}
\end{figure}

Figure~\ref{comassola0501} shows the images of comet 32P/Comas Sola in 2005 Jan. The 3.6 and 4.5 $\mu$m images are similar, with a roughly antisolar dust tail.

\clearpage

\subsection{Oort cloud comets}

\subsubsection{C/2002 T7 (LINEAR)}
\begin{figure}
\includegraphics[width=5in]{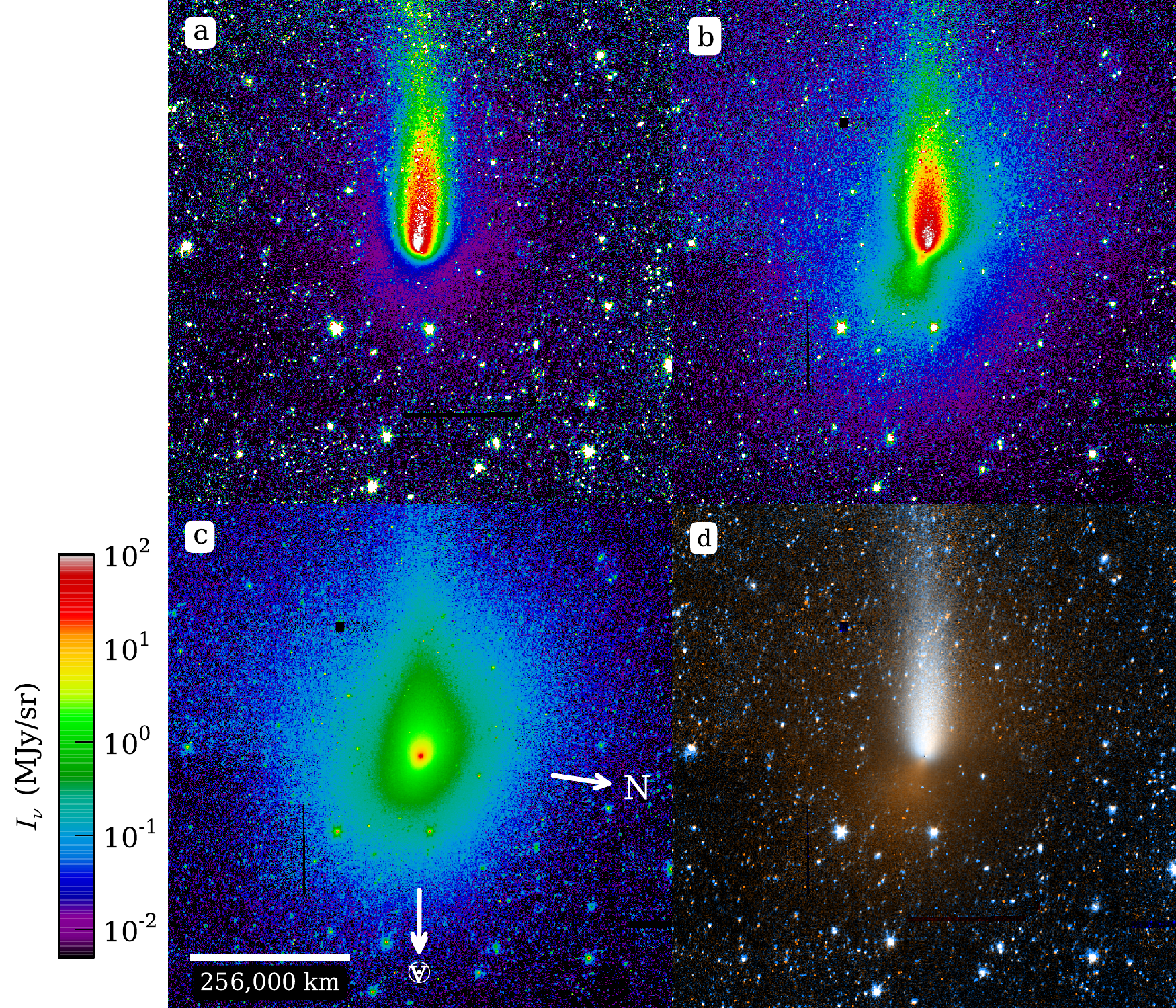}
\caption{
IRAC images of C/2002 T7 (LINEAR) on 2003 Dec 28. The panels are in the same format as in Figure~\protect \ref{tempel21009}.
While the total flux of the comet at 4.5 $\mu$m is consistent with being mostly dust emission, the morphology shows clear
distinction between the gas and dust, perhaps in part because the images of are high signal-to-noise (permitting detection of faint features).
Most notable is the very prominent jet on the sunward side of the comet, which is visible in gas only.
\label{c2002t70312}}
\end{figure}

Figure~\ref{c2002t70312} shows the images of C/2002 T7 in 203 Dec, pre-perihelion. The images are dominated by a dust tail and coma, in terms of total brightness, but there is a prominent gas jet (4.5 $\mu$m only) on the sunward side of the comet.

\subsubsection{C/2007 Q3 (Siding Spring)}
\begin{figure}
\includegraphics[width=5in]{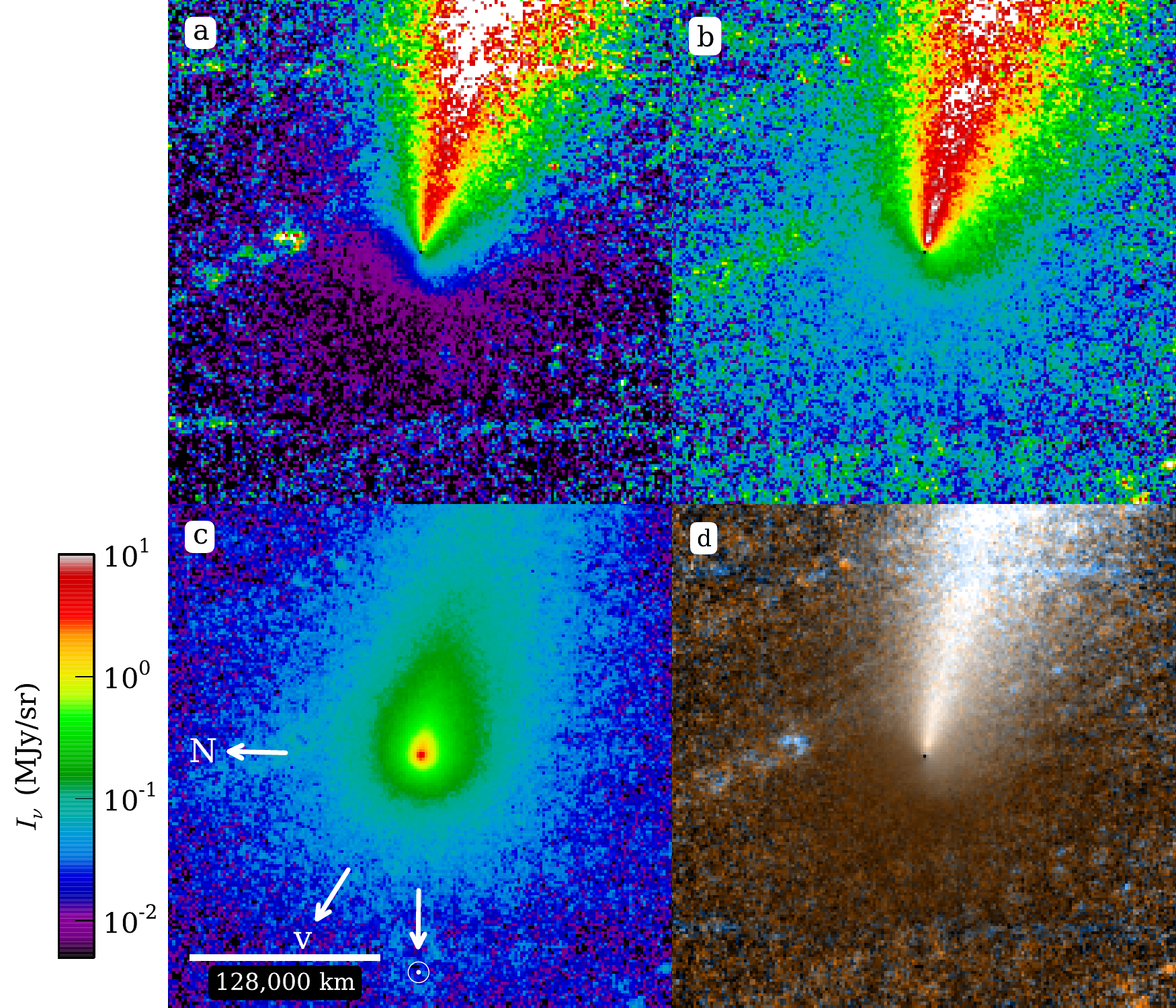}
\caption{
IRAC images of C/2007 Q3 (Siding Spring) on 2010 Aug 26. The panels are in the same format as in Figure~\protect \ref{tempel21009}.
Emission from this comet is dominated by dust, based on the similar morphologies at 3.6 and 4.5 $\mu$m, with a narrow dust tail directed roughly antisunward.
\label{c2007q31008}}
\end{figure}

Figure~\ref{c2007q31008} shows the images of C/2007 Q3 in 2010 Mar. The images are dominated by a dust tail.
However, the peak of the $\rho^{-1}$ normalized image is significantly offset from the center 
of the comet. This indicates dust production is occurring within the tail. The cause appears to
be a fragmentation event. On 2010 Mar 13, the day before our first observation of this comet, a
fragment was discovered approximately 6$''$ from the primary nucleus \citep{colas10}. 
This fragment, and any other smaller fragments produced in the preceding fragmentation event, would cause the observed
offset between the tail peak and the nucleus in the $\rho^{-1}$ normalized images. 

\subsubsection{C/2006 W3 (Christensen)}
\begin{figure}
\includegraphics{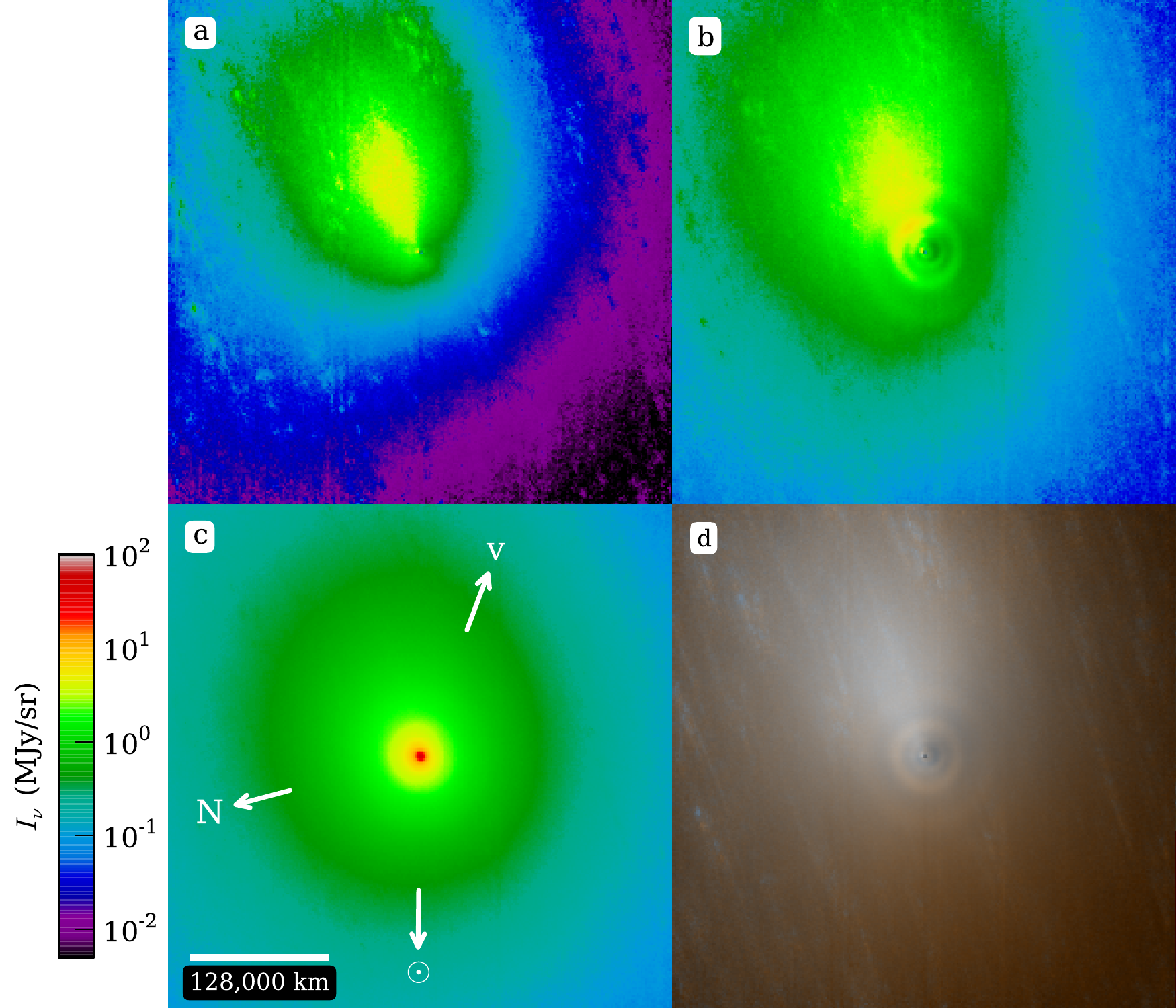}
\caption{
IRAC images of C/2006 W3 (Christensen) on 2009 Aug 6. The panels are in the same format as in Figure~\protect \ref{tempel21009}.
In addition to the prominent dust tail, which is bright at both wavelengths, the 4.5 $\mu$m image shows pronounced shell-like arcs.
\label{c2006w30908}}
\end{figure}

Figure~\ref{c2006w30908} show the images of C/2006 W3 in 2009 Aug, when the comet was 3.1 AU from the Sun.
The images show clear arc-like asymmetries, approximately as rings with centers slightly offset from the nucleus. These
rings are far too distant and from the nucleus, and too high in brightness, to be diffraction rings from the central
condensation (which is, in any event, an extended source due to the dust coma and not the nucleus itself). 
The observing sequence was repeated three times. The rings appear in each of the three images, at approximately
the same locations, but with a perceptible motion away from the nucleus in sequence from the first to second to third epoch. 
Using the images taken 155 min apart, for the arc at 35,300 km from the nucleus, the measured projected expansion velocity 
is  $0.46\pm 0.04$ km~s$^{-1}$. The inferred rotation rate for the nucleus is 21 hr.

The shell-like (or spiral) features from C/2006 W3 images are only in the 4.5 $\mu$m images, not at 3.6 $\mu$m. This suggests
that they are only present from CO$_2$ (or possibly CO) gas and not from dust. The segregation between the gas and dust
could have occurred post-ejection (and post decoupling), when the gas and dust feel different dynamical influences of solar radiation pressure;
the dust, specifically, gets swept back into the anti-solar tail.

\begin{figure}
\includegraphics[width=5in]{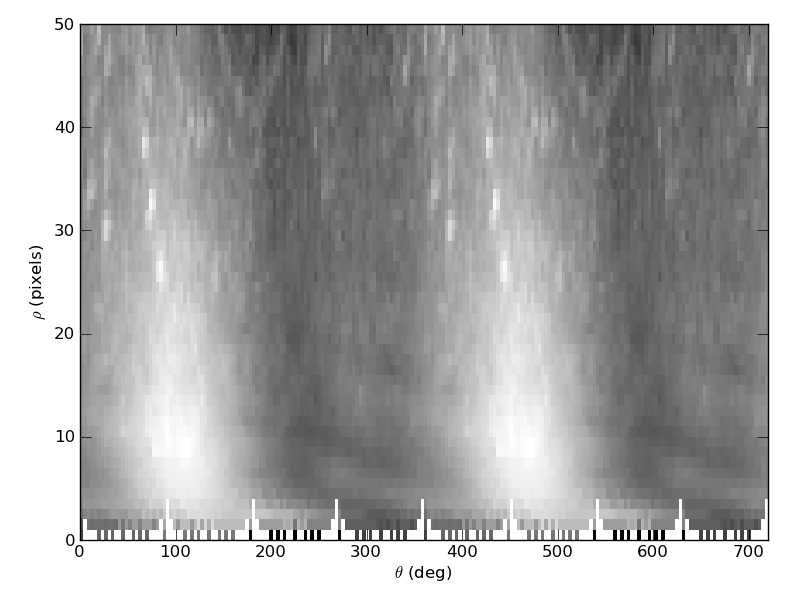}
\caption{
Cylindrical interpolation of the IRAC 4.5 $\mu$m image of C/2006 W3, versus azimuthal angle about the nucleus ($\theta$) and 
projected radial distance from the nucleus $\rho$. In such an image, a circular ring (or limb-brightened spherical shell) would appear as 
a horizontal feature; a radial spoke would appear as a vertical structure; and an Archimedean spiral would appear as a diagonal structure. 
The tail is roughly a vertical feature, since it is directed nearly anti-sunward in a line extending from the nucleus.
The fainter non-azimuthally-symmetric emission from the comet appears consistent with a spiral, with three wraps evident in the image.
\label{spiral}}
\end{figure}

Figure~\ref{spiral} shows an ($r$, $\theta$)  cylindrical projection of the image taken in 2010 Jun, when the comet was at 4.5 AU from the Sun. 
It appears that the faint extended asymmetries in the 
image may be more consistent with a spiral than with spherical shells. From the measured slope of
the spiral feature, an expansion velocity of $0.45\pm 0.03$ km~s$^{-1}$ is inferred; this is in excellent
agreement with the velocity from the observation taken at 3.1 AU.

Rotation periods are difficult to obtain for comets, in particular for dynamically new comets.
A short period could indicate a recent (off-axis) collision or
non-axisymmetric outgassing.
In their review, \citet{comrot04}\footnote{data from the Planetary Data System, Small Bodies Node, dataset EAR-C-COMPIL-5-COMET-NUC-ROTATION-V1.0}
show 24 comets with rotation periods measured. Of these, four are dynamically new: 
C/Hyakutake has a period of 6.2 hr;
C/Hale-Bopp has a period of 11 hr; 
C/Levy is reported twice, with periods of 17 and 8.4 hr; and
C/IRAS-Araki-Alcock has a period of 51 hr.
The period we measure for C/Christensen is in the range of these values.

\clearpage

\section{Conclusions}

We imaged 23 comets, some at 2 or 3 epochs, using {\it Spitzer}/IRAC at 3.6  and 4.5 $\mu$m wavelengths, allowing a distinction between dust and CO+CO$_2$ gas.
We assume CO$_2$ is the dominant source of the observed 4.5 $\mu$m excess flux seen in many comets, based on previous spectroscopic observations including
much higher abundance of CO$_2$ than CO; however,
in some cases including 29P/Schwassmann-Wachmann 1 and C/2006 W3, CO is the more abundant molecule.
The dust grains are ejected from the nucleus largely by H$_2$O sublimation for comets in the inner Solar System, but in many cases the CO$_2$ abundance is
comparable to that of H$_2$O. With higher volatility, CO$_2$ and CO it can dominate at larger distances from the Sun \citep{meech09}.

Our survey results do not show a simple transition between H$_2$O and CO$_2$ as a function of heliocentric distance, suggesting any such transition
occurs at $\sim 2.8$ AU or farther, consistent with an {\it Akari} survey of CO$_2$ and H$_2$O in comets. Instead, we find that some comets 
are `CO$_2$-rich' even at small heliocentric distance. It appears that cometary composition is heterogeneous, with marked differences between comets.
The close-up images of 103P/Hartley 2 from the {\it Deep Impact} spacecraft
 further show that CO$_2$-rich comet to have both types of volatiles, and it is easy to imagine that if the CO$_2$-rich
jets were inactive (as happens for 2P/Encke's primary jet due to its `winter' location on the surface relative to the comet's the pole orientation until close to perihelion),
that same comet would be H$_2$O dominated. Some of the observations we attribute to comets being `CO$_2$-rich' could possibly be due to enhanced CO. 
 A possibility is that CO, for example could be episodic, such as being enhanced during outbursts, and the observations we call `CO$_2$-rich' could be transient CO
enhancements. This hypothesis is not supported by our results because we observed a large number of comets at random times (not triggered
by optical brightness of the comets), and amateur observations did not report significant outbursts. The {\it Akari} results also show CO$_2$ is the dominant
emitter in the wavelength range of our survey \citep{ootsubo11}. Only 3 comets were actually detected in CO emission with {\it Akari}; in addition 
to the two mentioned in the previous paragraph, the third was C/2007 Q3, which we did not classify as `CO$_2$ rich'. The upper limits from
the {\it Akari} survey combined with our results further demonstrate that CO$_2$ is the most likely origin for the enhanced 4.5 $\mu$m brightness in
the comets we classify as `CO$_2$-rich'. Comets with CO$<$CO$_2$ from the {\it Akari} survey are classified as `CO$_2$-rich'; these include 118P and 88P.
Additionally, 81P has 4.5 $\mu$m excess (an intermediate amount that we did not classify as `CO$_2$-rich' but which is evident in a jet), despite having
CO$<0.2\times$CO$_2$.

The CO$_2$ images show several jets and spirals indicative of localized active areas on rotating nuclei. We find significantly different morphologies between
the dust and gas emission in many of the survey comets. 
The morphologies are reminiscent of CN-jets as observed at ground-based telescopes.
In some cases, gas jets are present with little or no dust; while in other cases the jets
are prominent at both wavelengths and dominated by dust. The wide range of morphologies and implied compositional variations testify to
a significant diversity of the nuclear properties and also heterogeneity of the individual nuclei.

{\bf Acknowledgements}

This work is based on observations made with the Spitzer Space
Telescope, which is operated by the Jet Propulsion Laboratory, California
Institute of Technology under a contract with NASA. Support for this work was
provided by NASA through an award issued by JPL/Caltech.

\bibliography{wtrbib}

\begin{thebibliography}{48}
\expandafter\ifx\csname natexlab\endcsname\relax\def\natexlab#1{#1}\fi
\expandafter\ifx\csname url\endcsname\relax
  \def\url#1{\texttt{#1}}\fi
\expandafter\ifx\csname urlprefix\endcsname\relax\def\urlprefix{URL }\fi

\bibitem[{{A'Hearn} et~al.(2011){A'Hearn}, {Belton}, {Delamere}, {Feaga},
  {Hampton}, {Kissel}, {Klaasen}, {McFadden}, {Meech}, {Melosh}, {Schultz},
  {Sunshine}, {Thomas}, {Veverka}, {Wellnitz}, {Yeomans}, {Besse}, {Bodewits},
  {Bowling}, {Carcich}, {Collins}, {Farnham}, {Groussin}, {Hermalyn}, {Kelley},
  {Kelley}, {Li}, {Lindler}, {Lisse}, {McLaughlin}, {Merlin}, {Protopapa},
  {Richardson}, and {Williams}}]{ahearn11epoxi}
{A'Hearn}, M.~F., {Belton}, M.~J.~S., {Delamere}, W.~A., {Feaga}, L.~M.,
  {Hampton}, D., {Kissel}, J., {Klaasen}, K.~P., {McFadden}, L.~A., {Meech},
  K.~J., {Melosh}, H.~J., {Schultz}, P.~H., {Sunshine}, J.~M., {Thomas}, P.~C.,
  {Veverka}, J., {Wellnitz}, D.~D., {Yeomans}, D.~K., {Besse}, S., {Bodewits},
  D., {Bowling}, T.~J., {Carcich}, B.~T., {Collins}, S.~M., {Farnham}, T.~L.,
  {Groussin}, O., {Hermalyn}, B., {Kelley}, M.~S., {Kelley}, M.~S., {Li},
  J.-Y., {Lindler}, D.~J., {Lisse}, C.~M., {McLaughlin}, S.~A., {Merlin}, F.,
  {Protopapa}, S., {Richardson}, J.~E., {Williams}, J.~L., Jun. 2011. {EPOXI at
  Comet Hartley 2}. Science 332, 1396--.

\bibitem[{{A'Hearn} et~al.(1995){A'Hearn}, {Millis}, {Schleicher}, {Osip}, and
  {Birch}}]{ahearn95}
{A'Hearn}, M.~F., {Millis}, R.~L., {Schleicher}, D.~G., {Osip}, D.~J., {Birch},
  P.~V., Dec. 1995. {The ensemble properties of comets: Results from narrowband
  photometry of 85 comets, 1976-1992.} \icarus 118, 223--270.

\bibitem[{{Bockelee-Morvan} et~al.(1990){Bockelee-Morvan}, {Crovisier}, and
  {Gerard}}]{bockelee90}
{Bockelee-Morvan}, D., {Crovisier}, J., {Gerard}, E., Nov. 1990. {Retrieving
  the coma gas expansion velocity in P/Halley, Wilson (1987 VII) and several
  other comets from the 18-cm OH line shapes}. \aap 238, 382--400.

\bibitem[{{Bockel{\'e}e-Morvan} et~al.(2010){Bockel{\'e}e-Morvan}, {Hartogh},
  {Crovisier}, {Vandenbussche}, {Swinyard}, {Biver}, {Lis}, {Jarchow},
  {Moreno}, {Hutsem{\'e}kers}, {Jehin}, {K{\"u}ppers}, {Lara}, {Lellouch},
  {Manfroid}, {de Val-Borro}, {Szutowicz}, {Banaszkiewicz}, {Bensch}, {Blecka},
  {Emprechtinger}, {Encrenaz}, {Fulton}, {Kidger}, {Rengel}, {Waelkens},
  {Bergin}, {Blake}, {Blommaert}, {Cernicharo}, {Decin}, {Encrenaz}, {de
  Graauw}, {Leeks}, {Medvedev}, {Naylor}, {Schieder}, and
  {Thomas}}]{bockelee10}
{Bockel{\'e}e-Morvan}, D., {Hartogh}, P., {Crovisier}, J., {Vandenbussche}, B.,
  {Swinyard}, B.~M., {Biver}, N., {Lis}, D.~C., {Jarchow}, C., {Moreno}, R.,
  {Hutsem{\'e}kers}, D., {Jehin}, E., {K{\"u}ppers}, M., {Lara}, L.~M.,
  {Lellouch}, E., {Manfroid}, J., {de Val-Borro}, M., {Szutowicz}, S.,
  {Banaszkiewicz}, M., {Bensch}, F., {Blecka}, M.~I., {Emprechtinger}, M.,
  {Encrenaz}, T., {Fulton}, T., {Kidger}, M., {Rengel}, M., {Waelkens}, C.,
  {Bergin}, E., {Blake}, G.~A., {Blommaert}, J.~A.~D.~L., {Cernicharo}, J.,
  {Decin}, L., {Encrenaz}, P., {de Graauw}, T., {Leeks}, S., {Medvedev}, A.~S.,
  {Naylor}, D., {Schieder}, R., {Thomas}, N., Jul. 2010. {A study of the
  distant activity of comet C/2006 W3 (Christensen) with Herschel and
  ground-based radio telescopes}. \aap 518, L149.

\bibitem[{{Bockel{\'e}e-Morvan} et~al.(2000){Bockel{\'e}e-Morvan}, {Lis},
  {Wink}, {Despois}, {Crovisier}, {Bachiller}, {Benford}, {Biver}, {Colom},
  {Davies}, {G{\'e}rard}, {Germain}, {Houde}, {Mehringer}, {Moreno}, {Paubert},
  {Phillips}, and {Rauer}}]{bockeleemorvan00}
{Bockel{\'e}e-Morvan}, D., {Lis}, D.~C., {Wink}, J.~E., {Despois}, D.,
  {Crovisier}, J., {Bachiller}, R., {Benford}, D.~J., {Biver}, N., {Colom}, P.,
  {Davies}, J.~K., {G{\'e}rard}, E., {Germain}, B., {Houde}, M., {Mehringer},
  D., {Moreno}, R., {Paubert}, G., {Phillips}, T.~G., {Rauer}, H., Jan. 2000.
  {New molecules found in comet C/1995 O1 (Hale-Bopp). Investigating the link
  between cometary and interstellar material}. \aap 353, 1101--1114.

\bibitem[{{Carter} et~al.(2012){Carter}, {Bodewits}, {Read}, and
  {Immler}}]{carter12}
{Carter}, J.~A., {Bodewits}, D., {Read}, A.~M., {Immler}, S., May 2012.
  {Simultaneous Swift X-ray and UV views of comet C/2007 N3 (Lulin)}. \aap 541,
  A70.

\bibitem[{{Chyba} and {Sagan}(1997)}]{ChybaSagan}
{Chyba}, C.~F., {Sagan}, C., 1997. {Comets as a source of prebiotic organic
  molecules for the early Earth.} In: {P.~J.~Thomas, C.~F.~Chyba, \&
  C.~P.~McKay} (Ed.), Comets and the Origin and Evolution of Life. pp.
  147--173.

\bibitem[{{Cochran} and {Schleicher}(1993)}]{cochran93}
{Cochran}, A.~L., {Schleicher}, D.~G., Sep. 1993. {Observational Constraints on
  the Lifetime of Cometary H$_{2}$O}. Icarus 105, 235--253.

\bibitem[{{Colangeli} et~al.(1999){Colangeli}, {Epifani}, {Brucato},
  {Bussoletti}, {de Sanctis}, {Fulle}, {Mennella}, {Palomba}, {Palumbo}, and
  {Rotundi}}]{colangeli103P}
{Colangeli}, L., {Epifani}, E., {Brucato}, J.~R., {Bussoletti}, E., {de
  Sanctis}, C., {Fulle}, M., {Mennella}, V., {Palomba}, E., {Palumbo}, P.,
  {Rotundi}, A., 1999. Infrared spectral observations of comet {103P/Hartley} 2
  by {ISOPHOT}. {\aap} 343, L87--L90.

\bibitem[{{Colas} et~al.(2010){Colas}, {Manzini}, {Howes}, and
  {Bryssinck}}]{colas10}
{Colas}, F., {Manzini}, F., {Howes}, N.~J., {Bryssinck}, E., Apr. 2010. {Comet
  C/2007 Q3 (Siding Spring)}. \iaucirc 9135, 2.

\bibitem[{{Combi} et~al.(2009){Combi}, {M{\"a}kinen}, {Bertaux}, {Lee}, and
  {Qu{\'e}merais}}]{combi09}
{Combi}, M.~R., {M{\"a}kinen}, J.~T.~T., {Bertaux}, J.-L., {Lee}, Y.,
  {Qu{\'e}merais}, E., Jun. 2009. {Water Production in Comets 2001 Q4 (NEAT)
  and 2002 T7 (LINEAR) Determined from SOHO/SWAN Observations}. \aj 137,
  4734--4743.

\bibitem[{{Cowan} and {Ahearn}(1979)}]{cowanahearn}
{Cowan}, J.~J., {Ahearn}, M.~F., Oct. 1979. {Vaporization of comet nuclei -
  Light curves and life times}. Moon and Planets 21, 155--171.

\bibitem[{{Crifo} and {Rodionov}(1997)}]{crifo97}
{Crifo}, J.~F., {Rodionov}, A.~V., Sep. 1997. {The Dependence of the
  Circumnuclear Coma Structure on the Properties of the Nucleus}. \icarus 129,
  72--93.

\bibitem[{{Crovisier} et~al.(1999){Crovisier}, {Encrenaz}, {Lellouch},
  {Bockel{\'e}e-Morvan}, {Altieri}, {Leech}, {Salama}, {Griffin}, {de Graauw},
  {van Dishoeck}, {Knacke}, and {Brooke}}]{crovisier99hartley2}
{Crovisier}, J., {Encrenaz}, T., {Lellouch}, E., {Bockel{\'e}e-Morvan}, D.,
  {Altieri}, B., {Leech}, K., {Salama}, A., {Griffin}, M.~J., {de Graauw}, T.,
  {van Dishoeck}, E.~F., {Knacke}, R., {Brooke}, T.~Y., Mar. 1999. {ISO
  spectroscopic observations of short-period comets}. In: {P.~Cox \&
  M.~Kessler} (Ed.), The Universe as Seen by ISO. Vol. 427 of ESA Special
  Publication. pp. 161--+.

\bibitem[{Crovisier et~al.(1997)Crovisier, Leech, Bockel\'ee-Morvan, Brooke,
  Hanner, Altieri, Keller, and Lellouch}]{crovisierHaleBopp}
Crovisier, J., Leech, K., Bockel\'ee-Morvan, D., Brooke, T.~Y., Hanner, M.~S.,
  Altieri, B., Keller, H.~U., Lellouch, E., 1997. The spectrum of comet
  {Hale-Bopp} ({C}/1995 {O}1) observed with the {Infrared} {Space}
  {Observatory} at 2.9 {AU} from the {Sun}. Science 275, 1905--1907.

\bibitem[{{Dello Russo} et~al.(1998){Dello Russo}, {DiSanti}, {Mumma},
  {Magee-Sauer}, and {Rettig}}]{dellorusso98}
{Dello Russo}, N., {DiSanti}, M.~A., {Mumma}, M.~J., {Magee-Sauer}, K.,
  {Rettig}, T.~W., Oct. 1998. {Carbonyl Sulfide in Comets C/1996 B2 (Hyakutake)
  and C/1995 O1 (Hale-Bopp): Evidence for an Extended Source in Hale-Bopp}.
  \icarus 135, 377--388.

\bibitem[{{DiSanti} et~al.(2001){DiSanti}, {Mumma}, {Dello Russo}, and
  {Magee-Sauer}}]{disanti01}
{DiSanti}, M.~A., {Mumma}, M.~J., {Dello Russo}, N., {Magee-Sauer}, K., Oct.
  2001. {Carbon Monoxide Production and Excitation in Comet C/1995 O1
  (Hale-Bopp): Isolation of Native and Distributed CO Sources}. \icarus 153,
  361--390.

\bibitem[{{Eberhardt}(1999)}]{eberhardt99}
{Eberhardt}, P., Oct. 1999. {Comet Halley's Gas Composition and Extended
  Sources: Results from the Neutral Mass Spectrometer on Giotto}. \ssr 90,
  45--52.

\bibitem[{{Fazio} et~al.(2004){Fazio}, {Hora}, {Allen}, {Ashby}, {Barmby},
  {Deutsch}, {Huang}, {Kleiner}, {Marengo}, {Megeath}, {Melnick}, {Pahre},
  {Patten}, {Polizotti}, {Smith}, {Taylor}, {Wang}, {Willner}, {Hoffmann},
  {Pipher}, {Forrest}, {McMurty}, {McCreight}, {McKelvey}, {McMurray}, {Koch},
  {Moseley}, {Arendt}, {Mentzell}, {Marx}, {Losch}, {Mayman}, {Eichhorn},
  {Krebs}, {Jhabvala}, {Gezari}, {Fixsen}, {Flores}, {Shakoorzadeh}, {Jungo},
  {Hakun}, {Workman}, {Karpati}, {Kichak}, {Whitley}, {Mann}, {Tollestrup},
  {Eisenhardt}, {Stern}, {Gorjian}, {Bhattacharya}, {Carey}, {Nelson},
  {Glaccum}, {Lacy}, {Lowrance}, {Laine}, {Reach}, {Stauffer}, {Surace},
  {Wilson}, {Wright}, {Hoffman}, {Domingo}, and {Cohen}}]{fazioirac}
{Fazio}, G.~G., {Hora}, J.~L., {Allen}, L.~E., {Ashby}, M.~L.~N., {Barmby}, P.,
  {Deutsch}, L.~K., {Huang}, J.-S., {Kleiner}, S., {Marengo}, M., {Megeath},
  S.~T., {Melnick}, G.~J., {Pahre}, M.~A., {Patten}, B.~M., {Polizotti}, J.,
  {Smith}, H.~A., {Taylor}, R.~S., {Wang}, Z., {Willner}, S.~P., {Hoffmann},
  W.~F., {Pipher}, J.~L., {Forrest}, W.~J., {McMurty}, C.~W., {McCreight},
  C.~R., {McKelvey}, M.~E., {McMurray}, R.~E., {Koch}, D.~G., {Moseley}, S.~H.,
  {Arendt}, R.~G., {Mentzell}, J.~E., {Marx}, C.~T., {Losch}, P., {Mayman}, P.,
  {Eichhorn}, W., {Krebs}, D., {Jhabvala}, M., {Gezari}, D.~Y., {Fixsen},
  D.~J., {Flores}, J., {Shakoorzadeh}, K., {Jungo}, R., {Hakun}, C., {Workman},
  L., {Karpati}, G., {Kichak}, R., {Whitley}, R., {Mann}, S., {Tollestrup},
  E.~V., {Eisenhardt}, P., {Stern}, D., {Gorjian}, V., {Bhattacharya}, B.,
  {Carey}, S., {Nelson}, B.~O., {Glaccum}, W.~J., {Lacy}, M., {Lowrance},
  P.~J., {Laine}, S., {Reach}, W.~T., {Stauffer}, J.~A., {Surace}, J.~A.,
  {Wilson}, G., {Wright}, E.~L., {Hoffman}, A., {Domingo}, G., {Cohen}, M.,
  Sep. 2004. {The Infrared Array Camera (IRAC) for the Spitzer Space
  Telescope}. \apjs 154, 10--17.

\bibitem[{{Feaga} et~al.(2007){Feaga}, {A'Hearn}, {Sunshine}, {Groussin}, and
  {Farnham}}]{feaga07}
{Feaga}, L.~M., {A'Hearn}, M.~F., {Sunshine}, J.~M., {Groussin}, O., {Farnham},
  T.~L., 2007. {Asymmetries in the distribution of H$_{2}$O and CO$_{2}$ in the
  inner coma of Comet 9P/Tempel 1 as observed by Deep Impact}. \icarus 191,
  134--145.

\bibitem[{{Feldman} et~al.(1997){Feldman}, {Festou}, {Tozzi}, and
  {Weaver}}]{feldman97}
{Feldman}, P.~D., {Festou}, M.~C., {Tozzi}, G.~P., {Weaver}, H.~A., Feb. 1997.
  {The CO 2/CO Abundance Ratio in 1P/Halley and Several Other Comets Observed
  by IUE and HST}. \apj 475, 829--+.

\bibitem[{{Fern{\'a}ndez} et~al.(2005){Fern{\'a}ndez}, {Lowry}, {Weissman},
  {Mueller}, {Samarasinha}, {Belton}, and {Meech}}]{fernandez05}
{Fern{\'a}ndez}, Y.~R., {Lowry}, S.~C., {Weissman}, P.~R., {Mueller}, B.~E.~A.,
  {Samarasinha}, N.~H., {Belton}, M.~J.~S., {Meech}, K.~J., May 2005. {New
  near-aphelion light curves of Comet 2P/Encke}. \icarus 175, 194--214.

\bibitem[{{Festou}(1981)}]{festou81}
{Festou}, M.~C., Feb. 1981. {The density distribution of neutral compounds in
  cometary atmospheres. I - Models and equations}. \aap 95, 69--79.

\bibitem[{{Festou}(1999)}]{festou99}
{Festou}, M.~C., Oct. 1999. {On the Existence of Distributed Sources in Comet
  Comae}. \ssr 90, 53--67.

\bibitem[{{Gunnarsson} et~al.(2002){Gunnarsson}, {Rickman}, {Festou},
  {Winnberg}, and {Tancredi}}]{gunnarsson02}
{Gunnarsson}, M., {Rickman}, H., {Festou}, M.~C., {Winnberg}, A., {Tancredi},
  G., Jun. 2002. {An Extended CO Source around Comet 29P/Schwassmann-Wachmann
  1}. \icarus 157, 309--322.

\bibitem[{{Huebner} et~al.(1992){Huebner}, {Keady}, and {Lyon}}]{huebner}
{Huebner}, W.~F., {Keady}, J.~J., {Lyon}, S.~P., Sep. 1992. {Solar photo rates
  for planetary atmospheres and atmospheric pollutants}. \apss 195, 1--289.

\bibitem[{{Kelley} et~al.(2013){Kelley}, {Lindler}, {Bodewits}, {A'Hearn},
  {Lisse}, {Kolokolova}, {Kissel}, and {Hermalyn}}]{kelley13}
{Kelley}, M.~S., {Lindler}, D.~J., {Bodewits}, D., {A'Hearn}, M.~F., {Lisse},
  C.~M., {Kolokolova}, L., {Kissel}, J., {Hermalyn}, B., Feb. 2013. {A
  distribution of large particles in the coma of Comet 103P/Hartley 2}. \icarus
  222, 634--652.

\bibitem[{{Kelley} and {Wooden}(2009)}]{kelleywooden}
{Kelley}, M.~S., {Wooden}, D.~H., Aug. 2009. {The composition of dust in
  Jupiter-family comets inferred from infrared spectroscopy}. \planss 57,
  1133--1145.

\bibitem[{Lamy et~al.(2004)Lamy, Toth, Fernandez, and Weaver}]{LamyNuc}
Lamy, P.~L., Toth, I., Fernandez, Y.~R., Weaver, H.~A., 2004. The sizes,
  shapes, albedos, and colors of cometary nuclei. In: Festou, M.~C., Keller,
  H.~U., Weaver, H.~A. (Eds.), Comets II. Tucson: Univ. of Arizona Press, pp.
  223--264.

\bibitem[{{Lowry} and {Weissman}(2007)}]{lowry07}
{Lowry}, S.~C., {Weissman}, P.~R., May 2007. {Rotation and color properties of
  the nucleus of Comet 2P/Encke}. \icarus 188, 212--223.

\bibitem[{{Meech} et~al.(2011){Meech}, {A'Hearn}, {Adams}, {Bacci}, {Bai},
  {Barrera}, {Battelino}, {Bauer}, {Becklin}, {Bhatt}, {Biver},
  {Bockel{\'e}e-Morvan}, {Bodewits}, {B{\"o}hnhardt}, {Boissier}, {Bonev},
  {Borghini}, {Brucato}, {Bryssinck}, {Buie}, {Canovas}, {Castellano},
  {Charnley}, {Chen}, {Chiang}, {Choi}, {Christian}, {Chuang}, {Cochran},
  {Colom}, {Combi}, {Coulson}, {Crovisier}, {Dello Russo}, {Dennerl}, {DeWahl},
  {DiSanti}, {Facchini}, {Farnham}, {Fern{\'a}ndez}, {Flor{\'e}n}, {Frisk},
  {Fujiyoshi}, {Furusho}, {Fuse}, {Galli}, {Garc{\'{\i}}a-Hern{\'a}ndez},
  {Gersch}, {Getu}, {Gibb}, {Gillon}, {Guido}, {Guillermo}, {Hadamcik},
  {Hainaut}, {Hammel}, {Harker}, {Harmon}, {Harris}, {Hartogh}, {Hashimoto},
  {H{\"a}usler}, {Herter}, {Hjalmarson}, {Holland}, {Honda}, {Hosseini},
  {Howell}, {Howes}, {Hsieh}, {Hsiao}, {Hutsem{\'e}kers}, {Immler}, {Jackson},
  {Jeffers}, {Jehin}, {Jones}, {Ovelar}, {Kaluna}, {Karlsson}, {Kawakita},
  {Keane}, {Keller}, {Kelley}, {Kinoshita}, {Kiselev}, {Kleyna}, {Knight},
  {Kobayashi}, {Kobulnicky}, {Kolokolova}, {Kreiny}, {Kuan}, {K{\"u}ppers},
  {Lacruz}, {Landsman}, {Lara}, {Lecacheux}, {Levasseur-Regourd}, {Li},
  {Licandro}, {Ligustri}, {Lin}, {Lippi}, {Lis}, {Lisse}, {Lovell}, {Lowry},
  {Lu}, {Lundin}, {Magee-Sauer}, {Magain}, {Manfroid}, {Mazzotta Epifani},
  {McKay}, {Melita}, {Mikuz}, {Milam}, {Milani}, {Min}, {Moreno}, {Mueller},
  {Mumma}, {Nicolini}, {Nolan}, {Nordh}, {Nowajewski}, {Odin Team}, {Ootsubo},
  {Paganini}, {Perrella}, {Pittichov{\'a}}, {Prosperi}, {Radeva}, {Reach},
  {Remijan}, {Rengel}, {Riesen}, {Rodenhuis}, {Rodr{\'{\i}}guez}, {Russell},
  {Sahu}, {Samarasinha}, {S{\'a}nchez Caso}, {Sandqvist}, {Sarid}, {Sato},
  {Schleicher}, {Schwieterman}, {Sen}, {Shenoy}, {Shi}, {Shinnaka}, {Skvarc},
  {Snodgrass}, {Sitko}, {Sonnett}, {Sosseini}, {Sostero}, {Sugita}, {Swinyard},
  {Szutowicz}, {Takato}, {Tanga}, {Taylor}, {Tozzi}, {Trabatti},
  {Trigo-Rodr{\'{\i}}guez}, {Tubiana}, {de Val-Borro}, {Vacca},
  {Vandenbussche}, {Vaubaillion}, {Velichko}, {Velichko}, {Vervack},
  {Vidal-Nunez}, {Villanueva}, {Vinante}, {Vincent}, {Wang}, {Wasserman},
  {Watanabe}, {Weaver}, {Weissman}, {Wolk}, {Wooden}, {Woodward}, {Yamaguchi},
  {Yamashita}, {Yanamandra-Fischer}, {Yang}, {Yao}, {Yeomans}, {Zenn}, {Zhao},
  and {Ziffer}}]{meech11hartley}
{Meech}, K.~J., {A'Hearn}, M.~F., {Adams}, J.~A., {Bacci}, P., {Bai}, J.,
  {Barrera}, L., {Battelino}, M., {Bauer}, J.~M., {Becklin}, E., {Bhatt}, B.,
  {Biver}, N., {Bockel{\'e}e-Morvan}, D., {Bodewits}, D., {B{\"o}hnhardt}, H.,
  {Boissier}, J., {Bonev}, B.~P., {Borghini}, W., {Brucato}, J.~R.,
  {Bryssinck}, E., {Buie}, M.~W., {Canovas}, H., {Castellano}, D., {Charnley},
  S.~B., {Chen}, W.~P., {Chiang}, P., {Choi}, Y.-J., {Christian}, D.~J.,
  {Chuang}, Y.-L., {Cochran}, A.~L., {Colom}, P., {Combi}, M.~R., {Coulson},
  I.~M., {Crovisier}, J., {Dello Russo}, N., {Dennerl}, K., {DeWahl}, K.,
  {DiSanti}, M.~A., {Facchini}, M., {Farnham}, T.~L., {Fern{\'a}ndez}, Y.,
  {Flor{\'e}n}, H.~G., {Frisk}, U., {Fujiyoshi}, T., {Furusho}, R., {Fuse}, T.,
  {Galli}, G., {Garc{\'{\i}}a-Hern{\'a}ndez}, D.~A., {Gersch}, A., {Getu}, Z.,
  {Gibb}, E.~L., {Gillon}, M., {Guido}, E., {Guillermo}, R.~A., {Hadamcik}, E.,
  {Hainaut}, O., {Hammel}, H.~B., {Harker}, D.~E., {Harmon}, J.~K., {Harris},
  W.~M., {Hartogh}, P., {Hashimoto}, M., {H{\"a}usler}, B., {Herter}, T.,
  {Hjalmarson}, A., {Holland}, S.~T., {Honda}, M., {Hosseini}, S., {Howell},
  E.~S., {Howes}, N., {Hsieh}, H.~H., {Hsiao}, H.-Y., {Hutsem{\'e}kers}, D.,
  {Immler}, S.~M., {Jackson}, W.~M., {Jeffers}, S.~V., {Jehin}, E., {Jones},
  T.~J., {Ovelar}, M.~d.~J., {Kaluna}, H.~M., {Karlsson}, T., {Kawakita}, H.,
  {Keane}, J.~V., {Keller}, L.~D., {Kelley}, M.~S., {Kinoshita}, D., {Kiselev},
  N.~N., {Kleyna}, J., {Knight}, M.~M., {Kobayashi}, H., {Kobulnicky}, H.~A.,
  {Kolokolova}, L., {Kreiny}, M., {Kuan}, Y.-J., {K{\"u}ppers}, M., {Lacruz},
  J.~M., {Landsman}, W.~B., {Lara}, L.~M., {Lecacheux}, A.,
  {Levasseur-Regourd}, A.~C., {Li}, B., {Licandro}, J., {Ligustri}, R., {Lin},
  Z.-Y., {Lippi}, M., {Lis}, D.~C., {Lisse}, C.~M., {Lovell}, A.~J., {Lowry},
  S.~C., {Lu}, H., {Lundin}, S., {Magee-Sauer}, K., {Magain}, P., {Manfroid},
  J., {Mazzotta Epifani}, E., {McKay}, A., {Melita}, M.~D., {Mikuz}, H.,
  {Milam}, S.~N., {Milani}, G., {Min}, M., {Moreno}, R., {Mueller}, B.~E.~A.,
  {Mumma}, M.~J., {Nicolini}, M., {Nolan}, M.~C., {Nordh}, H.~L., {Nowajewski},
  P.~B., {Odin Team}, {Ootsubo}, T., {Paganini}, L., {Perrella}, C.,
  {Pittichov{\'a}}, J., {Prosperi}, E., {Radeva}, Y.~L., {Reach}, W.~T.,
  {Remijan}, A.~J., {Rengel}, M., {Riesen}, T.~E., {Rodenhuis}, M.,
  {Rodr{\'{\i}}guez}, D.~P., {Russell}, R.~W., {Sahu}, D.~K., {Samarasinha},
  N.~H., {S{\'a}nchez Caso}, A., {Sandqvist}, A., {Sarid}, G., {Sato}, M.,
  {Schleicher}, D.~G., {Schwieterman}, E.~W., {Sen}, A.~K., {Shenoy}, D.,
  {Shi}, J.-C., {Shinnaka}, Y., {Skvarc}, J., {Snodgrass}, C., {Sitko}, M.~L.,
  {Sonnett}, S., {Sosseini}, S., {Sostero}, G., {Sugita}, S., {Swinyard},
  B.~M., {Szutowicz}, S., {Takato}, N., {Tanga}, P., {Taylor}, P.~A., {Tozzi},
  G.-P., {Trabatti}, R., {Trigo-Rodr{\'{\i}}guez}, J.~M., {Tubiana}, C., {de
  Val-Borro}, M., {Vacca}, W., {Vandenbussche}, B., {Vaubaillion}, J.,
  {Velichko}, F.~P., {Velichko}, S.~F., {Vervack}, Jr., R.~J., {Vidal-Nunez},
  M.~J., {Villanueva}, G.~L., {Vinante}, C., {Vincent}, J.-B., {Wang}, M.,
  {Wasserman}, L.~H., {Watanabe}, J., {Weaver}, H.~A., {Weissman}, P.~R.,
  {Wolk}, S., {Wooden}, D.~H., {Woodward}, C.~E., {Yamaguchi}, M., {Yamashita},
  T., {Yanamandra-Fischer}, P.~A., {Yang}, B., {Yao}, J.-S., {Yeomans}, D.~K.,
  {Zenn}, T., {Zhao}, H., {Ziffer}, J.~E., Jun. 2011. {EPOXI: Comet
  103P/Hartley 2 Observations from a Worldwide Campaign}. \apjl 734, L1+.

\bibitem[{{Meech} et~al.(1993){Meech}, {Belton}, {Mueller}, {Dicksion}, and
  {Li}}]{meech93}
{Meech}, K.~J., {Belton}, M.~J.~S., {Mueller}, B.~E.~A., {Dicksion}, M.~W.,
  {Li}, H.~R., Sep. 1993. {Nucleus properties of P/Schwassmann-Wachmann 1}. \aj
  106, 1222--1236.

\bibitem[{{Meech} et~al.(2009){Meech}, {Pittichov{\'a}}, {Bar-Nun}, {Notesco},
  {Laufer}, {Hainaut}, {Lowry}, {Yeomans}, and {Pitts}}]{meech09}
{Meech}, K.~J., {Pittichov{\'a}}, J., {Bar-Nun}, A., {Notesco}, G., {Laufer},
  D., {Hainaut}, O.~R., {Lowry}, S.~C., {Yeomans}, D.~K., {Pitts}, M., Jun.
  2009. {Activity of comets at large heliocentric distances pre-perihelion}.
  Icarus 201, 719--739.

\bibitem[{{Meech} and {Svoren}(2004)}]{meech04}
{Meech}, K.~J., {Svoren}, J., 2004. {Using cometary activity to trace the
  physical and chemical evolution of cometary nuclei}. In: {Festou, M.~C.,
  Keller, H.~U., \& Weaver, H.~A.} (Ed.), Comets II. pp. 317--335.

\bibitem[{{Murakami} et~al.(2007){Murakami}, {Baba}, {Barthel}, {Clements},
  {Cohen}, {Doi}, {Enya}, {Figueredo}, {Fujishiro}, {Fujiwara}, {Fujiwara},
  {Garcia-Lario}, {Goto}, {Hasegawa}, {Hibi}, {Hirao}, {Hiromoto}, {Hong},
  {Imai}, {Ishigaki}, {Ishiguro}, {Ishihara}, {Ita}, {Jeong}, {Jeong},
  {Kaneda}, {Kataza}, {Kawada}, {Kawai}, {Kawamura}, {Kessler}, {Kester},
  {Kii}, {Kim}, {Kim}, {Kobayashi}, {Koo}, {Kwon}, {Lee}, {Lorente}, {Makiuti},
  {Matsuhara}, {Matsumoto}, {Matsuo}, {Matsuura}, {M{\"u}ller}, {Murakami},
  {Nagata}, {Nakagawa}, {Naoi}, {Narita}, {Noda}, {Oh}, {Ohnishi}, {Ohyama},
  {Okada}, {Okuda}, {Oliver}, {Onaka}, {Ootsubo}, {Oyabu}, {Pak}, {Park},
  {Pearson}, {Rowan-Robinson}, {Saito}, {Sakon}, {Salama}, {Sato}, {Savage},
  {Serjeant}, {Shibai}, {Shirahata}, {Sohn}, {Suzuki}, {Takagi}, {Takahashi},
  {Tanab{\'e}}, {Takeuchi}, {Takita}, {Thomson}, {Uemizu}, {Ueno}, {Usui},
  {Verdugo}, {Wada}, {Wang}, {Watabe}, {Watarai}, {White}, {Yamamura},
  {Yamauchi}, and {Yasuda}}]{akarimission}
{Murakami}, H., {Baba}, H., {Barthel}, P., {Clements}, D.~L., {Cohen}, M.,
  {Doi}, Y., {Enya}, K., {Figueredo}, E., {Fujishiro}, N., {Fujiwara}, H.,
  {Fujiwara}, M., {Garcia-Lario}, P., {Goto}, T., {Hasegawa}, S., {Hibi}, Y.,
  {Hirao}, T., {Hiromoto}, N., {Hong}, S.~S., {Imai}, K., {Ishigaki}, M.,
  {Ishiguro}, M., {Ishihara}, D., {Ita}, Y., {Jeong}, W.-S., {Jeong}, K.~S.,
  {Kaneda}, H., {Kataza}, H., {Kawada}, M., {Kawai}, T., {Kawamura}, A.,
  {Kessler}, M.~F., {Kester}, D., {Kii}, T., {Kim}, D.~C., {Kim}, W.,
  {Kobayashi}, H., {Koo}, B.~C., {Kwon}, S.~M., {Lee}, H.~M., {Lorente}, R.,
  {Makiuti}, S., {Matsuhara}, H., {Matsumoto}, T., {Matsuo}, H., {Matsuura},
  S., {M{\"u}ller}, T.~G., {Murakami}, N., {Nagata}, H., {Nakagawa}, T.,
  {Naoi}, T., {Narita}, M., {Noda}, M., {Oh}, S.~H., {Ohnishi}, A., {Ohyama},
  Y., {Okada}, Y., {Okuda}, H., {Oliver}, S., {Onaka}, T., {Ootsubo}, T.,
  {Oyabu}, S., {Pak}, S., {Park}, Y.-S., {Pearson}, C.~P., {Rowan-Robinson},
  M., {Saito}, T., {Sakon}, I., {Salama}, A., {Sato}, S., {Savage}, R.~S.,
  {Serjeant}, S., {Shibai}, H., {Shirahata}, M., {Sohn}, J., {Suzuki}, T.,
  {Takagi}, T., {Takahashi}, H., {Tanab{\'e}}, T., {Takeuchi}, T.~T., {Takita},
  S., {Thomson}, M., {Uemizu}, K., {Ueno}, M., {Usui}, F., {Verdugo}, E.,
  {Wada}, T., {Wang}, L., {Watabe}, T., {Watarai}, H., {White}, G.~J.,
  {Yamamura}, I., {Yamauchi}, C., {Yasuda}, A., Oct. 2007. {The Infrared
  Astronomical Mission AKARI}. \pasj 59, 369.

\bibitem[{{Ootsubo} et~al.(2011){Ootsubo}, {Kawakita}, {Hamada}, {Kobayashi},
  {Yamaguchi}, and {Usui}}]{ootsubo11}
{Ootsubo}, T., {Kawakita}, H., {Hamada}, S., {Kobayashi}, H., {Yamaguchi}, M.,
  {Usui}, F., Oct. 2011. {Survey of CO2 in 18 comets with the Japanese Infrared
  Satellite AKARI}. In: EPSC-DPS Joint Meeting 2011. p. 369.

\bibitem[{{Ootsubo} et~al.(2012){Ootsubo}, {Kawakita}, {Hamada}, {Kobayashi},
  {Yamaguchi}, {Usui}, {Nakagawa}, {Ueno}, {Ishiguro}, {Sekiguchi}, {Watanabe},
  {Sakon}, {Shimonishi}, and {Onaka}}]{ootsubo2012}
{Ootsubo}, T., {Kawakita}, H., {Hamada}, S., {Kobayashi}, H., {Yamaguchi}, M.,
  {Usui}, F., {Nakagawa}, T., {Ueno}, M., {Ishiguro}, M., {Sekiguchi}, T.,
  {Watanabe}, J.-i., {Sakon}, I., {Shimonishi}, T., {Onaka}, T., Jun. 2012.
  {AKARI Near-infrared Spectroscopic Survey for CO$_{2}$ in 18 Comets}. \apj
  752, 15.

\bibitem[{{Pittichov{\'a}} et~al.(2008){Pittichov{\'a}}, {Woodward}, {Kelley},
  and {Reach}}]{pittichova08}
{Pittichov{\'a}}, J., {Woodward}, C.~E., {Kelley}, M.~S., {Reach}, W.~T., Sep.
  2008. {Ground-Based Optical and Spitzer Infrared Imaging Observations of
  Comet 21P/GIACOBINI-ZINNER}. \aj 136, 1127--1136.

\bibitem[{Reach et~al.(2000)Reach, Sykes, Lien, and Davies}]{reachencke00}
Reach, W.~T., Sykes, M.~V., Lien, D., Davies, J.~K., 2000. The formation of
  {Encke} meteoroids and dust trail. Icarus 148, 80--94.

\bibitem[{{Reach} et~al.(2009){Reach}, {Vaubaillon}, {Kelley}, {Lisse}, and
  {Sykes}}]{reach09sw3}
{Reach}, W.~T., {Vaubaillon}, J., {Kelley}, M.~S., {Lisse}, C.~M., {Sykes},
  M.~V., Oct. 2009. {Distribution and properties of fragments and debris from
  the split Comet 73P/Schwassmann-Wachmann 3 as revealed by Spitzer Space
  Telescope}. Icarus 203, 571--588.

\bibitem[{{Rodionov} et~al.(2002){Rodionov}, {Crifo}, {Szeg{\H o}}, {Lagerros},
  and {Fulle}}]{radionov02}
{Rodionov}, A.~V., {Crifo}, J.-F., {Szeg{\H o}}, K., {Lagerros}, J., {Fulle},
  M., Aug. 2002. {An advanced physical model of cometary activity}. \planss 50,
  983--1024.

\bibitem[{{Rubin} et~al.(2011){Rubin}, {Tenishev}, {Combi}, {Hansen},
  {Gombosi}, {Altwegg}, and {Balsiger}}]{rubin11}
{Rubin}, M., {Tenishev}, V.~M., {Combi}, M.~R., {Hansen}, K.~C., {Gombosi},
  T.~I., {Altwegg}, K., {Balsiger}, H., Jun. 2011. {Monte Carlo modeling of
  neutral gas and dust in the coma of Comet 1P/Halley}. \icarus 213, 655--677.

\bibitem[{{Samarasinha} et~al.(2004){Samarasinha}, {Mueller}, {Belton}, and
  {Jorda}}]{comrot04}
{Samarasinha}, N.~H., {Mueller}, B.~E.~A., {Belton}, M.~J.~S., {Jorda}, L.,
  2004. {Rotation of cometary nuclei}. In: {Festou}, M.~C., {Keller}, H.~U.,
  {Weaver}, H.~A. (Eds.), Comets II. pp. 281--299.

\bibitem[{{Santos-Sanz} et~al.(1997){Santos-Sanz}, {Sabalisck}, {Kidger},
  {Licandro}, {Serra-Ricart}, {Bellot Rubio}, {Casas}, {G{\'o}mez},
  {S{\'a}nchez Portero}, and {Osip}}]{santossanz97}
{Santos-Sanz}, P., {Sabalisck}, N., {Kidger}, M.~R., {Licandro}, J.,
  {Serra-Ricart}, M., {Bellot Rubio}, L.~R., {Casas}, R., {G{\'o}mez}, A.,
  {S{\'a}nchez Portero}, J., {Osip}, D., Jul. 1997. {A Comparison Between
  Near-Infrared And Visible Imaging Of The Inner Coma Of Comet Hale-Bopp At
  Perihelion}. Earth Moon and Planets 78, 235--241.

\bibitem[{{Tseng} et~al.(2007){Tseng}, {Bockel{\'e}e-Morvan}, {Crovisier},
  {Colom}, and {Ip}}]{tseng07}
{Tseng}, W.-L., {Bockel{\'e}e-Morvan}, D., {Crovisier}, J., {Colom}, P., {Ip},
  W.-H., May 2007. {Cometary water expansion velocity from OH line shapes}.
  \aap 467, 729--735.

\bibitem[{{Wiegert} and {Tremaine}(1999)}]{wiegerttremaine}
{Wiegert}, P., {Tremaine}, S., Jan. 1999. {The Evolution of Long-Period
  Comets}. \icarus 137, 84--121.

\bibitem[{{Woods} et~al.(1987){Woods}, {Feldman}, and {Dymond}}]{woods87}
{Woods}, T.~N., {Feldman}, P.~D., {Dymond}, K.~F., Nov. 1987. {The Atomic
  Carbon Distribution in the Coma of Comet p/ Halley}. \aap 187, 380--+.

\bibitem[{{Wright} et~al.(2010){Wright}, {Eisenhardt}, {Mainzer}, {Ressler},
  {Cutri}, {Jarrett}, {Kirkpatrick}, {Padgett}, {McMillan}, {Skrutskie},
  {Stanford}, {Cohen}, {Walker}, {Mather}, {Leisawitz}, {Gautier}, {McLean},
  {Benford}, {Lonsdale}, {Blain}, {Mendez}, {Irace}, {Duval}, {Liu}, {Royer},
  {Heinrichsen}, {Howard}, {Shannon}, {Kendall}, {Walsh}, {Larsen}, {Cardon},
  {Schick}, {Schwalm}, {Abid}, {Fabinsky}, {Naes}, and {Tsai}}]{wrightWISE}
{Wright}, E.~L., {Eisenhardt}, P.~R.~M., {Mainzer}, A.~K., {Ressler}, M.~E.,
  {Cutri}, R.~M., {Jarrett}, T., {Kirkpatrick}, J.~D., {Padgett}, D.,
  {McMillan}, R.~S., {Skrutskie}, M., {Stanford}, S.~A., {Cohen}, M., {Walker},
  R.~G., {Mather}, J.~C., {Leisawitz}, D., {Gautier}, III, T.~N., {McLean}, I.,
  {Benford}, D., {Lonsdale}, C.~J., {Blain}, A., {Mendez}, B., {Irace}, W.~R.,
  {Duval}, V., {Liu}, F., {Royer}, D., {Heinrichsen}, I., {Howard}, J.,
  {Shannon}, M., {Kendall}, M., {Walsh}, A.~L., {Larsen}, M., {Cardon}, J.~G.,
  {Schick}, S., {Schwalm}, M., {Abid}, M., {Fabinsky}, B., {Naes}, L., {Tsai},
  C.-W., Dec. 2010. {The Wide-field Infrared Survey Explorer (WISE): Mission
  Description and Initial On-orbit Performance}. \aj 140, 1868--1881.

\end{thebibliography}

\end{document}